\pdfoutput=1
\documentclass[aps,prl,superscriptaddress,floatfix]{revtex4}
\usepackage{epsfig, rotating, graphicx, color,tabularx,enumerate}

\newcommand{\ie}{{\it i.e.\ }}

\begin{document}

\title{Role models for complex networks}

\author{J\"org Reichardt}
\affiliation{Institute for Theoretical Physics, University of W\"urzburg, 
97074 W\"urzburg, Germany}
\author{Douglas R. White}
\affiliation{Department of Anthropology, Institute of Mathematical Behavioral Sciences, School of Social Sciences, University of California, Irvine, USA}

\begin{abstract} 
\noindent We present a framework for automatically decomposing (``block-modeling'') the functional classes of agents within a complex network. These classes are represented by the nodes of an image graph (``block model'') depicting the main patterns of connectivity and thus functional roles in the network. Using a first principles approach, we derive a measure for the fit of a network to any given image graph allowing objective hypothesis testing. From the properties of an optimal fit, we derive how to find the best fitting image graph directly from the network and present a criterion to avoid overfitting. The method can handle both two-mode and one-mode data, directed and undirected as well as weighted networks and allows for different types of links to be dealt with simultaneously. It is non-parametric and computationally efficient. The concepts of structural equivalence and modularity are found as special cases of our approach. We apply our method to the world trade network and analyze the roles individual countries play in the global economy. 
\end{abstract}

\maketitle

\noindent The analysis of the structural and statistical properties of complex networks is one of the major foci of complex systems science at the moment. In the context of social networks, the idea that the pattern of connectivity is related to the function of an agent in the  network is known as playing a ``role'' or assuming a ``position" \cite{BorgattiPositions, WassermanFaust}. Complex systems science has endorsed this idea. By investigating data from a wide range of sources encompassing the life sciences, ecology, information and social sciences as well as economics, researchers have shown that this intimate relation between topology and function indeed exists \cite{NewmanReview,BarabasiLarge,BarabasiCentrality,BarabasiNetBio}. Hence, understanding the topology of a network is a first step in understanding the function and eventually the dynamics of any network.

Of particular interest in recent years has been the possible decomposition of networks into largely \textit{independent} sub-parts called ``communities'' \cite{GirvanNewman}. As a community, one generally understands a group of nodes that is densely connected internally but sparsely connected externally. To sociologists the concept of community is known as ``cohesive subgroup'' \cite{WassermanFaust}, but the recent advancements have generalized its applicability much beyond sociology \cite{Wilkinson,PallaNature,RBJSTAT}. However, the sociological concept of roles in networks is much wider than mere cohesiveness as it specifically focuses on the \textit{inter-dependencies} between groups of nodes. Community structure, emphasizing the absence of dependencies between groups of nodes is only one special case.  

\begin{figure}[b]
\includegraphics[height=3cm]{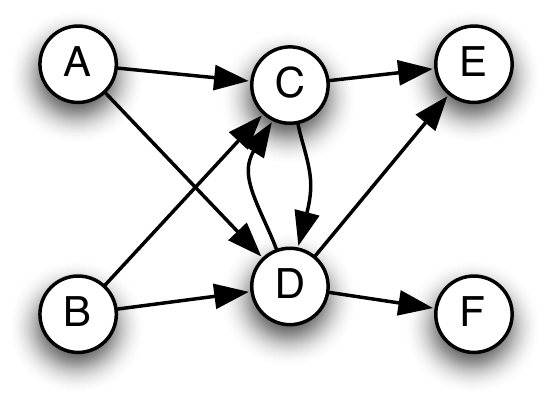}\hspace{0.5cm}
\includegraphics[width=4.5cm]{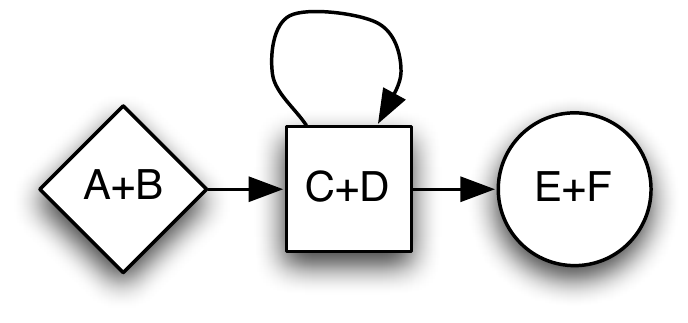}
\caption{Example network illustrating structural and regular equivalence. Nodes $A$ and $B$ have the same neighbors and are thus structurally equivalent and regularly equivalent. Nodes $C$ though $F$ form four different classes of structural equivalence but can be grouped into only two classes of regular equivalence as shown in the image graph or role model on the right.}
\label{Equival}
\end{figure}

The nodes in a network may be grouped into equivalence classes according to the role they play. Two basic concepts have been developed to formalize the assignments of roles individuals play in social networks: structural and regular equivalence. Both are illustrated in Figure 1. 
Two nodes are called structurally equivalent if they have the exact same neighbors \cite{LorrainWhite} . In Figure 1, 
only nodes $A$ and $B$ are structurally equivalent while all other nodes are structurally equivalent only to themselves. To relax this very strict criterion, regular equivalence was introduced \cite{WhiteReitz,EverettRegEq}. Two nodes are regularly equivalent if they are connected in the same way to equivalent others. Clearly, all nodes which are structurally equivalent must also be regularly equivalent, but not vice versa. The seemingly circular definition of regular equivalence is most easily understood in the following way: every class of regularly equivalent nodes is represented by a single node in an ``image graph''. The nodes in the image graph are connected (disconnected), if connections between nodes in the respective classes exist (are absent) in the original network. In Figure 1, 
nodes $A$ and $B$, $C$ and $D$ as well as $E$ and $F$ form three classes of regular equivalence. If the network in Figure 1 
represents the trade interactions on a market, we may interpret these 3 classes as producers, retailers and consumers, respectively. Producers sell to retailers, while retailers may sell to other retailers and consumers, which in turn only buy from retailers.  The image graph (also  ``block-'' or ``role model'') hence gives a bird's-eye view of the network by concentrating on the roles, \ie the functions, only. Note that no two nodes in the image graph may be structurally equivalent, otherwise the image graph is redundant.

Regular equivalence, though a looser concept than structural equivalence, is still very strict as it requires the nodes to play their roles \emph{exactly}, \ie each node must have at least one of the connections required and may not have any connection forbidden by the role model. In Figure 1, 
the link between $D$ and $E$ may be removed without changing the image graph, but an additional link from $A$ to $E$ would change the role model completely. Clearly, this is unsatisfactory in situations where the data is noisy or only approximate role models are desired for a very large data set.

Instead of requiring exact fit of every single node to the role model,
we require the fit of the network as a whole to the role model to be as
good as possible, such that perfect fit corresponds again to regular or
structural equivalence. 
Instead of requiring exact fit of every single node to the role model, we require the fit of the network as a whole to the role model to be as good as possible, such that perfect fit corresponds again to regular or structural equivalence, as set by an appropriate error function.

\begin{figure*}[t]
\hspace{-3mm}
\begin{tabular}{cccccccc}
a) \fbox{\includegraphics[height=1.75cm]{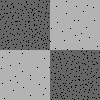}} &
\includegraphics[height=1.75cm]{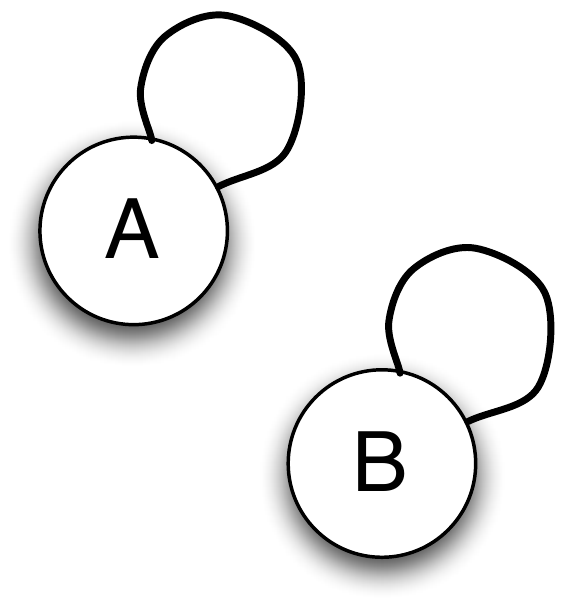} &
b) \fbox{\includegraphics[height=1.75cm]{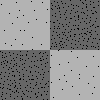}} &
\includegraphics[height=1.5cm]{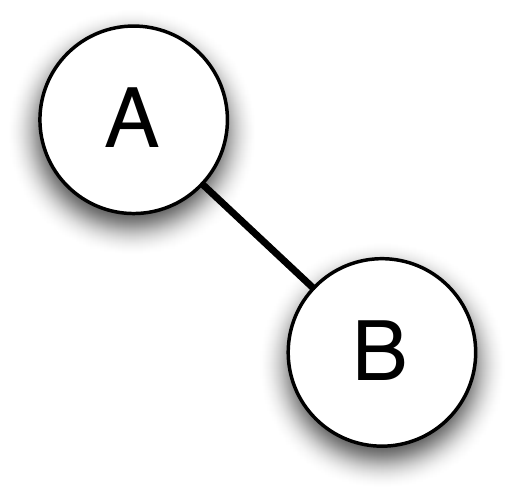} &
c) \fbox{\includegraphics[height=1.75cm]{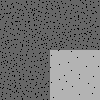}} &
\includegraphics[height=1.75cm]{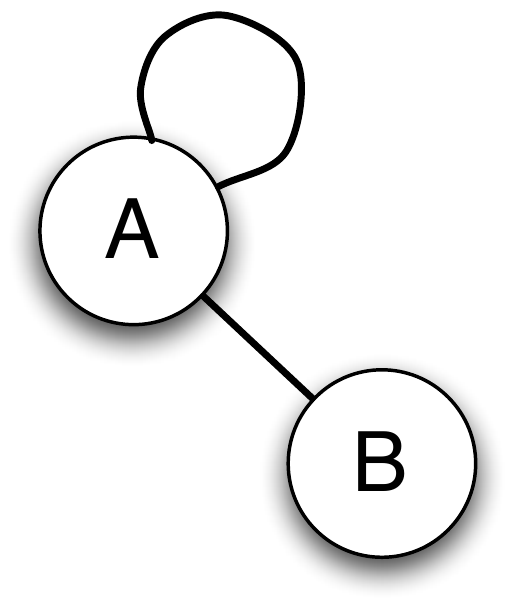} & & \\
d) \fbox{\includegraphics[height=1.75cm]{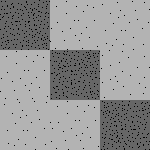}} &
\includegraphics[height=1.75cm]{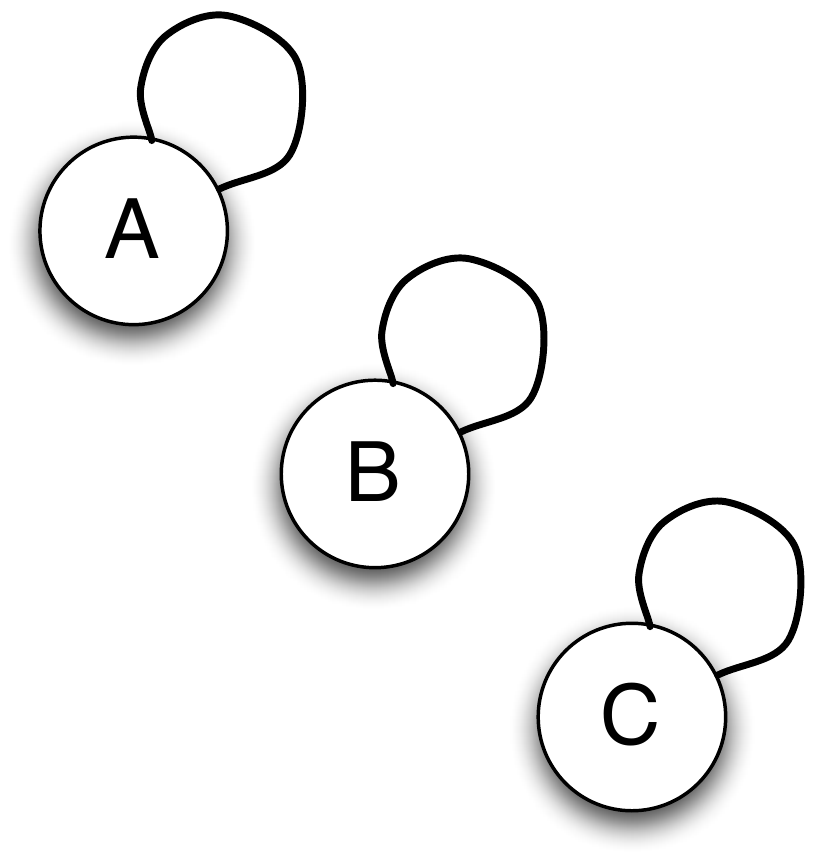} &
e) \fbox{\includegraphics[height=1.75cm]{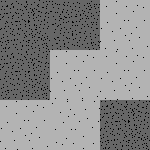}} &
\includegraphics[height=1.75cm]{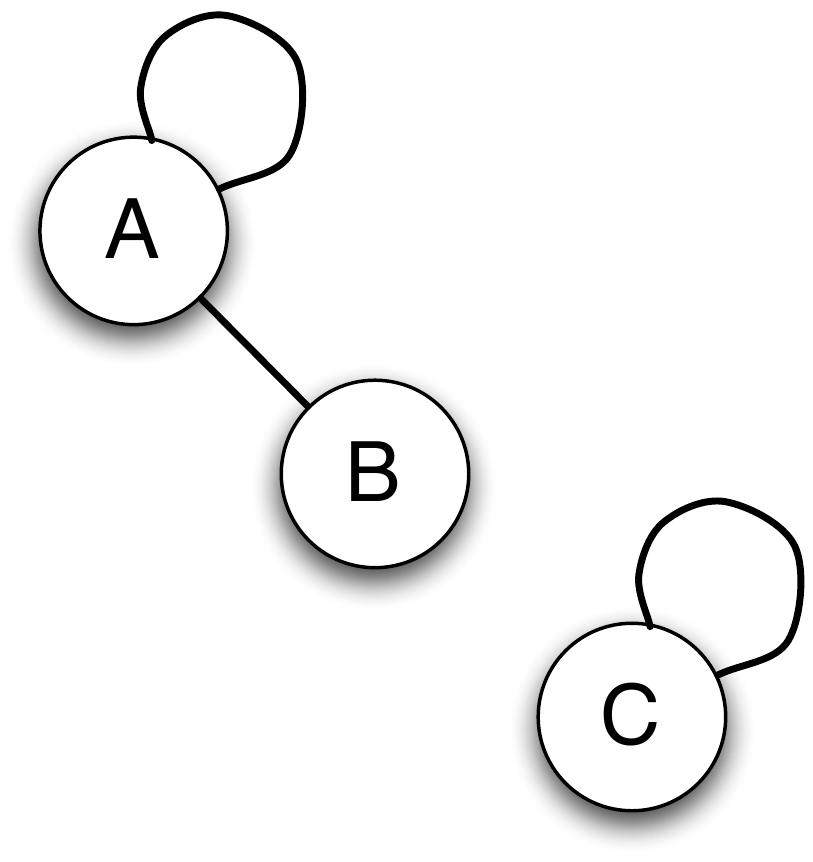} &
f) \fbox{\includegraphics[height=1.75cm]{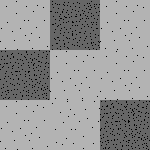}} &
\includegraphics[height=1.75cm]{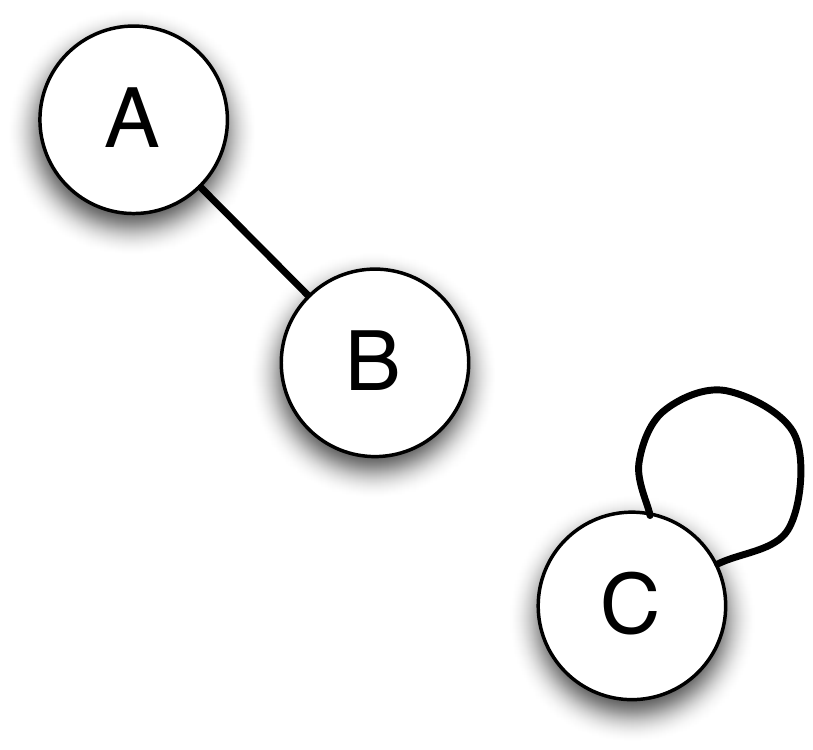} &
g) \fbox{\includegraphics[height=1.75cm]{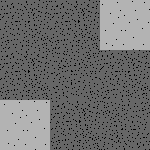}} &
\includegraphics[height=1.75cm]{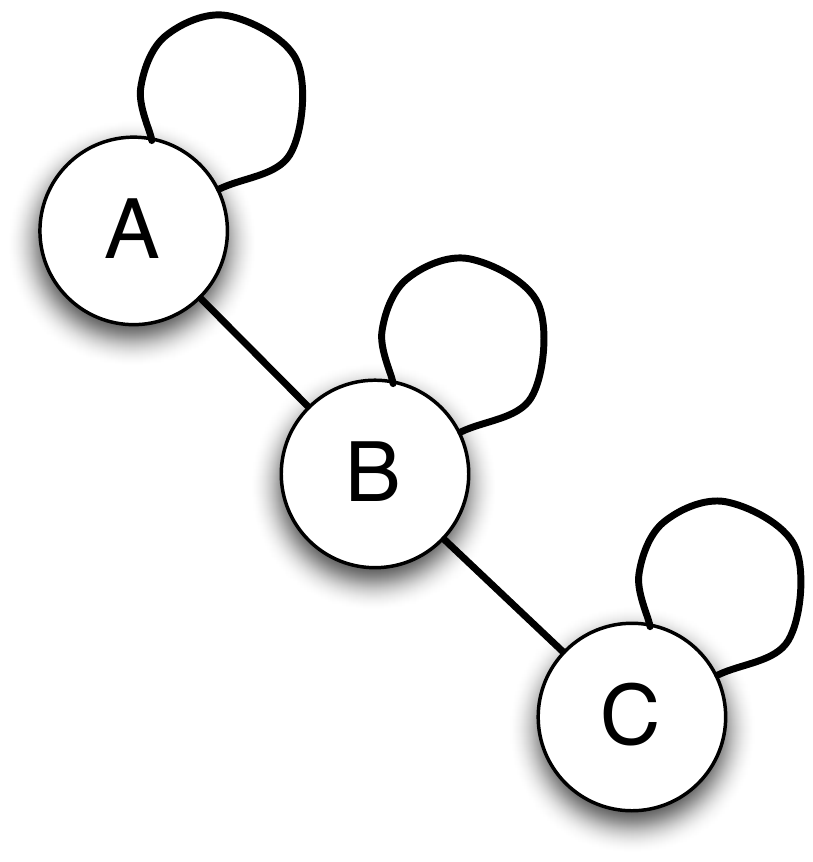} \\
h) \fbox{\includegraphics[height=1.75cm]{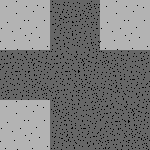}} &
\includegraphics[height=1.75cm]{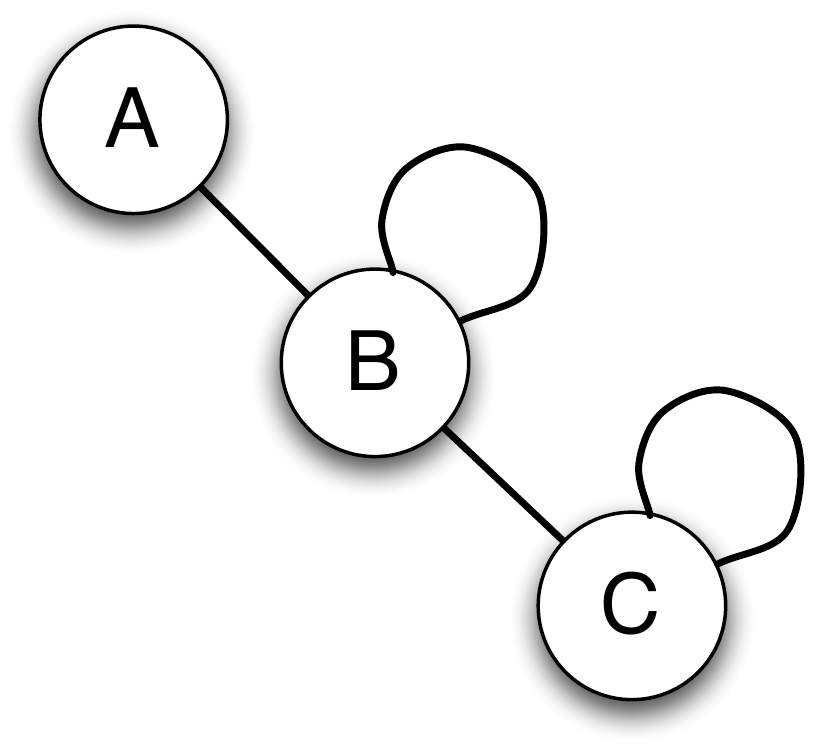} &
i) \fbox{\includegraphics[height=1.75cm]{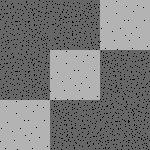}} &
\includegraphics[height=1.75cm]{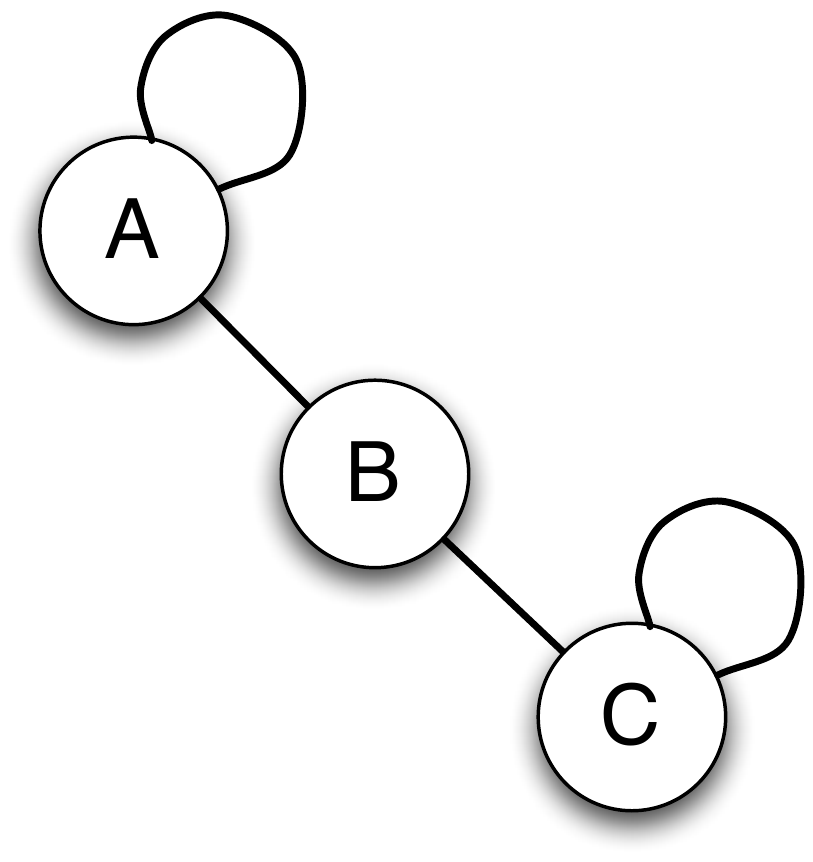} &
j) \fbox{\includegraphics[height=1.75cm]{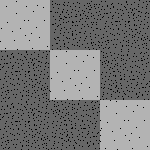}} &
\includegraphics[height=1.75cm]{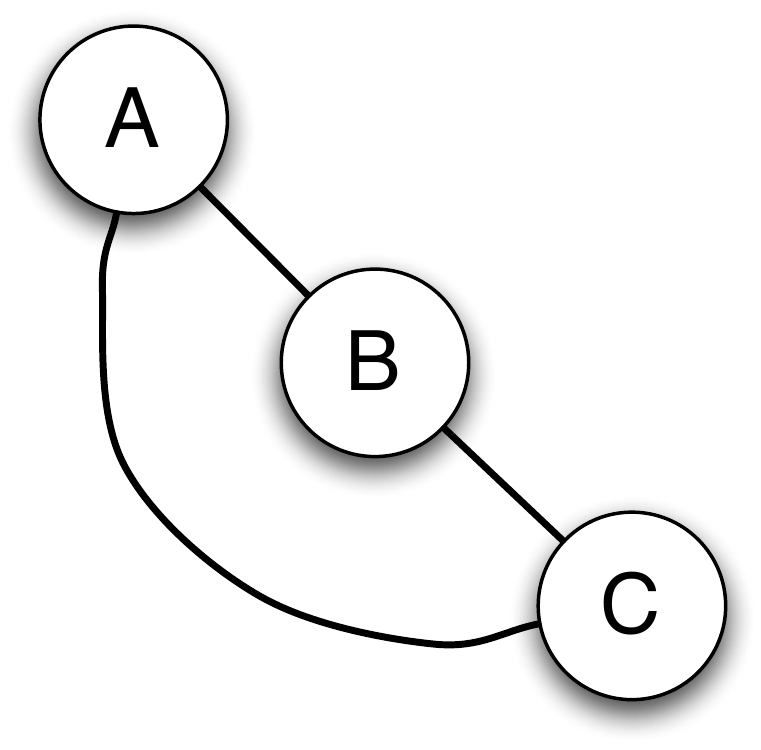} &
k) \fbox{\includegraphics[height=1.75cm]{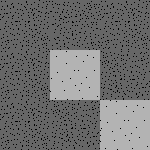}} &
\includegraphics[height=1.75cm]{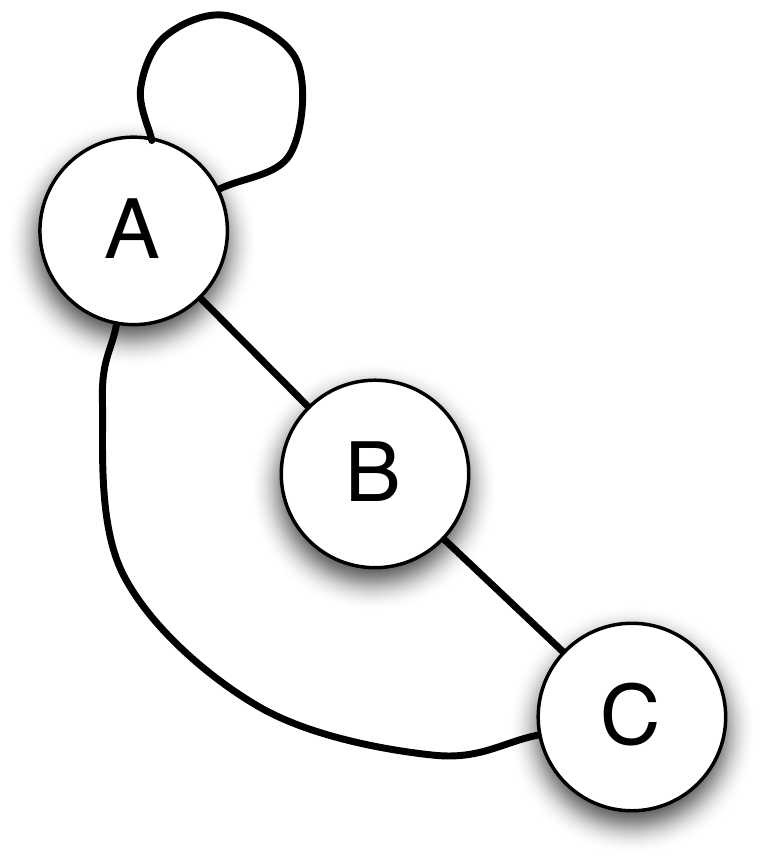}\\
\end{tabular}
\caption{Example adjacency matrices and corresponding image graphs with two and three roles. Nodes with the same pattern of connectivity appear as blocks in the adjacency matrix and are represented by a single node in the image graph. Background shading of matrices reflects link density in blocks. We show only those three role models which are not isomorphic and which cannot be reduced to a block model of two roles only. The two-role-models can be understood as a) modular structure, where nodes connect primarily to nodes of the same role, b) bipartition, with connections primarily between nodes of different type and c) a core-periphery structure with nodes of type A (the core) connecting preferentially among themselves and to nodes of type B (the periphery). The three role models can be seen as combinations of these three basic structures plus the possibility of having intermediates.}
\label{BlockModels}
\end{figure*}

We approach the problem in the following way: First, we assume a given image graph and assignment of roles to nodes. We derive a quality function $\mathcal{Q}^B$ as an objective measure of fit between the image and the network under this assignment of roles. We then consider that assignment of roles which maximizes this quality function. The higher $\mathcal{Q}^B$, the better the given image graph can describe the connection structure of the original network. The concepts of modularity introduced by Newman \cite{NewmanModularity} and structural equivalence are found as special cases for particular image graphs. We then consider the general properties of an assignment of nodes into roles which yields the highest $\mathcal{Q}^B$ across \emph{all possible} given image graphs with a certain number of roles. This suggests a transformation of the quality function enabling us to find this optimal assignment of nodes into roles directly from the network and read off the image graph afterwards. The block models we find are characterized by a maximum deviation (both, positive and negative) of the link weight that meets in a given block from the expectations based on the paired row/column totals that meet in this block. Further, we give a criterion for the selection of the optimal number of roles to avoid over-fitting. Finally, we will apply this technique to detect the roles individual countries play in the global trade network.


\section{Fitting a network to a given image graph}
Suppose we are given a hypothetical image graph with $q$ roles in form of its $q\times q$ adjacency matrix $B_{rs}$. For any assignment of roles $\sigma_i\in \{1,..,q\}$ to the nodes $i$ of a network with $N$ nodes and $M$ edges represented by its adjacency matrix $A_{ij}$, we measure the quality of the fit to the image graph as:
\begin{equation}
\mathcal{Q}^B(\{\sigma\})=\frac{1}{M}\left(\sum_{i\neq j}a_{ij}A_{ij}B_{\sigma_i\sigma_j}+b_{ij}(1-A_{ij})(1-B_{\sigma_i\sigma_j})\right).
\label{Qfunction}
\end{equation}
That is, we reward the matching to the image graph of an edge ($A_{ij}=1$) going from node $i$ to node $j$ with some contribution $a_{ij}$, if links going from nodes of type $\sigma_i$ to nodes of type $\sigma_j$ are allowed, \ie $B_{\sigma_i\sigma_j}=1$. Also, we reward a missing edge ($A_{ij}=0$) matching one the image graph  with some contribution $b_{ij}$, if such an edges is forbidden ($B_{\sigma_i\sigma_j}=0$). When the rewards allow missing edges in the network matching to edges in the image graph, the optimal fit will recover regular equivalence. We do not do so in this paper, and reserve that analysis for a comparative treatment of these two options.

A configuration $\{\sigma\}$ of roles $\sigma_i$ which maximizes (\ref{Qfunction}) constitutes an optimal fit of the network to a given image graph. We can further simplify (\ref{Qfunction}) to 
\begin{equation}
\mathcal{Q}^B(\{\sigma\})=\frac{1}{M}\sum_{i\neq j}\left((a_{ij}+b_{ij})A_{ij}-b_{ij}\right)B_{\sigma_i\sigma_j},
\label{Qfunction2}
\end{equation}
droppeding all terms not depending on $\{\sigma\}$. 

As an example, Figure \ref{BlockModels} shows a few adjacency matrices of undirected networks and the corresponding image graphs with two and three roles. The adjacency matrices are ordered such that rows (columns) corresponding to nodes in the same role are next to each other. Due to the similar connection pattern of nodes of the same role, blocks of high link density appear in the adjacency matrices. These make the term ``block modeling'' \cite{DoreianBook} for grouping nodes according to similar connection patterns intuitive and we will differentiate between a block model as the adjacency matrix in a particular order on one side and its image graph on the other. Note that only $a)$ and $d)$ represent community structures, but the spectrum of possible topologies is much, much wider.  

By introducing the abbreviations $e_{rs}=1/M\sum_{i\neq j}(a_{ij}+b_{ij})A_{ij}\delta_{\sigma_i,r}\delta_{\sigma_j,s}$ and $[e_{rs}]=1/M\sum_{i\neq j}b_{ij}\delta_{\sigma_i,r}\delta_{\sigma_j,s}$ we can write (\ref{Qfunction2}) in a very convenient form:
\begin{eqnarray}
\mathcal{Q}^B(\{\sigma\}) & = &\sum_{r,s}^q(e_{rs}-[e_{rs}])B_{rs}\label{QfunctionSimple1}\\
& = & -\sum_{r,s}^q(e_{rs}-[e_{rs}])(1-B_{rs})+C.
\label{QfunctionSimple2}
\end{eqnarray}
Here,  the sums run over role indices instead of node indices and we have a constant term $C=\sum_{r,s}^q(e_{rs}-[e_{rs}])$. Note the equivalence of counting matches/mismatches to allowed links ($B_{rs}=1$) and forbidden links ($B_{rs}=0$).

Since generally there are not as many edges in a network as there are missing ones, we'd like to balance the contribution of present and absent edges $a_{ij}$ and $b_{ij}$ to $\mathcal{Q}^B$. We want $\sum_{i\neq j}a_{ij}A_{ij}=\sum_{i\neq j}b_{ij}(1-A_{ij})$, which also makes the constant $C$ in the above equation zero. To achieve this, a sensible choice is $a_{ij}=1-p_{ij}$ and $b_{ij}=p_{ij}$ provided that $\sum_{i\neq j}A_{ij}=\sum_{i\neq j}p_{ij}$. Other choices are possible \cite{RBPRL,RBPRE}. Then one may interpret $p_{ij}$ as the probability for the nodes $i$ and $j$ being connected. This choice of $a_{ij}$ and $b_{ij}$ allows to interpret $e_{rs}$ as the fraction of edges connecting nodes in groups $r$ and $s$ and $[e_{rs}]$ as the expectated fraction of edges running between $r$ to $s$. Our choice of weights aims at optimizing for structural equivalence for the present analysis. Using other settings of $a_{ij}$ and $b_{ij}$ we can tune the quality function to optimize for regular equivalence. 

The simplest choice for $p_{ij}$ is $p_{ij}=p$ which is a suitable choice for undirected networks with a Poissonian degree distribution. A more refined choice adequate to a broader range of degree distributions and especially for directed networks is $p_{ij}=k_i^{out}k_j^{in}/M$ which makes $[e_{rs}]=K_r^{out}K_s^{in}/M^2$, where $k_i^{in/out}$ is the in/out-degree of node $i$. The sum of in/out-degrees of all nodes in role $s$ is denoted by $K_s^{in/out}$. Also this has the nice property that  $[e_{rs}]=a_ra_s$ with $a_s=\sum_{r}e_{rs}$, \ie the expectation value is calculated as the product of the marginals of $e_{rs}$. For the remainder of this paper, we will use this choice of weights. Note that using an image graph with self-links only ($B_{rs}=\delta_{rs}$) with these weights, we recover the Newman modularity \cite{NewmanModularity}.

With this quality function at hand, we can find the assignment of roles to nodes simply by optimizing it in order to maximize $\mathcal{Q}^B$. The function (\ref{Qfunction2}) is computationally easy to implement for a given image graph. The difference in $\mathcal{Q}^B$ for a change of node $i$ from role $\alpha$ to role $\phi$ is:
\begin{samepage}
\begin{eqnarray*}
\Delta \mathcal{Q}^B(\sigma_i=\alpha\to\phi) & = &\frac{1}{M}\sum_s(B_{\phi s}-B_{\alpha s})(k^{out}_{i\to s}-[k^{out}_{i \to s}])\\ & + &\frac{1}{M}\sum_r(B_{r\phi}-B_{r\alpha})(k^{in}_{r\to i}-[k^{in}_{r\to i}]).
\label{DeltaQ}
\end{eqnarray*}
\end{samepage}
Here $k^{out}_{i\to s}=\sum_{j\neq i}(a_{ij}+b_{ij})A_{ij}\delta_{\sigma_j,s}$ denotes the number of links node $i$ has to nodes in role $s$  and $[k^{out}_{i\to s}]=\sum_{j\neq i} b_{ij}\delta_{\sigma_j,s}$ denotes the respective expectation value. For undirected networks, the two contributions of incoming and outgoing links are of course equal. Hence, a local updating scheme needs to assess the $k_i$ neighbors of node $i$ and then to determine which of the $q$ roles is best for this node, which takes $\mathcal{O}(q^2)$ operations. Thus a local update needs $\mathcal{O}(\langle k\rangle +q^2)$ operations and can be implemented efficiently on sparse graphs as long as the number of roles is much smaller than the number of nodes in the network. Local search heuristics capable of escaping local optima such as simulated annealing can then be used to find the desired globally optimal assignment of roles to nodes. Naturally, the optimal assignment of roles to nodes is characterized by  $\Delta \mathcal{Q}(\sigma_i=\alpha\to\phi\neq\alpha)\leq 0$, \ie every node assumes its best-fitting role, provided all other nodes do not change.  

So far, we have dealt with directed, unweighed one mode networks. Weighted networks \cite{ZibernaWeighted} can be dealt with by considering a weighted adjacency matrix and setting $k_i=\sum_{j\neq i}A_{ij}$. Two mode data \cite{DoreianTwoMode} can be seen as directed networks with one part of the nodes having only outgoing links and the other part of the nodes having only incoming links.


\section{The optimal fit between network and image graph}
Let us first consider the maximum achievable value of $\mathcal{Q}^B$ for any image graph with any number of roles. From (\ref{Qfunction2}) we see that every allowed edge ($A_{ij}=1$) in the network contributes $a_{ij}$ and every missing edge ($A_{ij}=0$) that would be allowed contributes $-b_{ij}$ to $\mathcal{Q}^B$. The maximum of $\mathcal{Q}^B$ is thus achieved when every edge in the network is allowed and there are no missing edges that would be allowed. The minimal image graph $B_{rs}$ which can achieve this is simply that which depicts the connectivities of the classes of structural equivalence in the network. This makes the maximum sensible number of roles in the image graph $q_{max}$ the number of structural equivalence classes. We can calculate $\mathcal{Q}^B_{max}$ even without knowledge of the structural equivalence classes simply by replacing $B_{rs}$ by $A_{ij}$ in (\ref{Qfunction2})
\begin{equation}
\mathcal{Q}_{max} = \frac{1}{M}\sum_{ij}a_{ij}A_{ij}=\frac{1}{M}\sum_{i\neq j}\left(1-\frac{k^{out}_ik^{in}_j}{M}\right)A_{ij},
\label{Qmax}
\end{equation} 
where we have set $a_{ij}=1-p_{ij}$ with our preferred form of $p_{ij}$.

Let us now consider the properties of an image graph with $q$ roles and a corresponding assignment of roles to nodes which achieve the highest $\mathcal{Q}^B$ across all image graphs with the same number of roles.  From (\ref{QfunctionSimple1}) we see immediately that $\mathcal{Q}^B$ is maximal when every addend $(e_{rs}-[e_{rs}])B_{rs}$ is maximized. If $B_{rs}=1$ then $(e_{rs}-[e_{rs}])$ cannot be negative. Likewise, we see from  (\ref{QfunctionSimple2})  that if $B_{rs}=0$ then $(e_{rs}-[e_{rs}])$ cannot be positive. This means that for the best fitting image graph, we have more links than expected between nodes in roles connected in the image graph.  Further, we have less links than expected between nodes in roles disconnected in the image graph. These two observations are in fact equivalent due to the equivalence of (\ref{QfunctionSimple1}) and (\ref{QfunctionSimple2}).


\section{Deriving the best block model from the data}
A comparison of the optimal $\mathcal{Q}^B$ across all possible image graphs is impractical as their number grows exponentially fast with the number of roles $q$. Essentially all graphs with $q$ nodes (not counting isomorphisms) would need to be considered. Following the discussion in the last section, we
will show how the best image graph can be found in a single step. 

Recall that the best possible fit of the network to an image graph is characterized by $(e_{rs}-[e_{rs}])\geq0$ for all allowed links $B_{rs}=1$ and by $(e_{rs}-[e_{rs}])\leq0$ for all forbidden links $B_{rs}=0$. This suggests a simple way to eliminate the need for a given image graph by considering the following quality function
\begin{equation}
\mathcal{Q}^*(\{\sigma\})=\frac{1}{2}\sum_{r,s}^q\|e_{rs}-[e_{rs}]\|.
\label{QwoBM}
\end{equation}
The factor $1/2$ enters to make the scores of $\mathcal{Q}^B$ and $\mathcal{Q}^*$ comparable. 
From the assignment of roles that maximizes (\ref{QwoBM}), we can read off the image graph simply by setting $B_{rs}=1$,   if $(e_{rs}-[e_{rs}])>0$ and $B_{rs}=0$,  if $(e_{rs}-[e_{rs}])\leq 0$.
The function (\ref{QwoBM}) is steadily increasing with the number of possible roles $q$ until it reaches its maximum value $Q_{max}$ when $q$ equals the number of structural equivalence classes in the network. For  $q$ roles, this allows to compare $Q^*(q)/Q_{max}$ for the actual data and a randomized version and to use this comparison as a basis for the selection of the optimal number of roles in the image graph in order to avoid over-fitting of the data. 

A comparison of the image graphs and role assignments found independently for different numbers of roles then also allows for the detection of possibly hierarchical or overlapping organization of the role structure in the network.


\section{Role decomposition of world trade patterns}
As an example application we investigate a data set for the year 2000 from the United Nations commodity trade data base \cite{Mahutga}. Independent research \cite{SmithNemeth,SmithWhite} has shown that the 55 commodities that make up the bulk of world trade, when factor analyzed, form five major groups, and that commodities are highly correlated within each group. 
\begin{figure*}[t]
\begin{tabular}{ccccc}
\multicolumn{5}{c}{
\begin{minipage}[b][5cm][t]{7.2cm}
	\begin{minipage}[b][3.6cm][t]{4.9cm}
		\includegraphics[height=3.5cm]{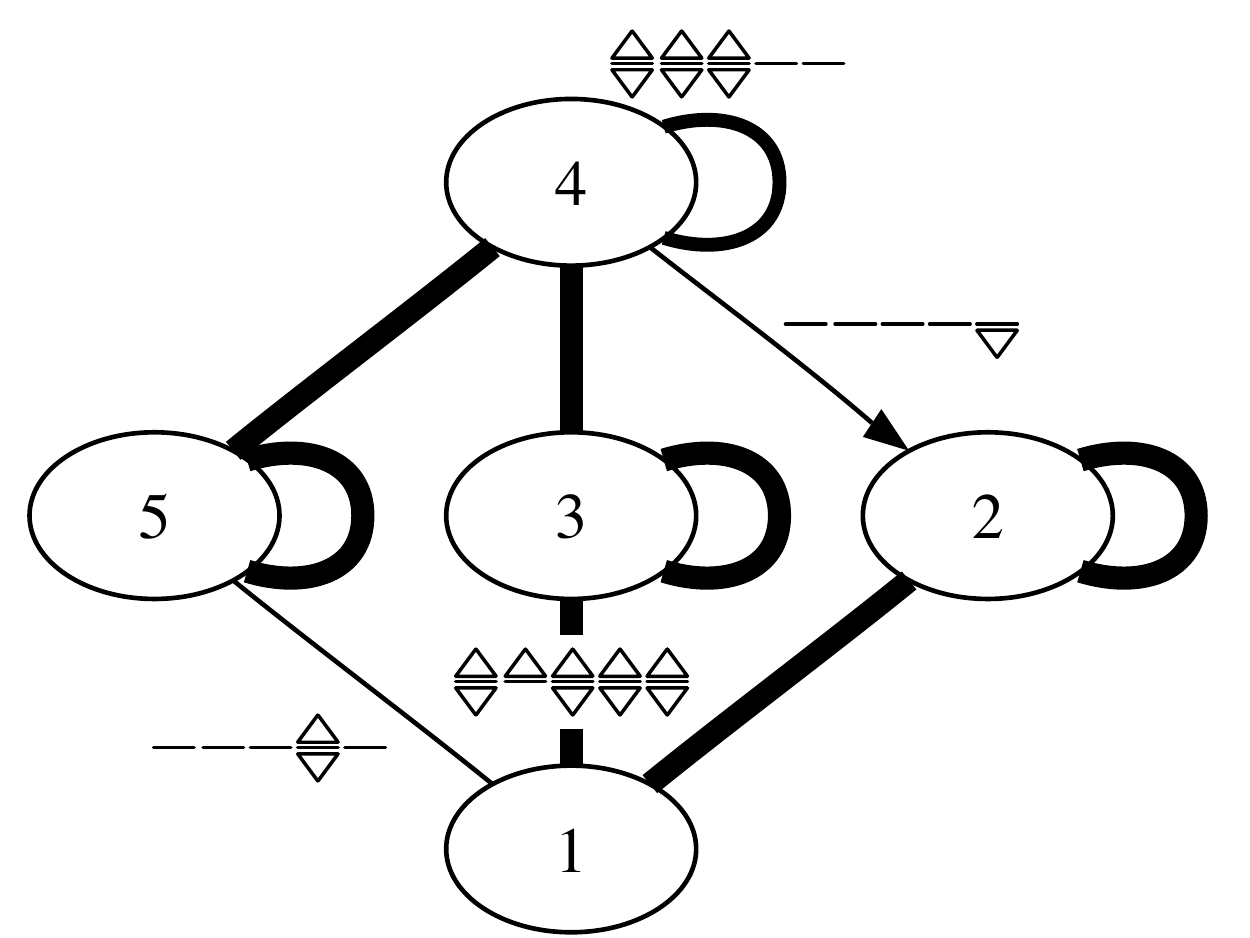}
	\end{minipage}
	\begin{minipage}[b][3.6cm][t]{2.2cm}\begin{footnotesize}\begin{flushleft}
		1: Ctrl. Europe\\
		2: East. Europe, North Afr.\\
		3: Africa, Poly., Mid. East\\
		4: North Am., Japan, SE Asia\\
		5: Middle and South America
		\end{flushleft}\end{footnotesize}
	\end{minipage}\\
	\begin{minipage}[b][1.3cm]{7.cm}
	\hspace{1.6cm}\includegraphics[height=1.2cm]{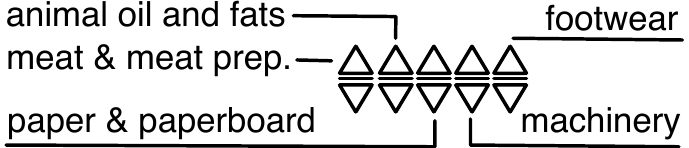}
	\end{minipage}
\end{minipage}
\includegraphics[height=5cm]{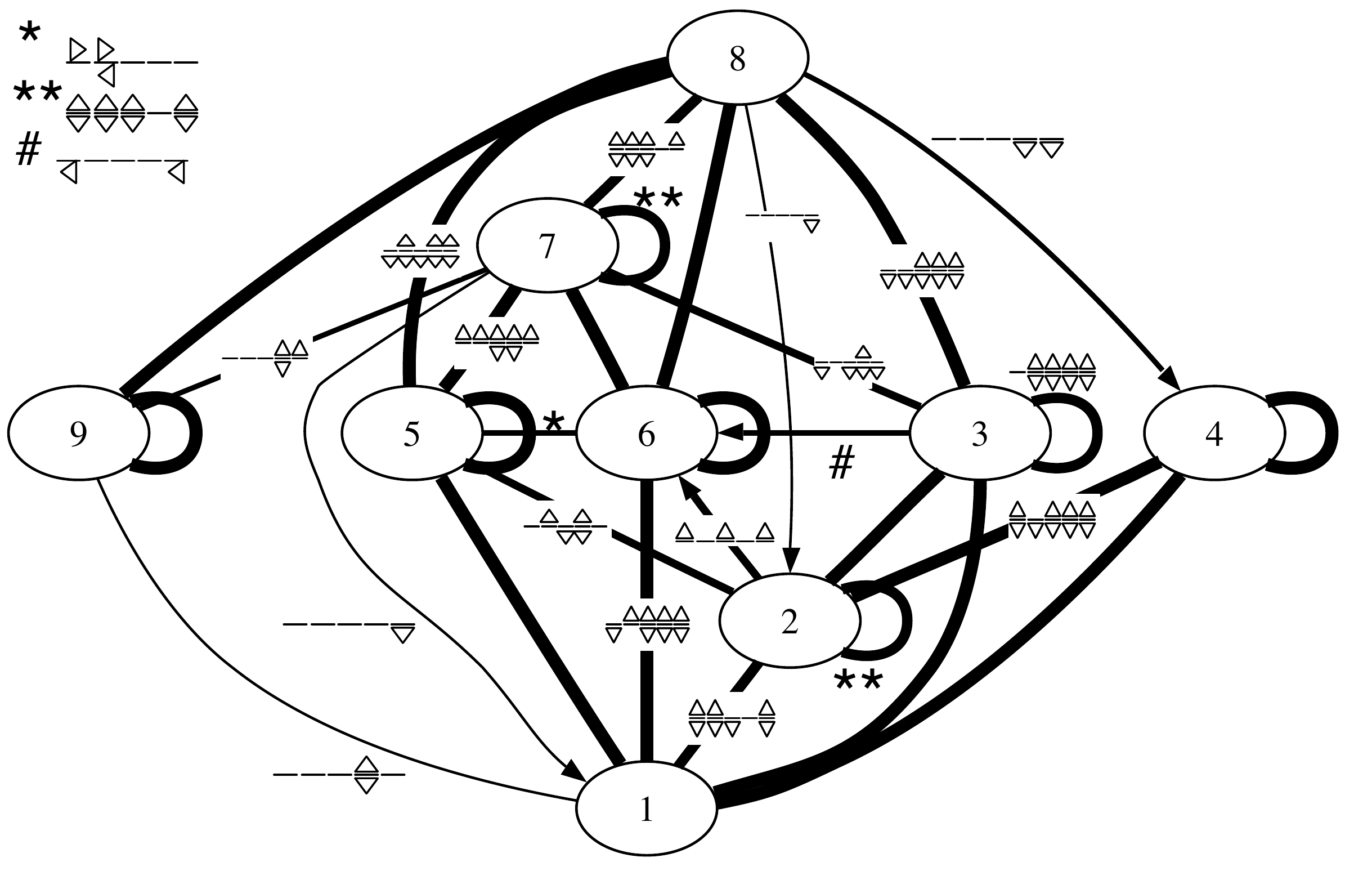}
\begin{minipage}[b][5cm][t]{2.1cm}\begin{footnotesize}\begin{flushleft}
1: Ctrl. Europe\\
2: 1st Peri. EU\\ 
3: 2nd Peri. EU\\
4: East. Europe\\
5: Africa and Mid. East\\
6: Polynesia\\
7: SE Asia\\
8: North Am, Japan\\
9: Middle and South America\\
\end{flushleft}\end{footnotesize}
\end{minipage}
}\\
\includegraphics[height=3.55cm]{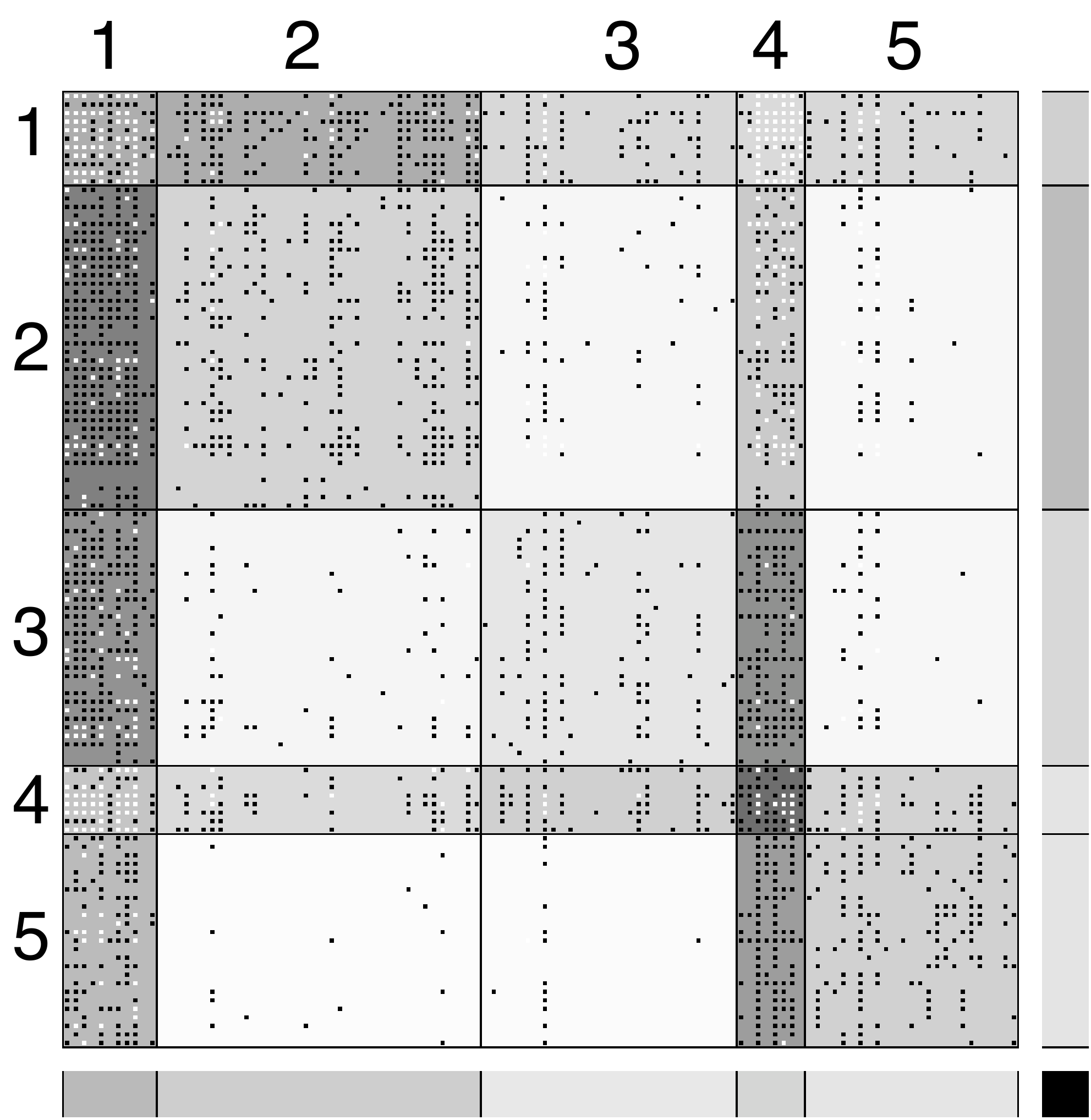} &
\includegraphics[height=3.2cm]{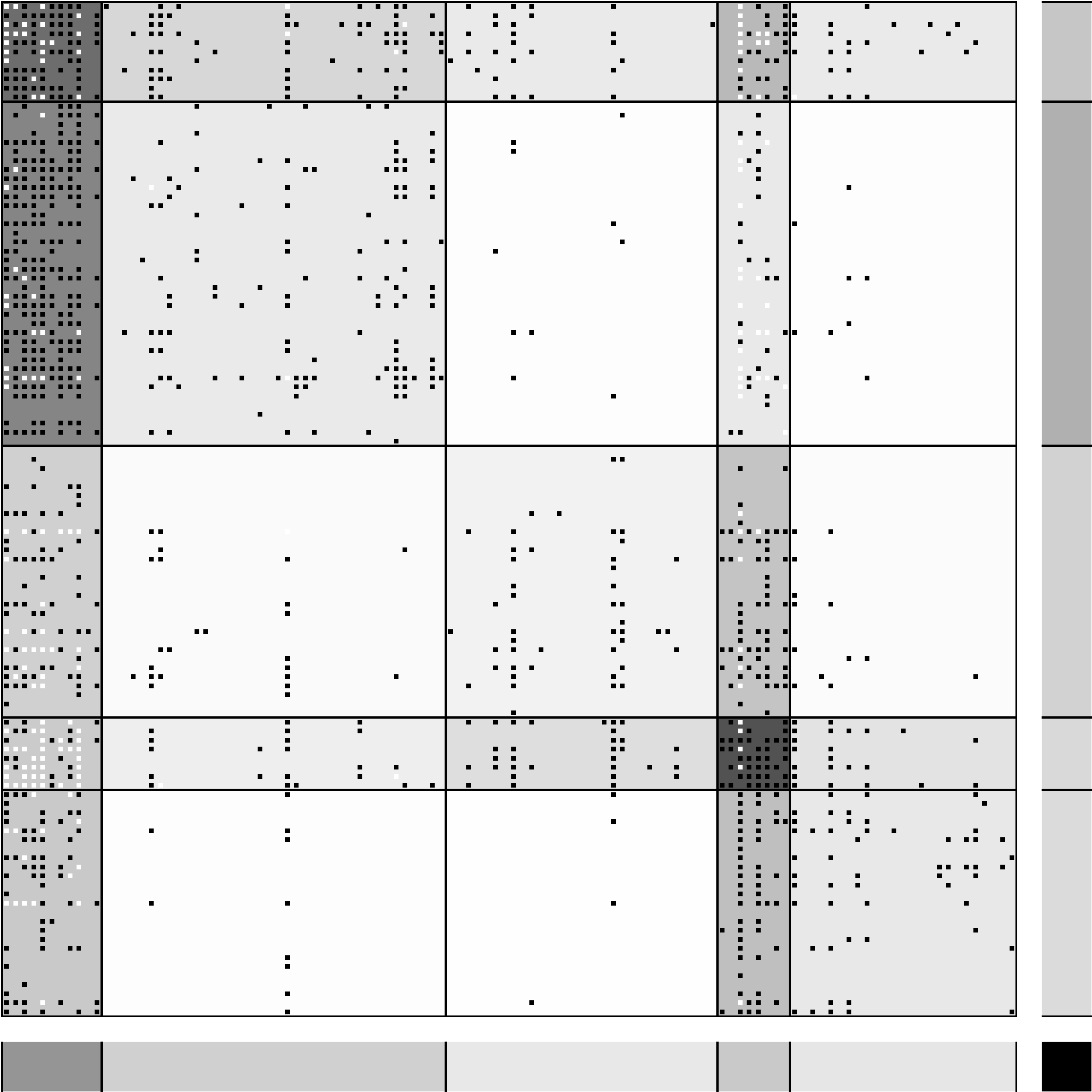} &
\includegraphics[height=3.2cm]{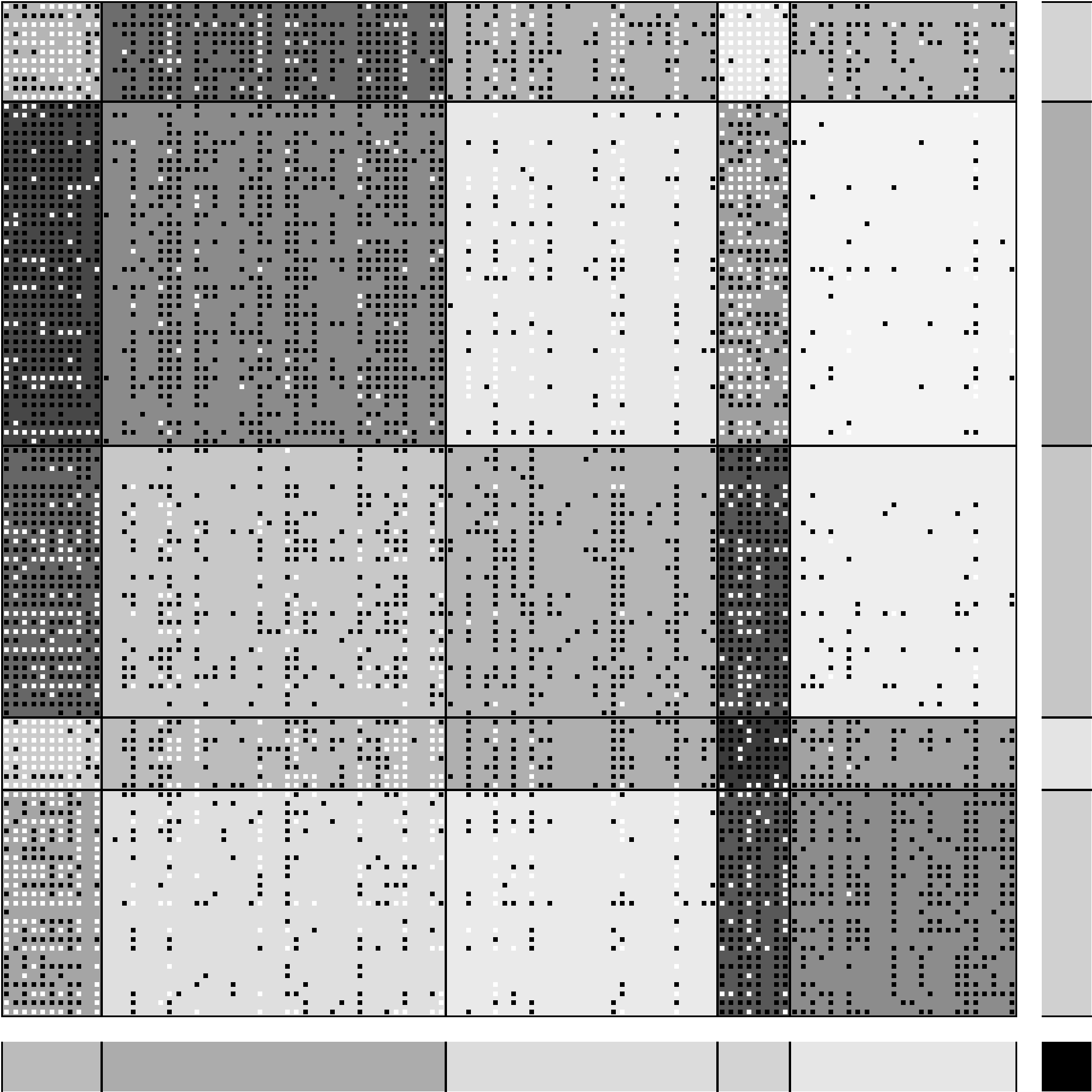} &
\includegraphics[height=3.2cm]{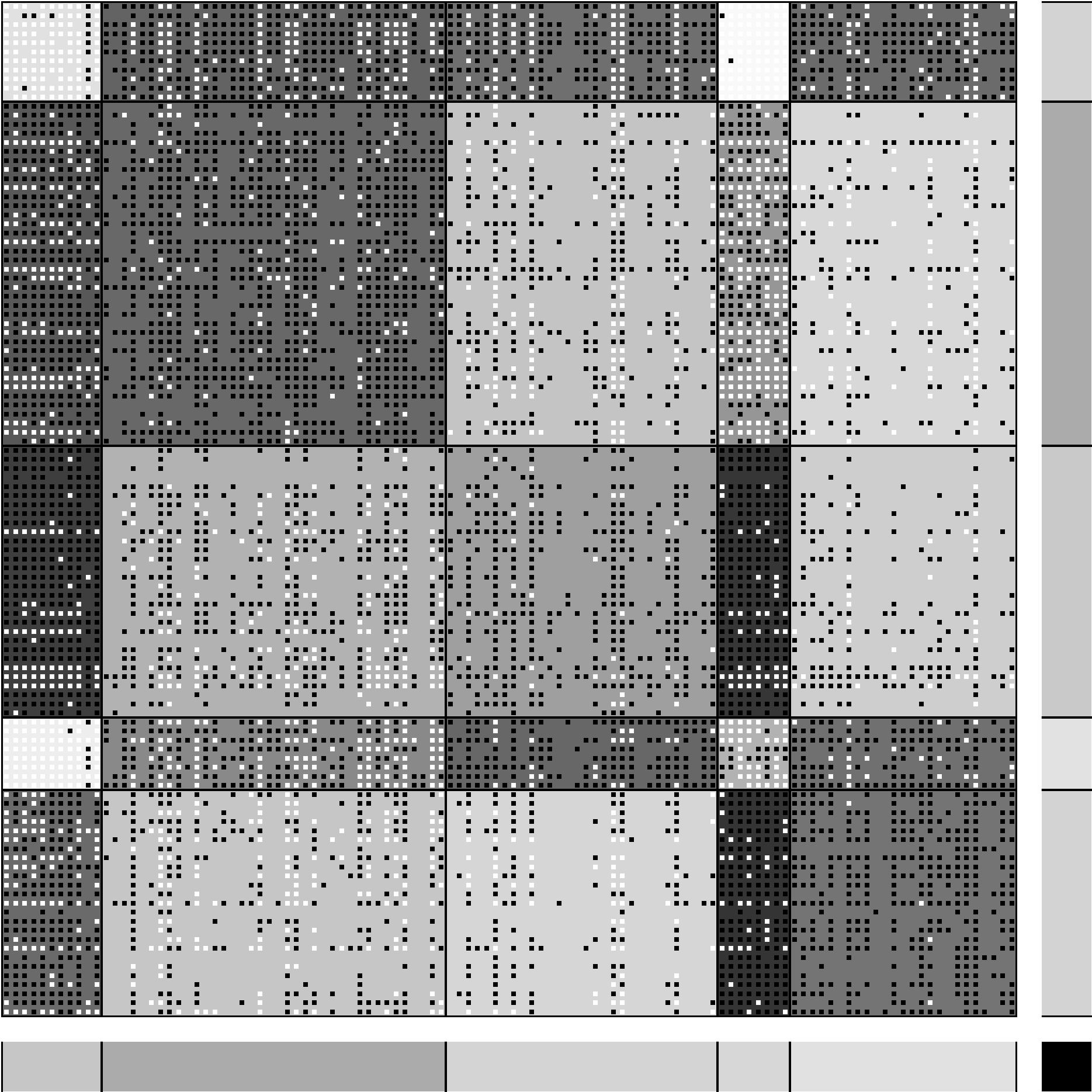} &
\includegraphics[height=3.2cm]{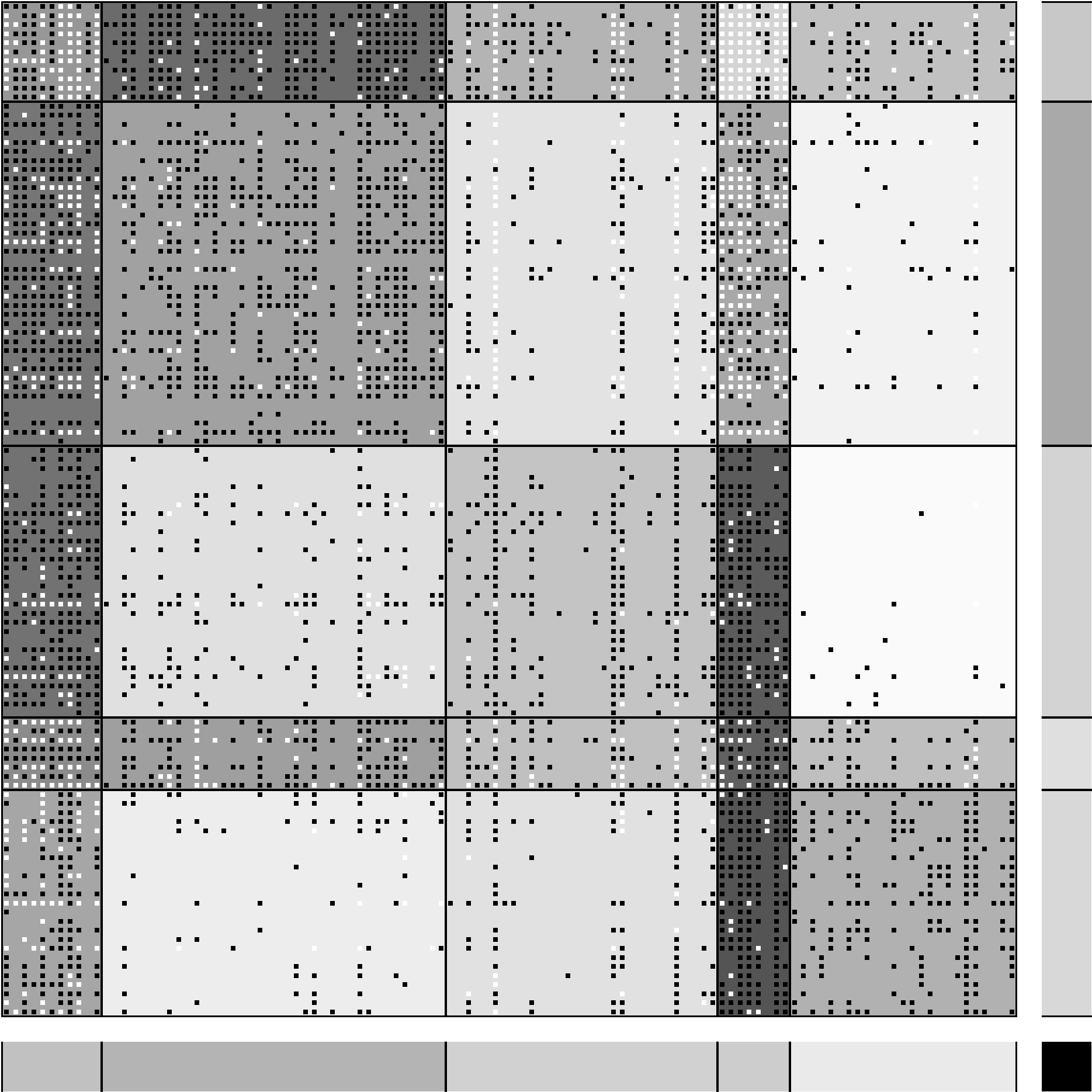} \\
\includegraphics[height=3.55cm]{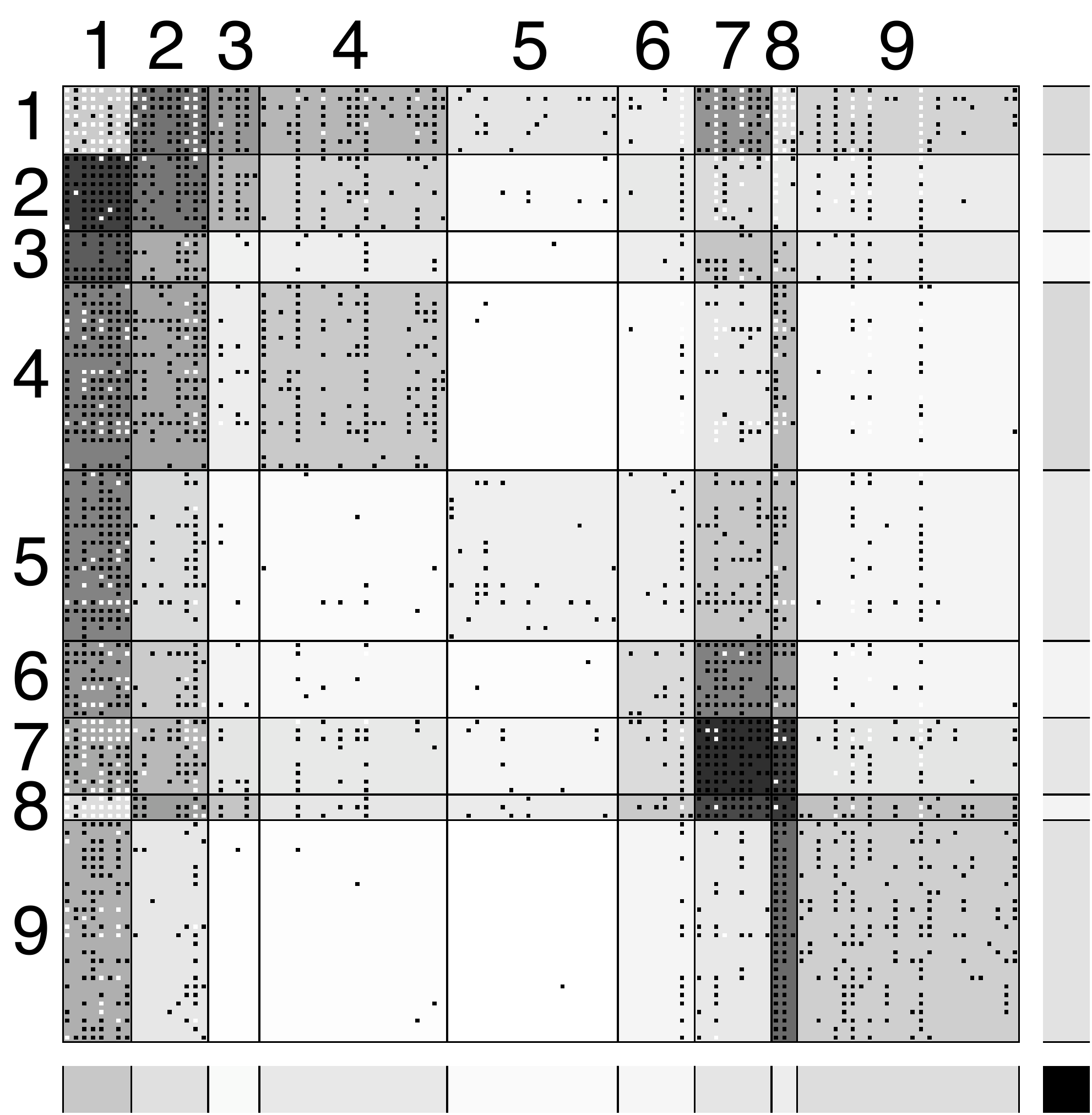} &
\includegraphics[height=3.2cm]{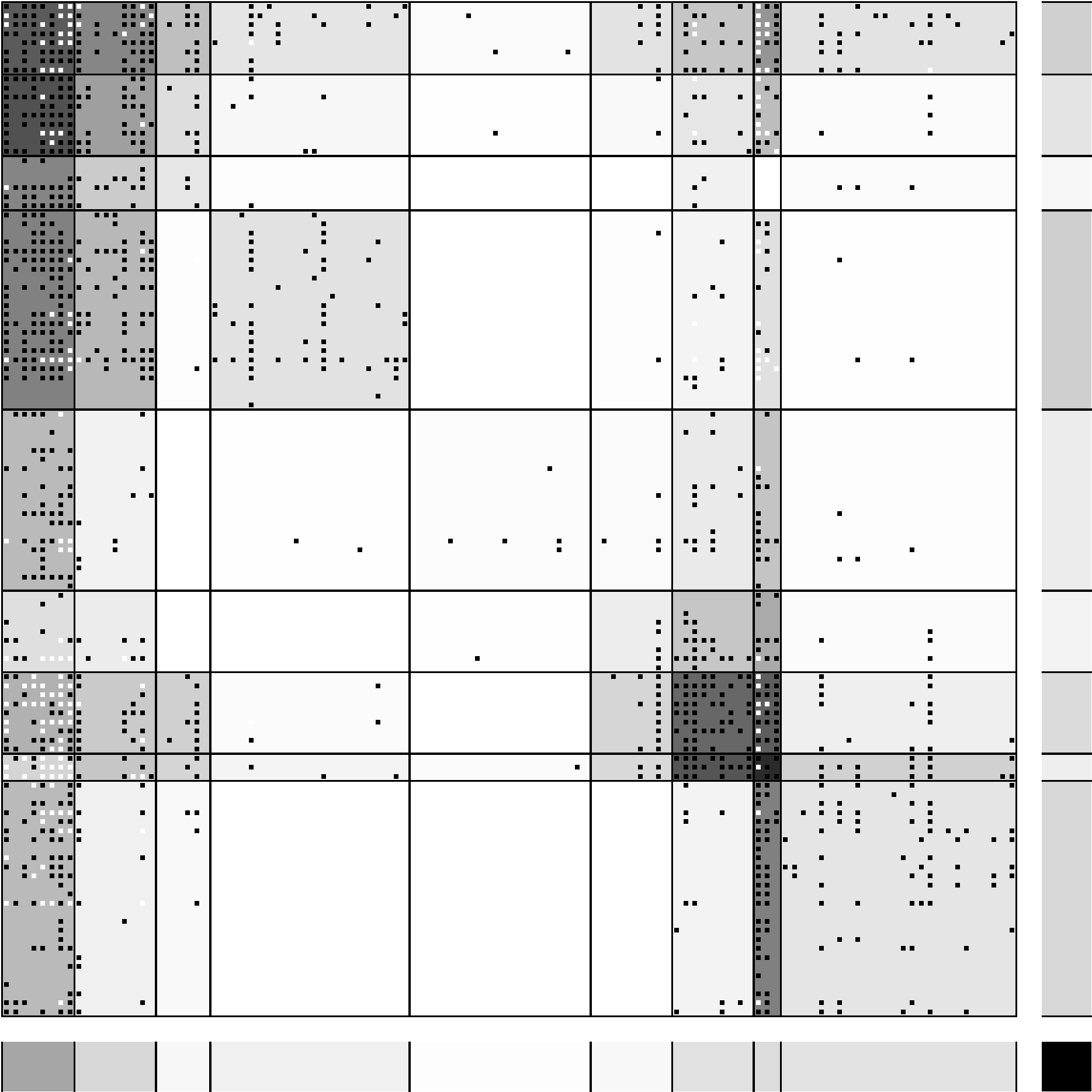} &
\includegraphics[height=3.2cm]{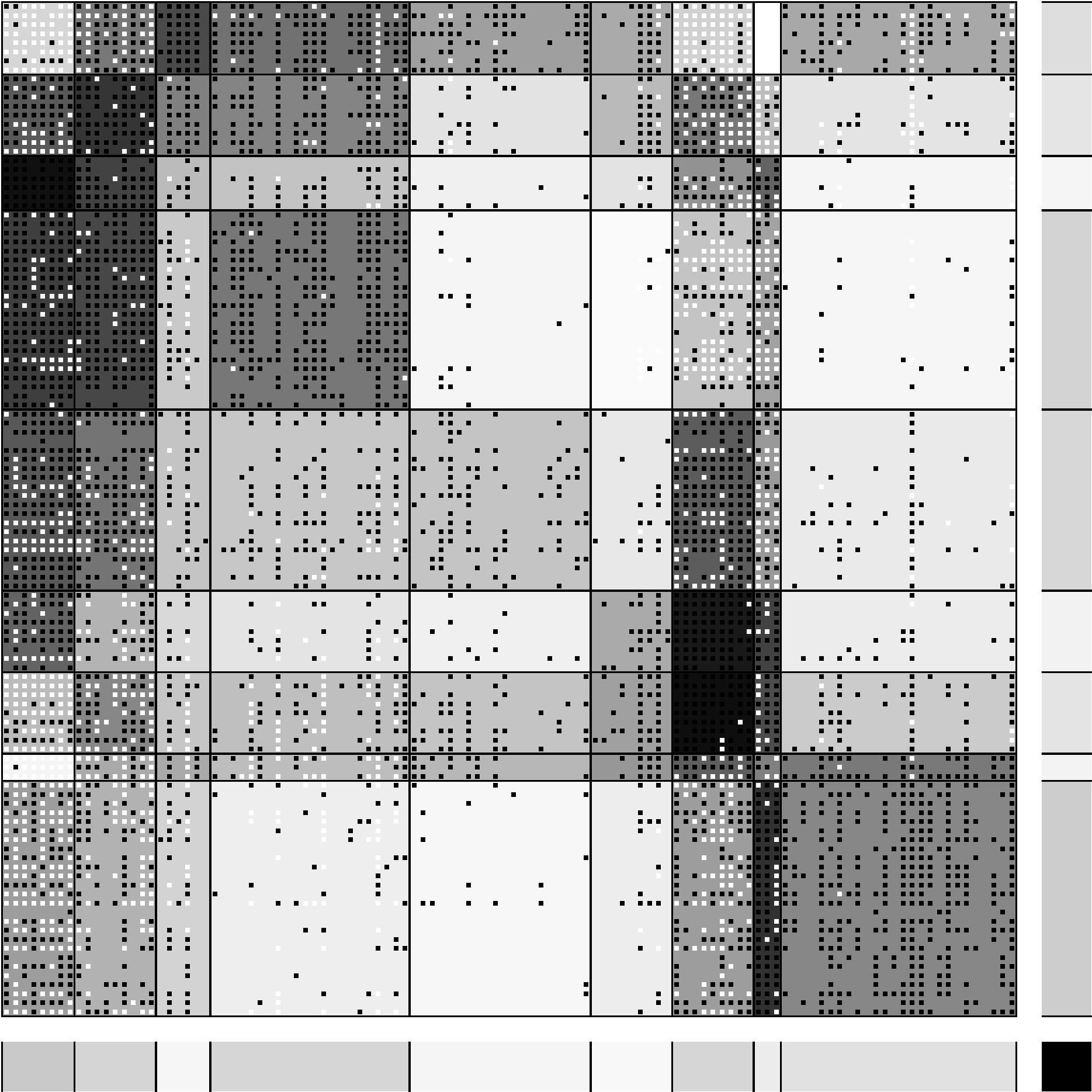} &
\includegraphics[height=3.2cm]{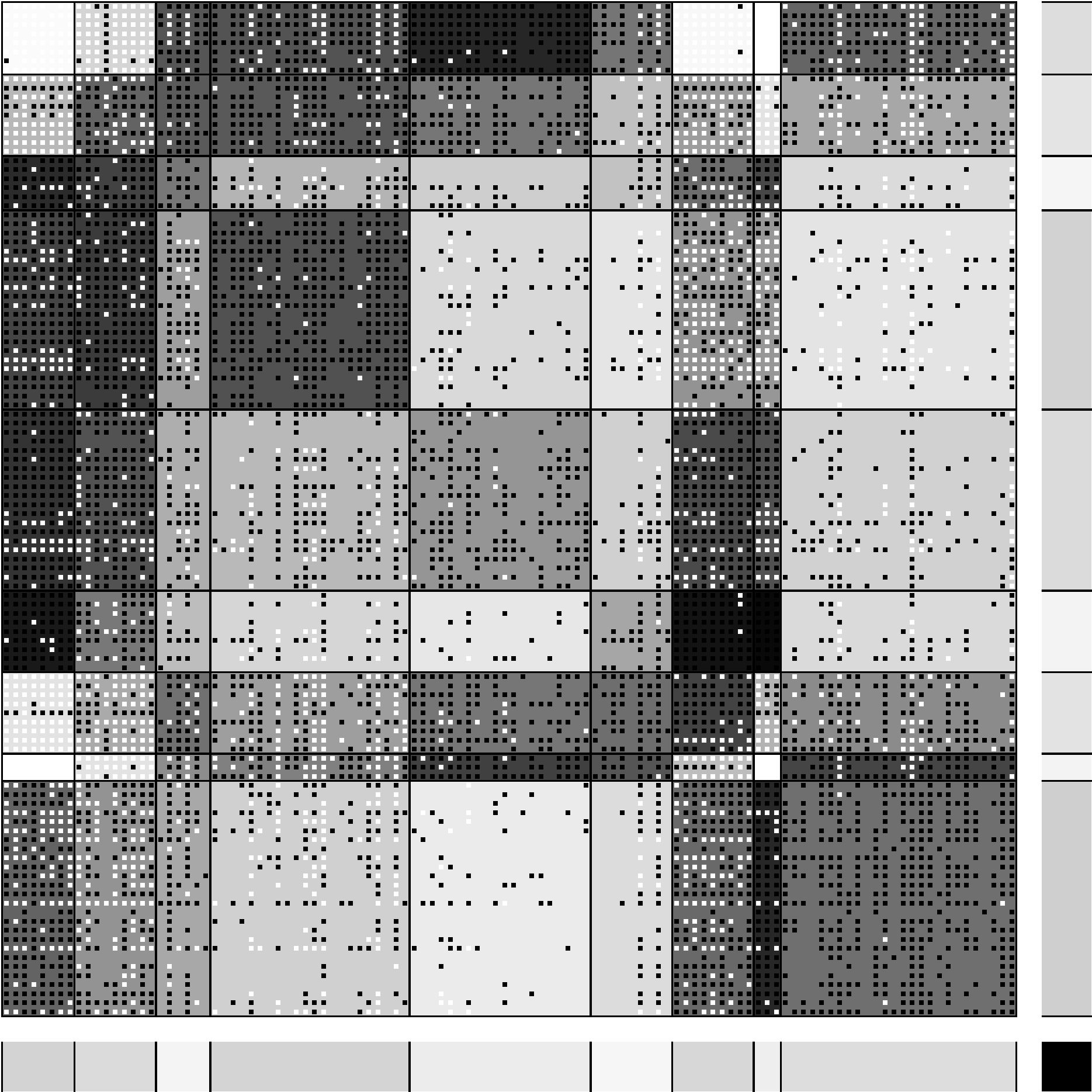} &
\includegraphics[height=3.2cm]{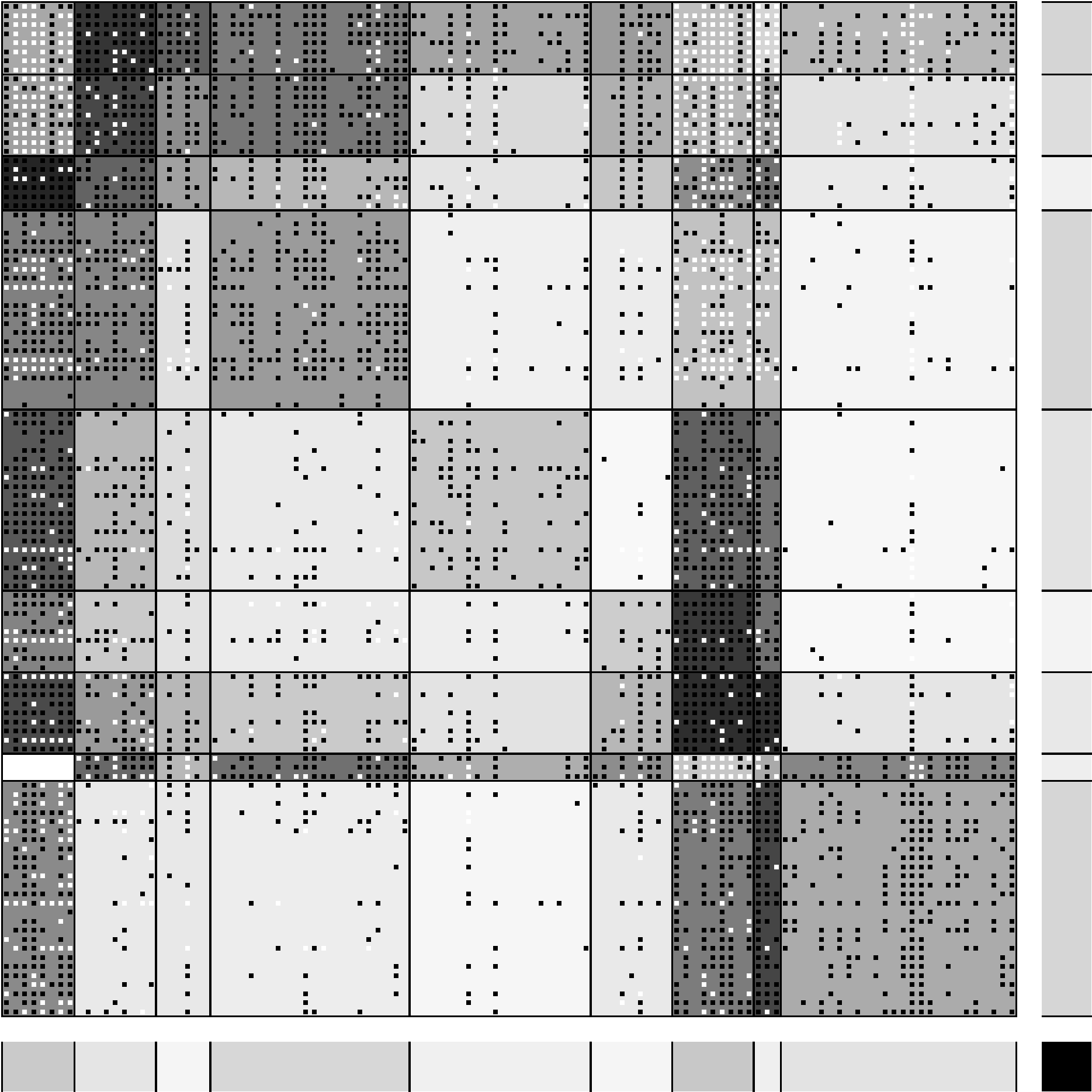} \\
meat \& meat prepartions & animal oil \& fats & paper \& paperboard & machinery & footwear\\
\end{tabular}
\caption{Consensus image graphs and block matrix plots for the 5 commodities studied at $q=5$ and $q=9$ roles. Note the high symmetry of the image graphs. Triangle labels indicate commodity and direction of the flow of goods. Unlabeled links carry all five commodities in both directions. Side and bottom bars encode the marginal fraction of import and export of the total traded volume for each block in gray scale, respectively. Black dots indicate trade greater than expected from the marginals for \emph{pairs} of countries, white dots smaller than expected. Background shading of blocks corresponds to density of black dots in block. See Table 1 
for individual countries grouped in each block and text for details. The ordering of blocks in the matrices is suggested by the proximity oder of the splitting diagram as depicted in Figure \ref{SupportSplitting} in the Appendix.}
\label{NineRoleBMs}
\end{figure*}
They are differentiated by proportions of production with extraction, capital-intensive or labor-intensive processing.The five groups are a) food products and by-products, b) simple extractive, c) sophisticated extractive, d) high technology and heavy manufacture and e) low wage/light manufacture. Representative for each of these groups, we chose one commodity each and obtained 5 different networks of commodity trade. The five commodities are a) meat and meat preparations, b) animal oil and fats, c) paper, paperboard and articles of pulp, d) machinery and e) footwear. The data set is based on the volumes of import as reported by 112 countries to the UN in 2000.
The only pretreatment applied to the data was to take the logarithm of the trade volumes which preserves the relative strength of trade volumes but reduces the effect that the fit of high volume countries alone dominates the quality of the role models.

Since the five different commodities had been found to be largely independent \cite{SmithNemeth} and also have different overall volumes, we do not simply sum the volumes but extend (\ref{QwoBM}) in order to accommodate for different types of links in the network. Instead of performing the same analysis for the different commodities independently and trying to form a consensus {\it a posteriori}, we include the different kinds of traded goods at the same time in the model finding process. The quantity that we maximize is:
 \begin{equation}
\mathcal{Q}^*(\{\sigma\})=\frac{1}{2}\sum_{c}\sum_{r,s}^q\|e^c_{rs}-[e^c_{rs}]\|.
\label{QwoBMmultiCom}
\end{equation}

Here, the first sum runs over the different commodities $c$ and every country $i$ is assigned exactly one role $\sigma_i$ from $\sigma_i\in\{1,...,q\}$ which it assumes in all block models. Further, $e^c_{rs}$ is the fraction of the log of the total volume of commodity $c$ imported by countries in roles $r$ from those in role $s$. As before,  $[e^c_{rs}]$ is the corresponding expectation value based on the marginals. Once an assignment of roles to countries has been found that maximizes (\ref{QwoBMmultiCom}), we can read off the five different image graphs $B_{rs}^c$ directly from the terms $e^c_{rs}-[e^c_{rs}]$ as before. The different models can then be overlaid easily as the same countries are assigned into the same roles for all of them. The computational effort for this multi commodity block modeling is still moderate as it increases over that for the case of one link type only by a factor of the number of different commodities. 

\begin{figure}[t]
\includegraphics[width=7cm]{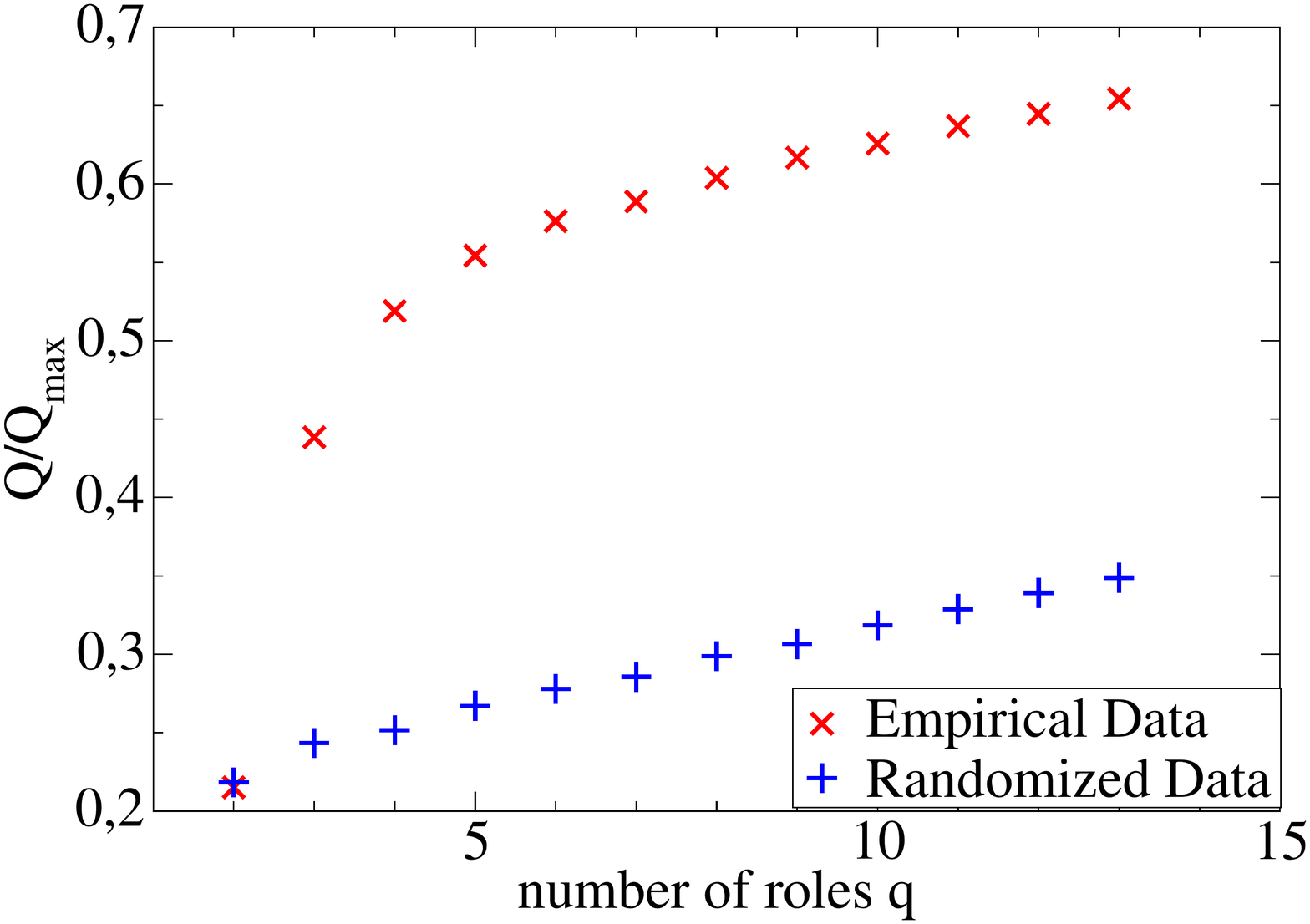}
\includegraphics[width=7cm]{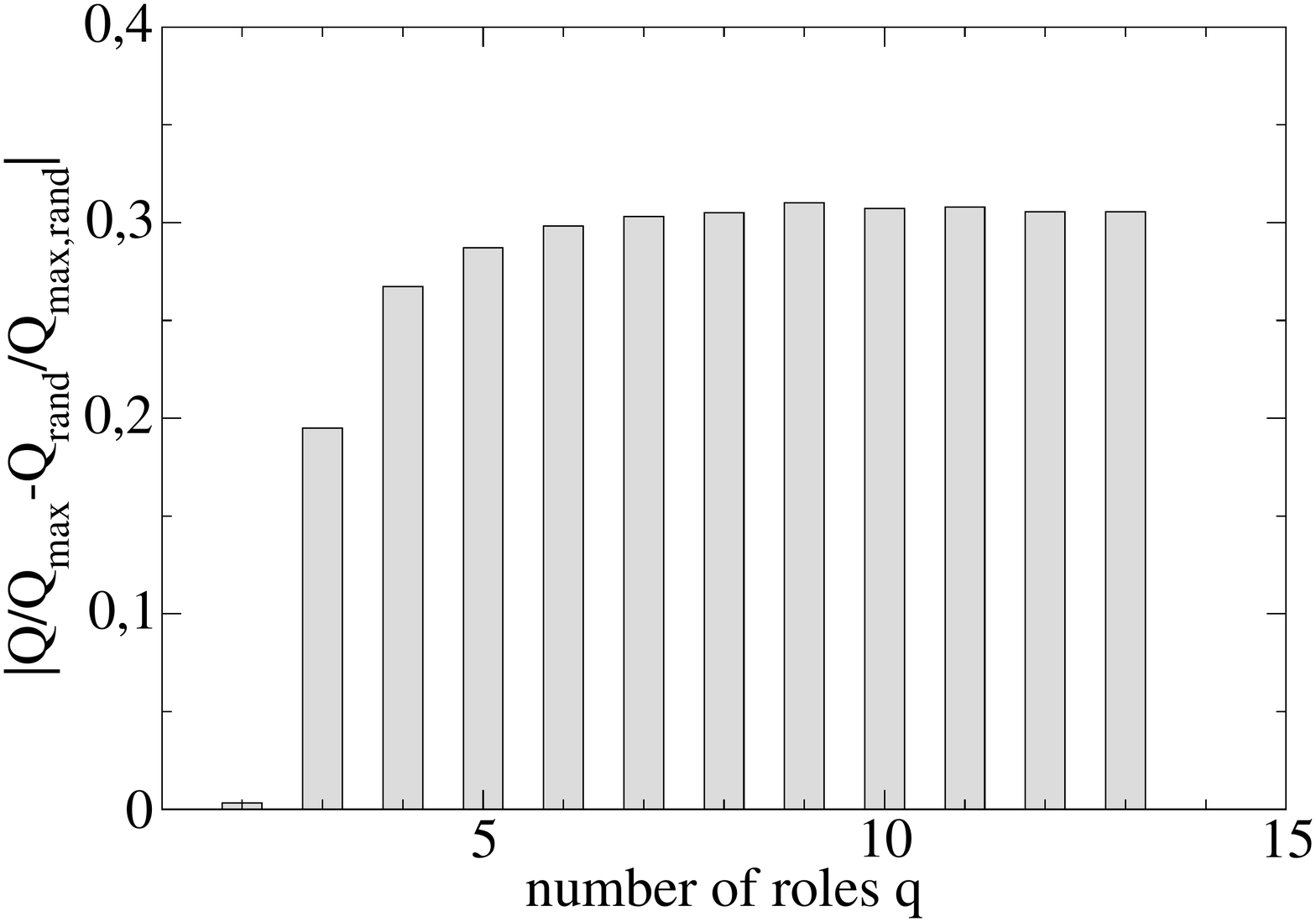}
\caption{{\bf Left:} Average of $\mathcal{Q}^*(q)/\mathcal{Q}_{max}$ over five commodities for the world trade network as a function of the number of roles $q$ in the block model. Red (x) denote the actual empirical data, blue (+) denote the results averaged over randomly rewired versions of the empirical data as a null model. While the randomized data shows a linear increase of $\mathcal{Q}^*/\mathcal{Q}_{max}$ with the number of roles, the empirical data exibits a strong increase for smaller numbers of $q$ and then also turns into a linear regime. {\bf Right:} Difference between $\mathcal{Q}^*/\mathcal{Q}_{max}$ for the empirical data and the randomized data. At $q=5$ we observe the transition to the linear regime. At $q=9$ the largest difference between empirical data and the random null model occurs capturing 60\% of $\mathcal{Q}_{max}$ with only 8\% of the total number of structural equivalence classes needed to achieve this maximum. }
\label{SelectOptq}
\end{figure}

Before discussing the block models we obtain, we need to determine the optimal number of roles. We calculate $\mathcal{Q}^c_{max}$ for each of the five commodities separately according to (\ref{Qmax}). For different number of roles $q$, we then maximize (\ref{QwoBMmultiCom}) and find $\mathcal{Q}^*(q)/Q_{max}$ averaged over the five commodities. This is necessary since we can define $Q_{max}$ only for a single link type and (\ref{QwoBMmultiCom}) aims at constructing a consensus model for all link types. This average value tells us what fraction of the total link structure we mimic in our image graph. As a random null model, we created randomized versions of the empirical data by rewiring the original network but keeping the number of connections constant for each node and link type. This holds the marginals roughly constant but rewires the network topology. Then, the same procedure as for the empirical data was used to obtain $\mathcal{Q}^*(q)_{rnd}/Q_{max,rnd}$ which is also averaged over several realizations of the disorder. In the left part of Figure \ref{SelectOptq} we compare the values of  $\mathcal{Q}^*(q)/Q_{max}$ for the empirical data and the randomized data. While the randomized data shows a linear increase with the number of roles from the beginning, the empirical data shows a strong increase at small numbers of roles and then also changes into a linear regime. The right part of Figure \ref{SelectOptq}  shows the difference in the ratio $\mathcal{Q}^*(q)/Q_{max}$ of empirical and randomized data. Though every block model from $q=2$ to $q=112$ has its own merit, after all, the countries do all have individuality, two points may be chosen as particularly meaningful: Either the number of roles at which we observe the transition to a linear increase in $\mathcal{Q}^*(q)/Q_{max}$  which happens at $q=5$ or the point at which we observe the largest difference to the randomized data at $q=9$. An alternative approach to select the optimal number of roles would be to use the minimum description length of the block model as suggested in Ref. \cite{RosvallMixture}. 

Figure \ref{NineRoleBMs} shows the image graphs and block matrix plots for five and nine roles. Additional models with fewer and intermediate numbers of roles can be found in the Appendix. Note the progression of differentiation as more and more roles are included. Already at $q=5$ we observe a structure that can be seen as a coarse-grained version of the model with 9 roles, with the models in between mediating the transition. 
Inspection of Table 1 
shows for all the block models that the progressive refinements in Figure  \ref{NineRoleBMs} induce a fair approximation to a hierarchical clustering of roles. This is not required by the model and rather than split as the number of roles increases, memberships merge from different blocks about 8\% of the time, including cases where two roles keep their identity but contribute overlapping members to form a third. 

A pattern of geographical proximities appears in an ordering of partitions that minimizes distances between sets that are merged or split. This unique compact layout of the splitting/merging diagram (Figure \ref{SupportSplitting} in Appendix) is used to order the partitions in Table 1. 
Countries in the same group tend to be located in each other's proximity. Geographical position is thus a strong factor determining the structural equivalence roles among countries. One reason is of course that geographical proximity means that such countries have similar geographical conditions and hence similar conditions for agriculture, mining etc. Another is that geographically close countries often form localized trade alliances. 

Additional to geographic proximity, the second striking feature of these models, based on optimizing structural equivalence, is that there exists considerable symmetry in the way the world trade is organized. Symmetry of the image graphs suggests that there are also regular equivalences across regions that organize the role structures across different regions \cite{SmithWhite}. Let us consider the $q=9$ model. Thus, on one hand, there is the region around the Pacific with the US, Canada and Japan (8) in a central position, South America (9) as an out-group and South East Asia (7) as a sub-center. On the other hand, we see the core of the European Union (1) in an equally central position as the US, Canada and Japan (8), however, with Eastern Europe and the former Soviet Union (4) assuming the position that South America (9) takes on across the Atlantic. 
\begin{table}[ht]
\caption{Assignment of countries in models with two to nine roles. The horizontal lines separate the $q=9$ different roles of the most detailed block model from Figure \ref{NineRoleBMs}. Note how the blocks form an almost perfect hierarchy in the way that successive blocks split apart although this is \textit{not} required by the algorithm. This is also shown by the splitting diagram in Figure \ref{SupportSplitting} in the Appendix which further suggests the order of the groups of countries in this table.}
\label{CountryTable}
\tabcolsep1mm
\renewcommand{\arraystretch}{0.6}
\begin{tiny}
\begin{tabular}{|r|l|c|c|c|c|c|c|c|c|}
\hline
Group Label &	Country	&	q=2	&	q=3	&	q=4	&	q=5	&	q=6	&	q=7	&	q=8	&	q=9	\\
\hline
&	Belgium-Luxembourg	&	1	&	1	&	1	&	1	&	1	&	1	&	1	&	1	\\
&	France	&	1	&	1	&	1	&	1	&	1	&	1	&	1	&	1	\\
&	Germany	&	1	&	1	&	1	&	1	&	1	&	1	&	1	&	1	\\
&	Italy	&	1	&	1	&	1	&	1	&	1	&	1	&	1	&	1	\\
Core EU &	Netherlands	&	1	&	1	&	1	&	1	&	1	&	1	&	1	&	1	\\
&	Spain	&	1	&	1	&	1	&	1	&	1	&	1	&	1	&	1	\\
&	Switzerland	&	1	&	1	&	1	&	1	&	1	&	1	&	1	&	1	\\
&	United Kingdom	&	1	&	1	&	1	&	1	&	1	&	1	&	1	&	1	\\
\hline																		
&	Denmark	&	1	&	1	&	1	&	1	&	1	&	1	&	2	&	2	\\
&	Sweden	&	1	&	1	&	1	&	1	&	1	&	1	&	2	&	2	\\
&	Austria	&	1	&	1	&	2	&	2	&	2	&	2	&	2	&	2	\\
1st Peri. EU &	Turkey	&	1	&	2	&	2	&	2	&	2	&	1	&	2	&	2	\\
&	Greece	&	1	&	2	&	2	&	2	&	2	&	2	&	2	&	2	\\
&	Norway	&	1	&	2	&	2	&	2	&	2	&	2	&	2	&	2	\\
&	Finland	&	2	&	2	&	2	&	2	&	2	&	2	&	3	&	2	\\
&	Ireland	&	2	&	2	&	2	&	2	&	2	&	2	&	2	&	2	\\
&	Cyprus	&	2	&	2	&	2	&	2	&	2	&	2	&	2	&	2	\\
\hline																		
&	Portugal	&	2	&	2	&	2	&	2	&	2	&	2	&	2	&	3	\\
&	Andorra	&	2	&	2	&	2	&	2	&	2	&	2	&	3	&	3	\\
2nd Peri. EU  &	Iceland	&	2	&	2	&	2	&	2	&	2	&	2	&	3	&	3	\\
&	Israel	&	2	&	2	&	2	&	2	&	2	&	2	&	3	&	3	\\
&	Greenland	&	2	&	2	&	3	&	2	&	2	&	2	&	3	&	3	\\
&	Malta	&	2	&	2	&	2	&	2	&	2	&	2	&	4	&	3	\\
\hline																		
&	Russian Federation	&	1	&	1	&	2	&	2	&	2	&	2	&	2	&	4	\\
&	Czech Rep.	&	1	&	2	&	2	&	2	&	2	&	2	&	2	&	4	\\
&	Turkmenistan	&	1	&	2	&	2	&	2	&	2	&	2	&	3	&	4	\\
&	Albania	&	2	&	2	&	2	&	2	&	2	&	2	&	3	&	4	\\
&	Armenia	&	2	&	2	&	2	&	2	&	2	&	2	&	3	&	4	\\
&	Azerbaijan	&	2	&	2	&	2	&	2	&	2	&	2	&	3	&	4	\\
&	Belarus	&	2	&	2	&	2	&	2	&	2	&	2	&	3	&	4	\\
&	Bulgaria	&	2	&	2	&	2	&	2	&	2	&	2	&	3	&	4	\\
&	Estonia	&	2	&	2	&	2	&	2	&	2	&	2	&	3	&	4	\\
East. Europe&	Georgia	&	2	&	2	&	2	&	2	&	2	&	2	&	3	&	4	\\
&	Hungary	&	2	&	2	&	2	&	2	&	2	&	2	&	3	&	4	\\
&	Iran	&	2	&	2	&	2	&	2	&	2	&	2	&	3	&	4	\\
&	Kazakhstan	&	2	&	2	&	2	&	2	&	2	&	2	&	3	&	4	\\
&	Latvia	&	2	&	2	&	2	&	2	&	2	&	2	&	3	&	4	\\
&	Lithuania	&	2	&	2	&	2	&	2	&	2	&	2	&	3	&	4	\\
&	Poland	&	2	&	2	&	2	&	2	&	2	&	2	&	3	&	4	\\
&	Rep. of Moldova	&	2	&	2	&	2	&	2	&	2	&	2	&	3	&	4	\\
&	Romania	&	2	&	2	&	2	&	2	&	2	&	2	&	3	&	4	\\
&	Serbia and Montenegro	&	2	&	2	&	2	&	2	&	2	&	2	&	3	&	4	\\
&	Slovakia	&	2	&	2	&	2	&	2	&	2	&	2	&	3	&	4	\\
&	Tajikistan	&	2	&	2	&	2	&	2	&	2	&	2	&	3	&	4	\\
&	Syria	&	2	&	2	&	2	&	2	&	2	&	3	&	4	&	4	\\
\hline																		
&	Saudi Arabia	&	1	&	1	&	1	&	1	&	3	&	3	&	4	&	5	\\
&	Algeria	&	2	&	2	&	2	&	2	&	2	&	3	&	4	&	5	\\
&	Morocco	&	2	&	2	&	2	&	2	&	2	&	3	&	4	&	5	\\
&	Tunisia	&	2	&	2	&	2	&	2	&	2	&	3	&	4	&	5	\\
&	Bahrain	&	2	&	2	&	3	&	3	&	3	&	3	&	4	&	5	\\
&	Comoros	&	2	&	2	&	3	&	3	&	3	&	3	&	4	&	5	\\
&	Cote d'Ivoire	&	2	&	2	&	3	&	3	&	3	&	3	&	4	&	5	\\
&	Ethiopia	&	2	&	2	&	3	&	3	&	3	&	3	&	4	&	5	\\
&	Ghana	&	2	&	2	&	3	&	3	&	3	&	3	&	4	&	5	\\
Africa, Mid. East &	Guinea	&	2	&	2	&	3	&	3	&	3	&	3	&	4	&	5	\\
&	Jordan	&	2	&	2	&	3	&	3	&	3	&	3	&	4	&	5	\\
&	Nigeria	&	2	&	2	&	3	&	3	&	3	&	3	&	4	&	5	\\
&	Oman	&	2	&	2	&	3	&	3	&	3	&	3	&	4	&	5	\\
&	Senegal	&	2	&	2	&	3	&	3	&	3	&	3	&	4	&	5	\\
&	Togo	&	2	&	2	&	3	&	3	&	3	&	3	&	4	&	5	\\
&	Burundi	&	2	&	3	&	3	&	3	&	3	&	3	&	4	&	5	\\
&	Kenya	&	2	&	3	&	3	&	3	&	3	&	3	&	4	&	5	\\
&	Mauritius	&	2	&	3	&	3	&	3	&	3	&	3	&	4	&	5	\\
&	Pakistan	&	2	&	3	&	3	&	3	&	3	&	3	&	4	&	5	\\
&	Uganda	&	2	&	3	&	3	&	3	&	3	&	3	&	4	&	5	\\
\hline																		
&	China, Macao SAR	&	2	&	3	&	3	&	3	&	3	&	4	&	5	&	6	\\
&	French Polynesia	&	2	&	3	&	3	&	3	&	3	&	4	&	5	&	6	\\
&	Maldives	&	2	&	3	&	3	&	3	&	3	&	4	&	5	&	6	\\
&	Nepal	&	2	&	3	&	3	&	3	&	3	&	4	&	5	&	6	\\
Polynesia &	New Caledonia	&	2	&	3	&	3	&	3	&	3	&	4	&	5	&	6	\\
&	New Zealand	&	2	&	3	&	3	&	3	&	3	&	4	&	5	&	6	\\
&	Papua New Guinea	&	2	&	3	&	3	&	3	&	3	&	4	&	5	&	6	\\
&	Philippines	&	2	&	3	&	3	&	3	&	3	&	4	&	5	&	6	\\
&	Vanuatu	&	2	&	3	&	3	&	3	&	3	&	4	&	5	&	6	\\
\hline																		
&	Malaysia	&	2	&	3	&	3	&	3	&	3	&	5	&	6	&	7	\\
&	Indonesia	&	2	&	3	&	3	&	3	&	4	&	5	&	6	&	7	\\
&	Singapore	&	2	&	3	&	3	&	3	&	4	&	5	&	6	&	7	\\
&	South Africa	&	1	&	3	&	3	&	3	&	4	&	5	&	6	&	7	\\
SE Asia&	Thailand	&	1	&	3	&	3	&	3	&	4	&	5	&	6	&	7	\\
&	Australia	&	1	&	3	&	3	&	4	&	4	&	5	&	6	&	7	\\
&	China	&	1	&	3	&	1	&	4	&	4	&	5	&	6	&	7	\\
&	China, Hong Kong SAR	&	1	&	3	&	3	&	4	&	4	&	5	&	6	&	7	\\
&	Rep. of Korea	&	1	&	3	&	3	&	4	&	4	&	5	&	6	&	7	\\
\hline																		
&	Japan	&	1	&	3	&	3	&	4	&	5	&	6	&	7	&	8	\\
North Am, Japan &	Canada	&	1	&	3	&	4	&	4	&	5	&	6	&	7	&	8	\\
&	USA 	&	1	&	3	&	4	&	4	&	5	&	6	&	7	&	8	\\
\hline																		
&	Brazil	&	1	&	3	&	4	&	4	&	6	&	7	&	8	&	9	\\
&	Argentina	&	1	&	3	&	4	&	5	&	6	&	7	&	8	&	9	\\
&	Barbados	&	1	&	3	&	4	&	5	&	6	&	7	&	8	&	9	\\
&	Honduras	&	1	&	3	&	4	&	5	&	6	&	7	&	8	&	9	\\
&	Panama	&	1	&	3	&	4	&	5	&	6	&	7	&	8	&	9	\\
&	Bolivia	&	2	&	3	&	4	&	5	&	6	&	7	&	8	&	9	\\
&	Chile	&	2	&	3	&	4	&	5	&	6	&	7	&	8	&	9	\\
&	Colombia	&	2	&	3	&	4	&	5	&	6	&	7	&	8	&	9	\\
&	Costa Rica	&	2	&	3	&	4	&	5	&	6	&	7	&	8	&	9	\\
&	Dominica	&	2	&	3	&	4	&	5	&	6	&	7	&	8	&	9	\\
&	Ecuador	&	2	&	3	&	4	&	5	&	6	&	7	&	8	&	9	\\
South Ameriac &	El Salvador	&	2	&	3	&	4	&	5	&	6	&	7	&	8	&	9	\\
&	Guatemala	&	2	&	3	&	4	&	5	&	6	&	7	&	8	&	9	\\
&	Jamaica	&	2	&	3	&	4	&	5	&	6	&	7	&	8	&	9	\\
&	Mexico	&	2	&	3	&	4	&	5	&	6	&	7	&	8	&	9	\\
&	Montserrat	&	2	&	3	&	4	&	5	&	6	&	7	&	8	&	9	\\
&	Nicaragua	&	2	&	3	&	4	&	5	&	6	&	7	&	8	&	9	\\
&	Paraguay	&	2	&	3	&	4	&	5	&	6	&	7	&	8	&	9	\\
&	Peru	&	2	&	3	&	4	&	5	&	6	&	7	&	8	&	9	\\
&	Saint Kitts and Nevis	&	2	&	3	&	4	&	5	&	6	&	7	&	8	&	9	\\
&	Saint Lucia	&	2	&	3	&	4	&	5	&	6	&	7	&	8	&	9	\\
&	Saint Vincent and the Grenadines	&	2	&	3	&	4	&	5	&	6	&	7	&	8	&	9	\\
&	Suriname	&	2	&	3	&	4	&	5	&	6	&	7	&	8	&	9	\\
&	Trinidad and Tobago	&	2	&	3	&	4	&	5	&	6	&	7	&	8	&	9	\\
&	Uruguay	&	2	&	3	&	4	&	5	&	6	&	7	&	8	&	9	\\
&	Venezuela	&	2	&	3	&	4	&	5	&	6	&	7	&	8	&	9	\\
\hline
\end{tabular}
\end{tiny}
\label{CountryTable}
\end{table}
Scandinavia and some peripheral European countries such as Ireland, Austria, Greece and also Turkey (2) are for the core EU states (1) what South East Asia (7) is for the North America and Japan (8). In the middle of all, we find the African and Middle Eastern countries (5), Polynesia (6) and a second group of peripheral European countries (3) which are Greenland, Iceland, Portugal, Andorra, Malta and Israel in approximately equal positions.

Competing models for regular and structural equivalence blockmodeling can be explored through other methods sensitive to how different equivalences can be defined by constraints that differ block by block, as in generalized blockmodeling \cite{DoreianBook}. Our model can parsimoniously evaluate at a global level, however, how regular and structural equivalence role models differ, simply by changing our quality function. We reserve that comparison for a future paper.
It is also interesting to observe the gradual refinement of the roles, for instance when concentrating on the core EU countries. For small numbers of roles, countries such as Denmark, Sweden, Austria and Norway are grouped together with them, but with more roles available, they are moved into their own groups to merge with countries such as Cyprus, Finland and Ireland which had been in more peripheral positions from the start. Such behavior can be interpreted as showing, with greater refinement in the role structure, the intermediary positions between the clear role of the core EU states and the more peripheral countries. 

\section{Conclusion}
The proposed framework for block modeling is a density-based measure but not, as in some earlier methods \cite{WassermanFaust}, based on a notion of high/low densities within position-to-position blocks compared to {\it global} densities. Rather, its partitions are based on the marginal expectations from the paired row-column positional totals that meet in a given block, \ie where links are concentrated.  We thus take a different approach than the parameter rich mixture model approaches in Refs. \cite{WassermanAPost,RosvallMixture,LeichtMixture,KempLatent,KempLatent2,NowickiSnijders}. Our method allows the use of weighted data sets of multiple link types and results from a generalization of proven sociological concepts containing them as limiting cases.

As we showed for the cross sectional snapshot of the United Nations commodity trade database recording trade flows between countries on an annual basis, these partitions obtained are valid and capable of producing new insights and theories of role structure and dynamics. 

The second distinctive and positive feature of density generalized block modeling is that 
successive partitions are not necessarily sequential hierarchical subclusters but may be overlapping. This is a major advantage of this modeling perspective that has been almost totally neglected in the 
previous traditions of block modeling and network-based role analysis. This limitation in prior perspectives has prevented block modeling from modeling the fact that actors do not usually take on a single role but an intersection of roles. The proposed framework can help to recover some of these intersections through the overlapping partitions that occur with different granularities of roles. 

What density generalized block modeling contributes are new measurement, structural, and 
potentially dynamic perspectives on problems of explanation, assuming that models are 
constructed as time series of how networks evolve. 

We'd like to thank Jeroen Bruggeman and Scott White for useful discussions and David Smith and Matthew Mahutga also for sharing the UN comtrade data with us. 

\clearpage

\begin{samepage}
\appendix
\subsection{Appendix}
\begin{figure*}[h]
\begin{tabular}{ccccc}
\multicolumn{5}{r}{
\includegraphics[height=2cm]{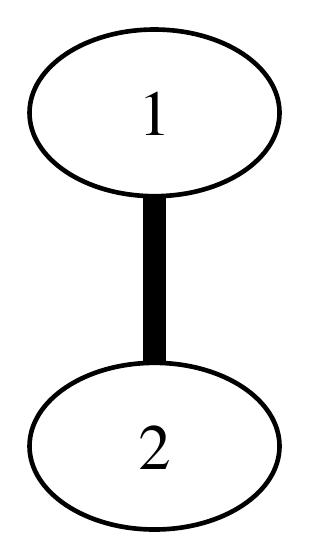}
\hspace{2.5cm}
\includegraphics[height=1.2cm]{Diamonds}
}\\
\includegraphics[height=3.25cm]{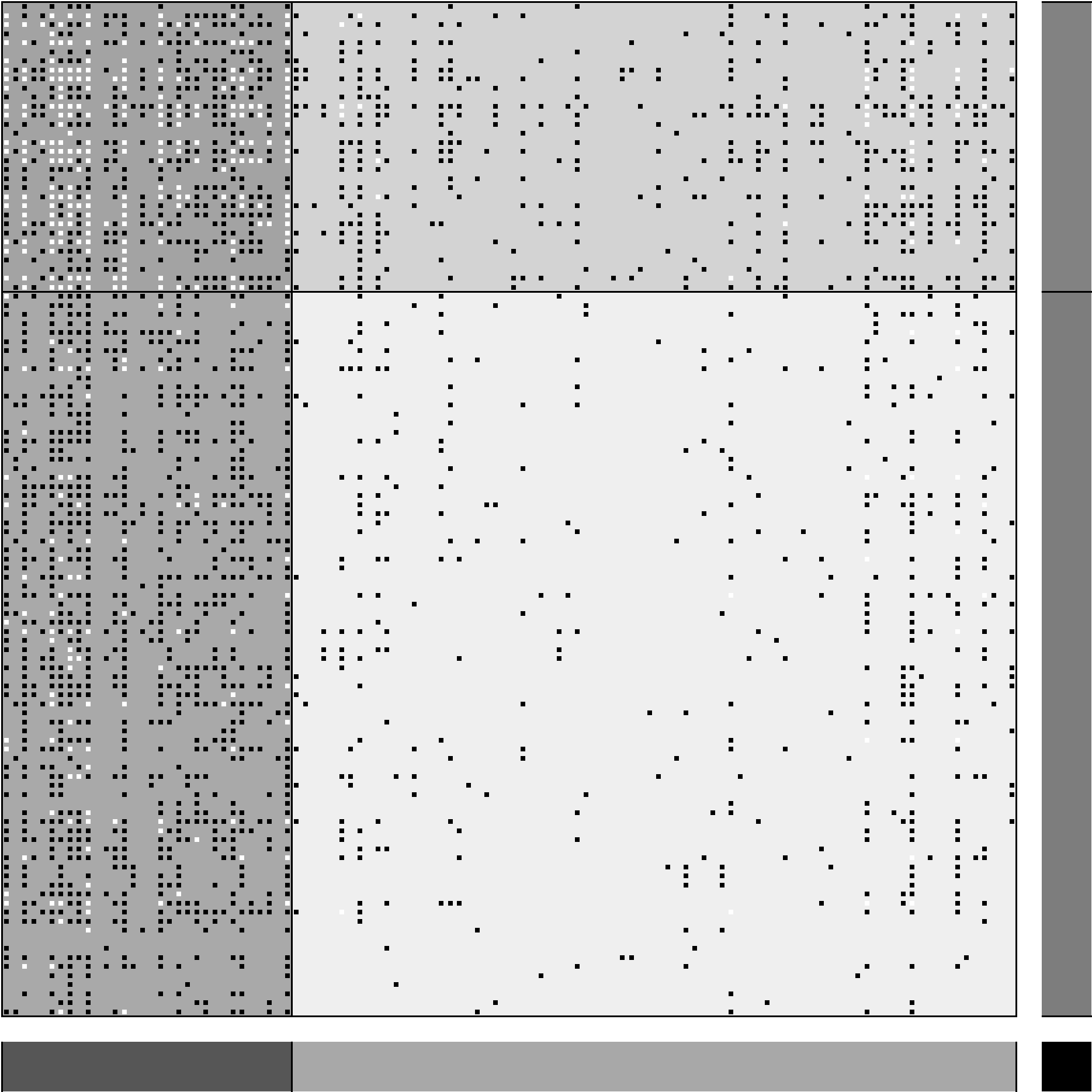} &
\includegraphics[height=3.25cm]{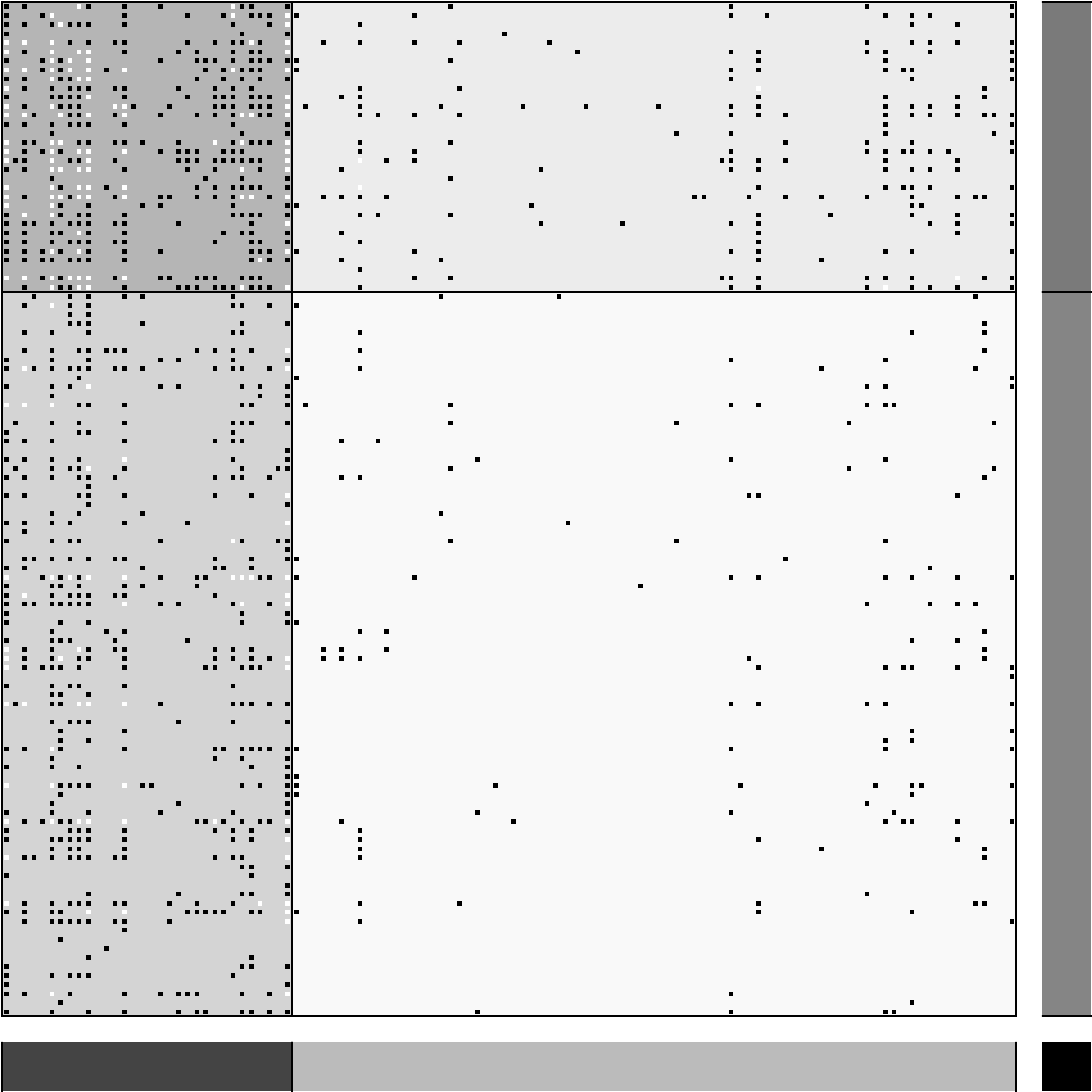} &
\includegraphics[height=3.25cm]{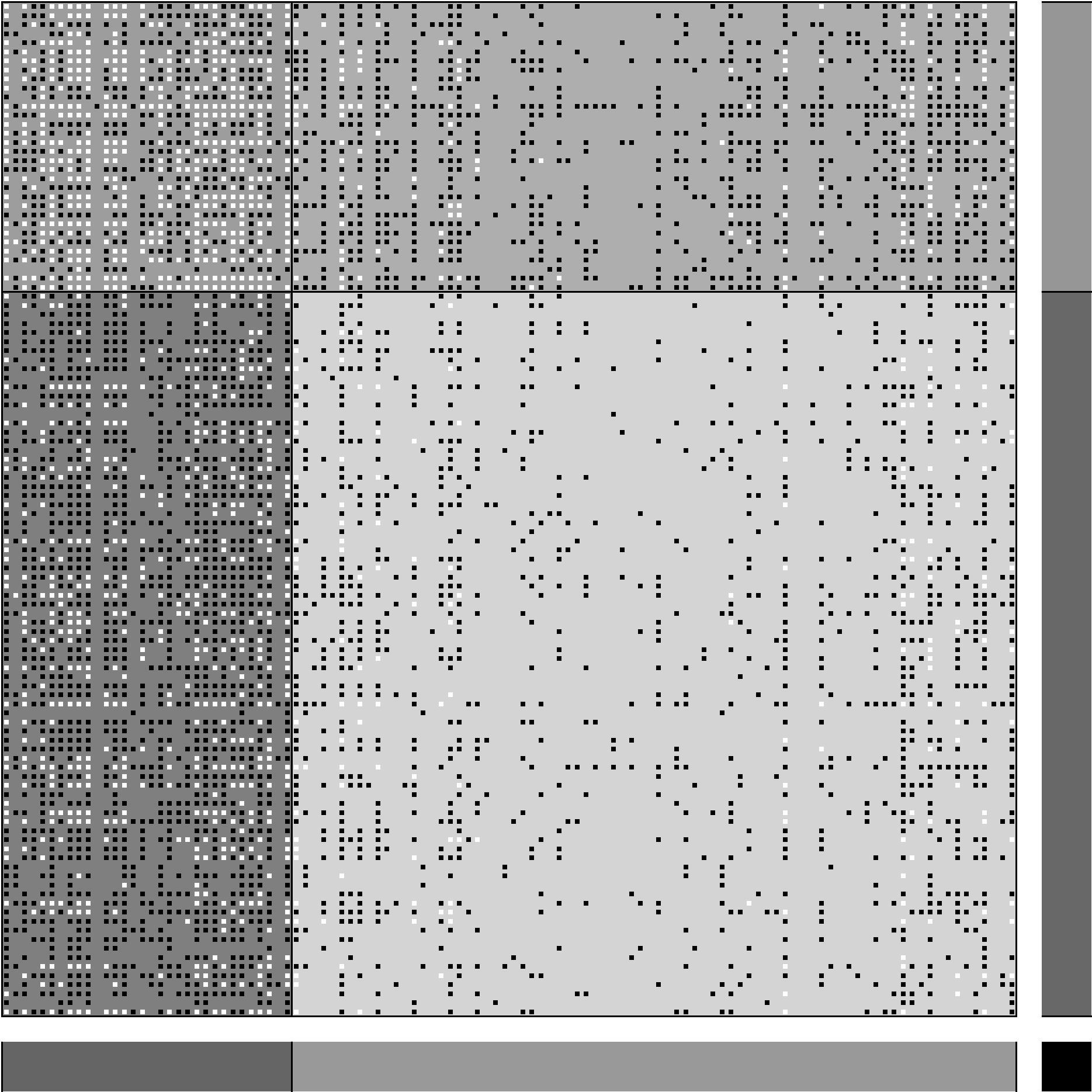} &
\includegraphics[height=3.25cm]{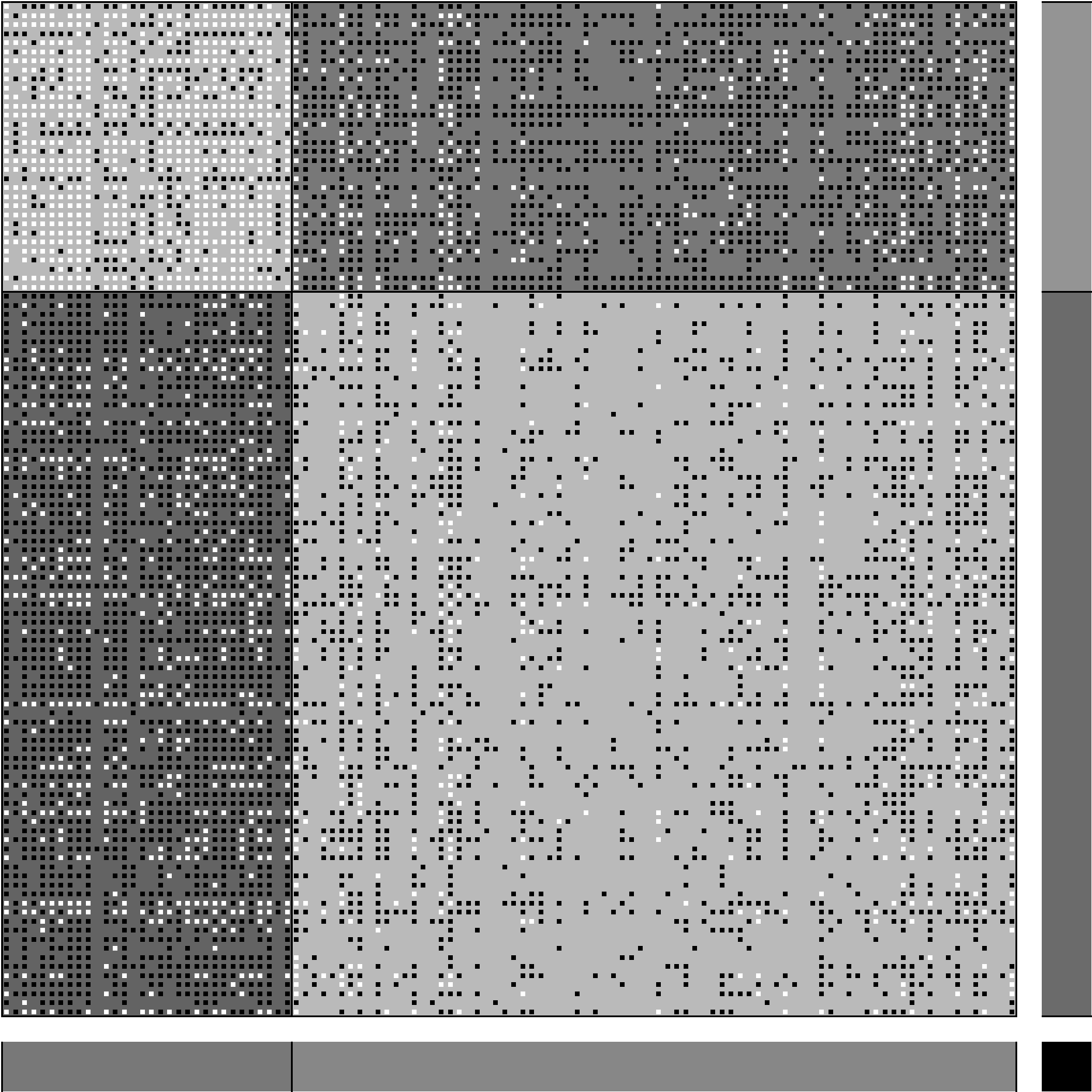} &
\includegraphics[height=3.25cm]{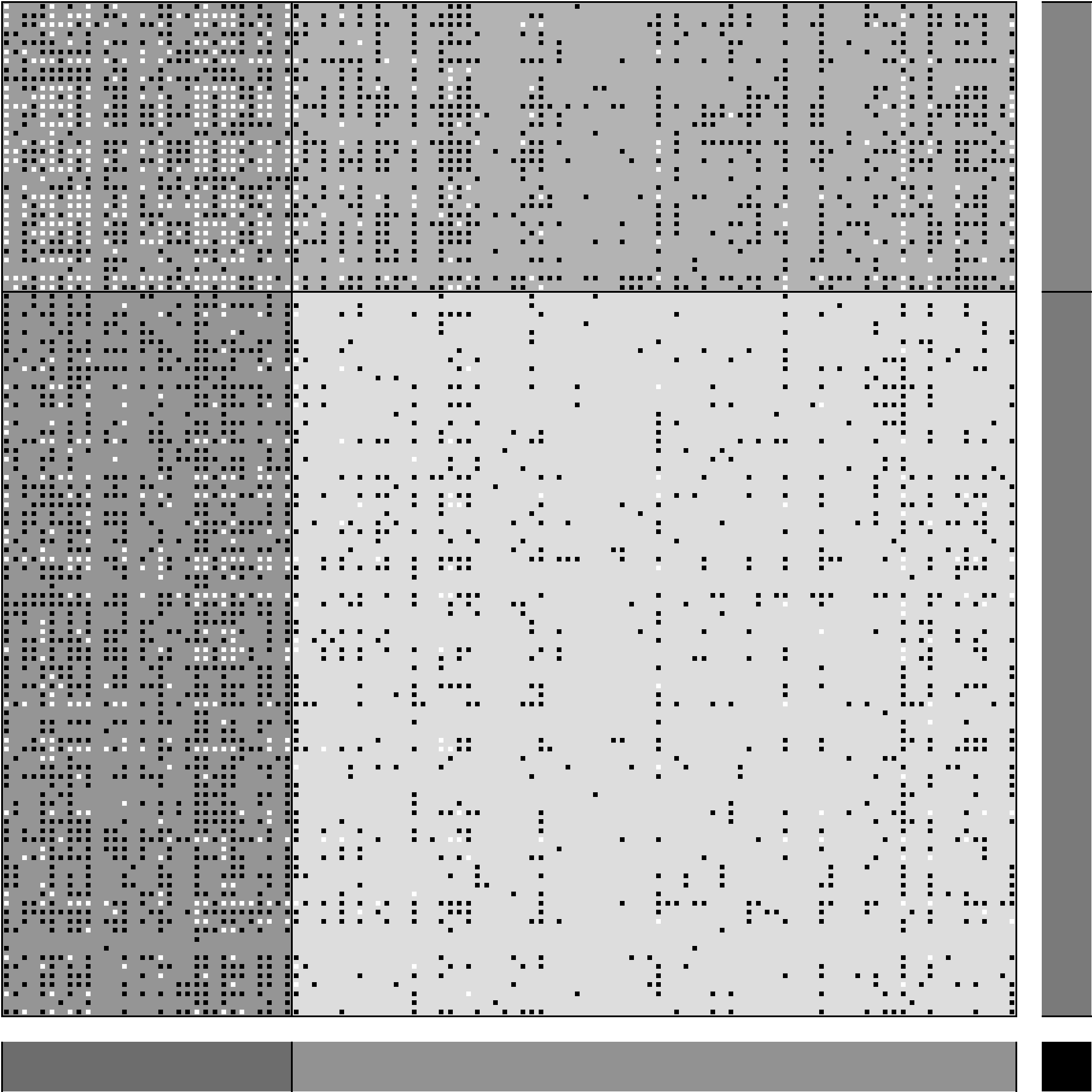} \\
meat \& meat prepartions & animal oil \& fats & paper \& paperboard & machinery & footwear\vspace{0.5cm}\\
\multicolumn{5}{r}{
\includegraphics[height=2cm]{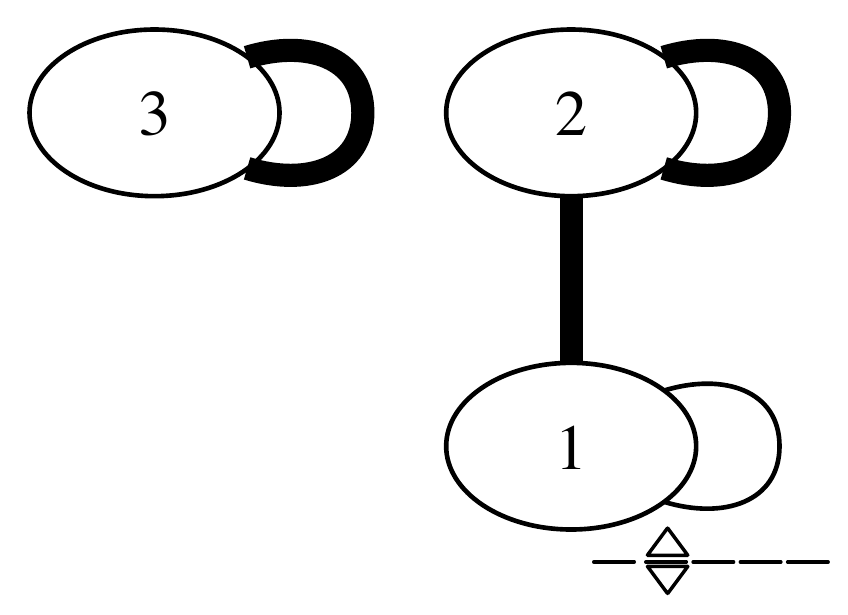}
\hspace{2cm}
\includegraphics[height=1.2cm]{Diamonds}
}\\
\includegraphics[height=3.25cm]{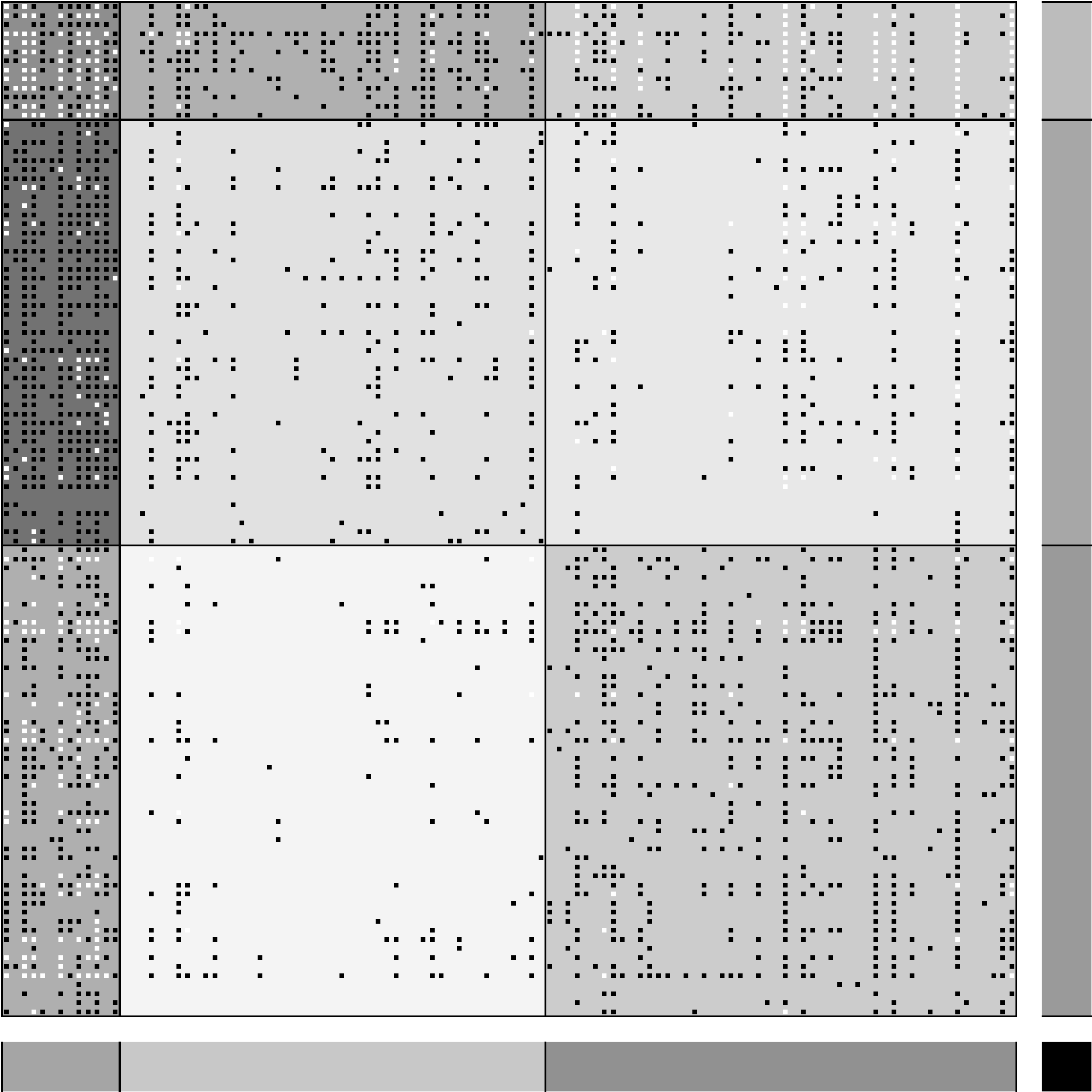} &
\includegraphics[height=3.25cm]{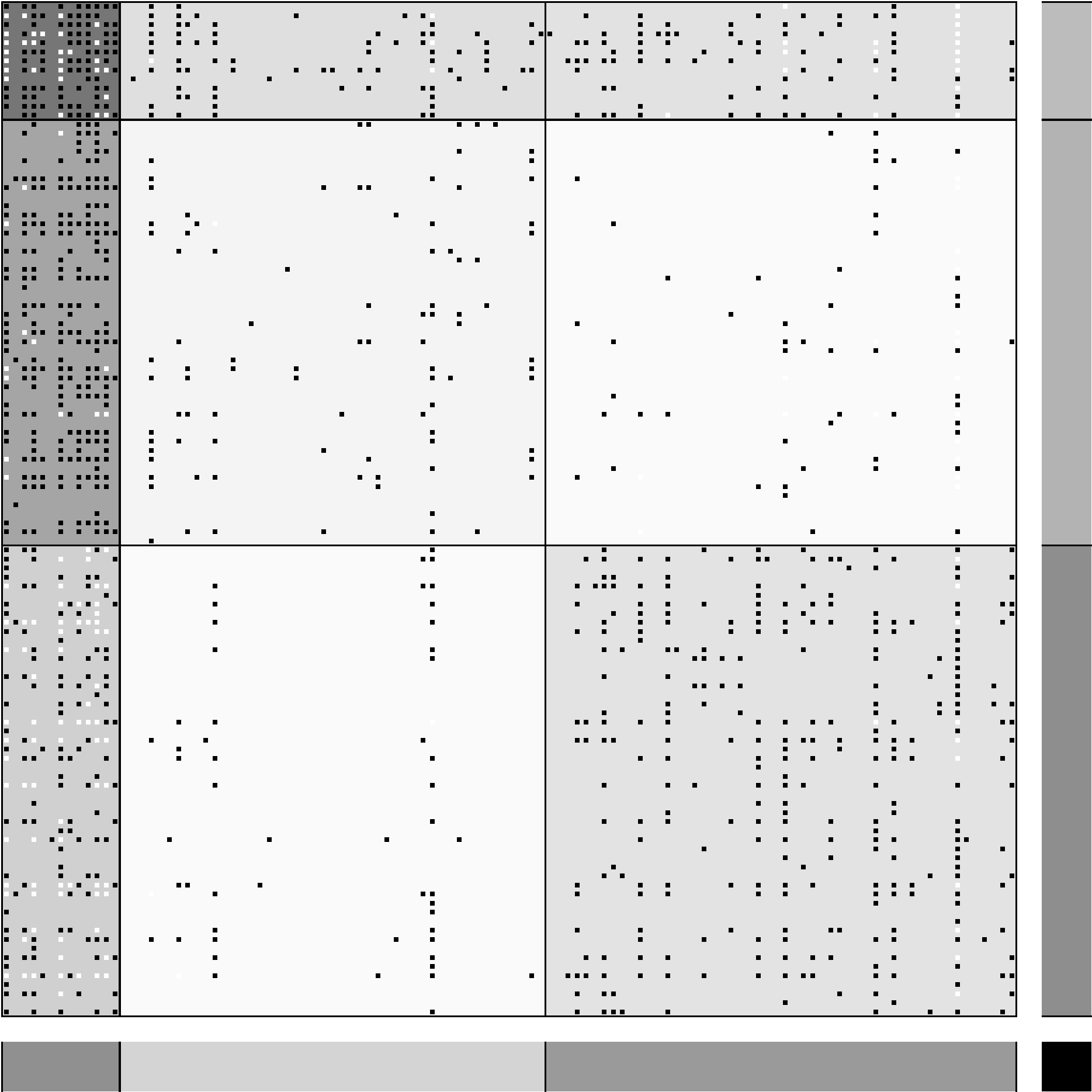} &
\includegraphics[height=3.25cm]{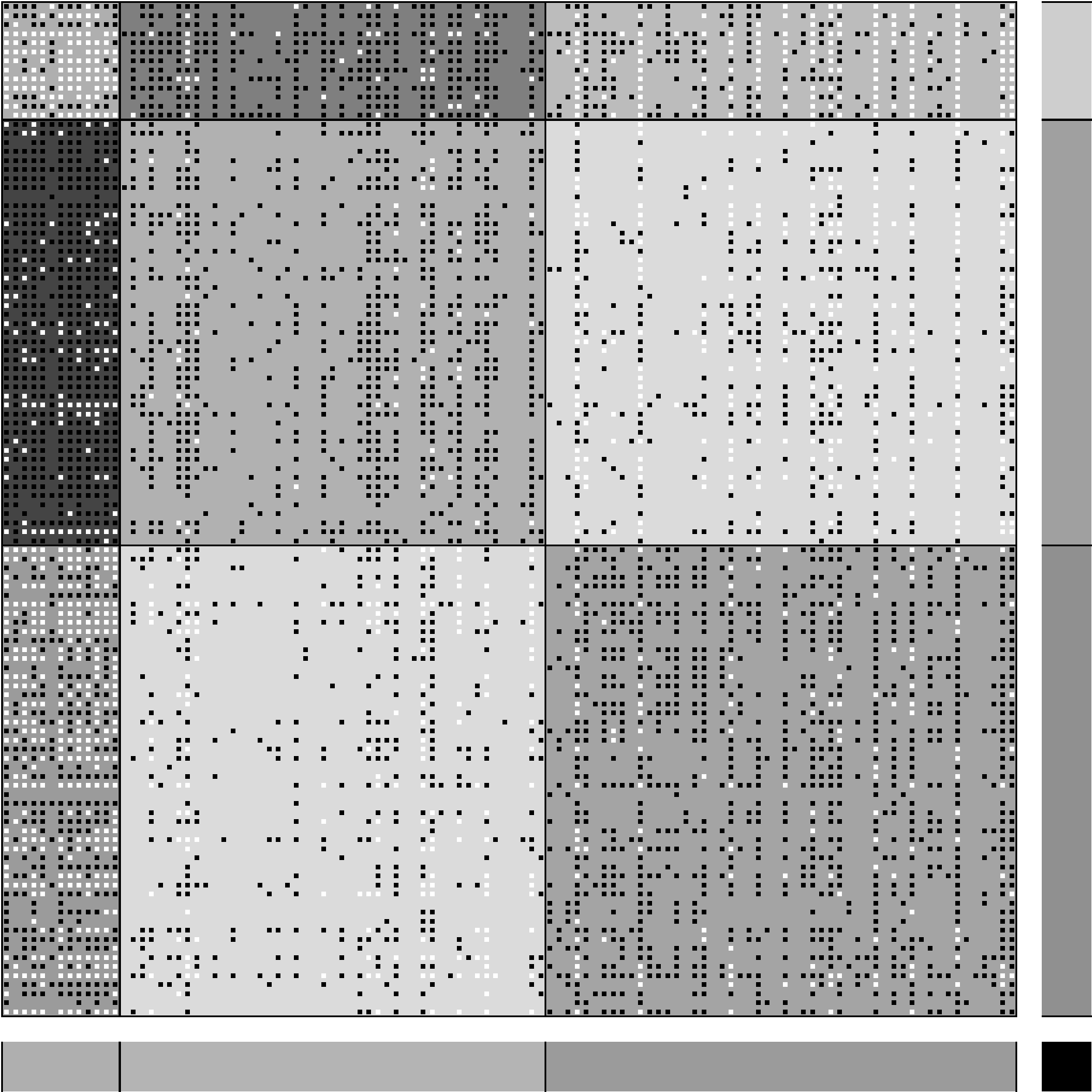} &
\includegraphics[height=3.25cm]{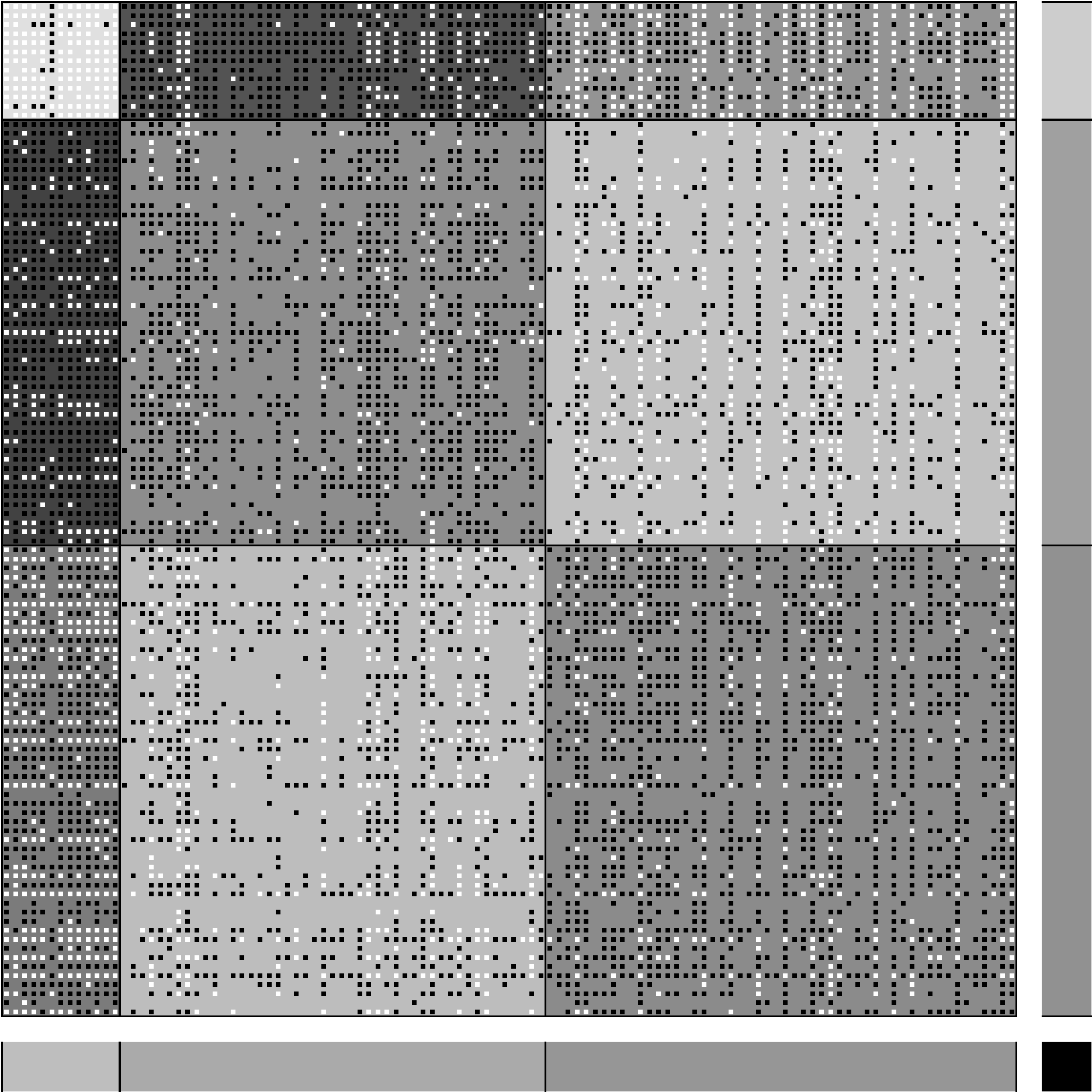} &
\includegraphics[height=3.25cm]{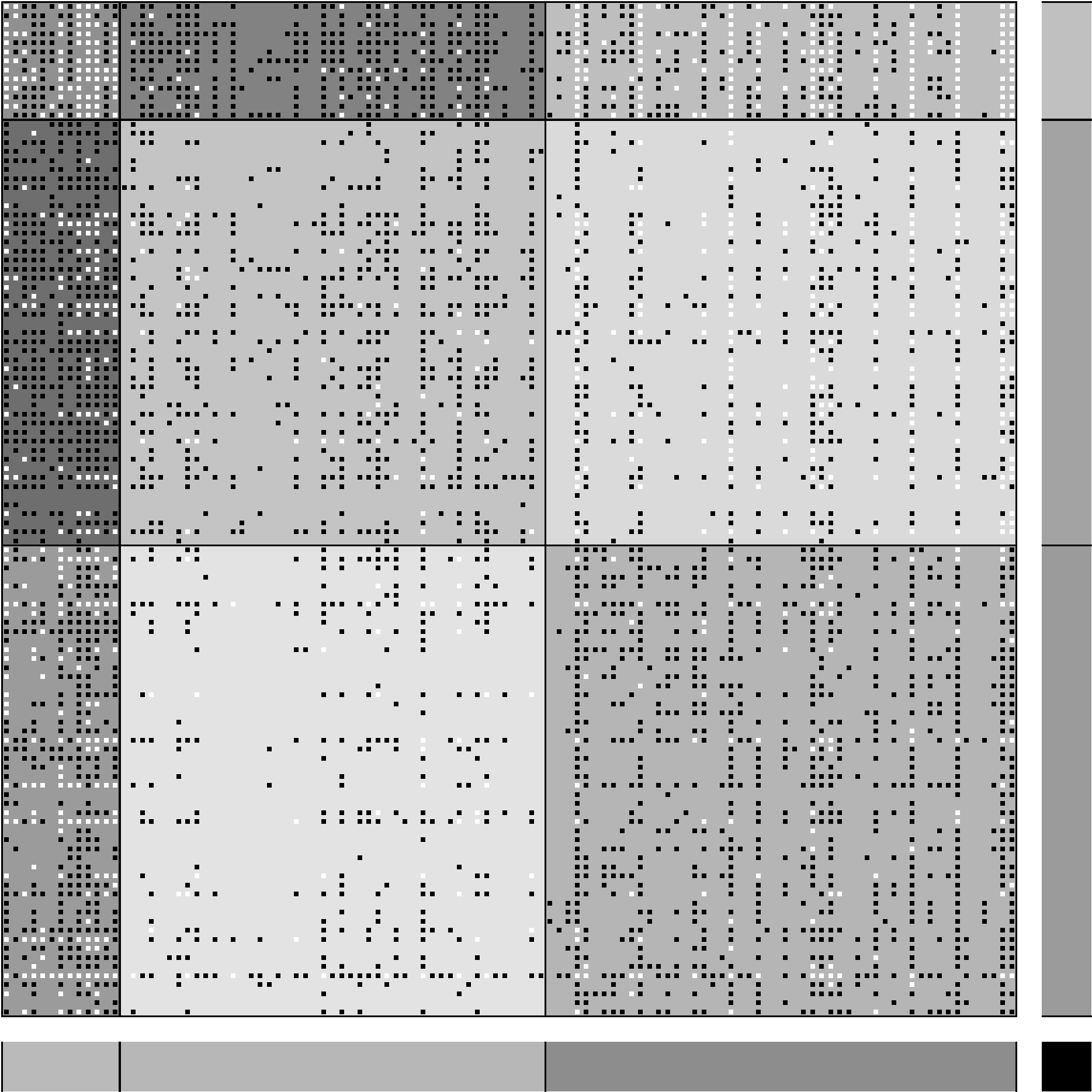} \\
meat \& meat prepartions & animal oil \& fats & paper \& paperboard & machinery & footwear\vspace{0.5cm}\\
\multicolumn{5}{r}{
\includegraphics[height=2cm]{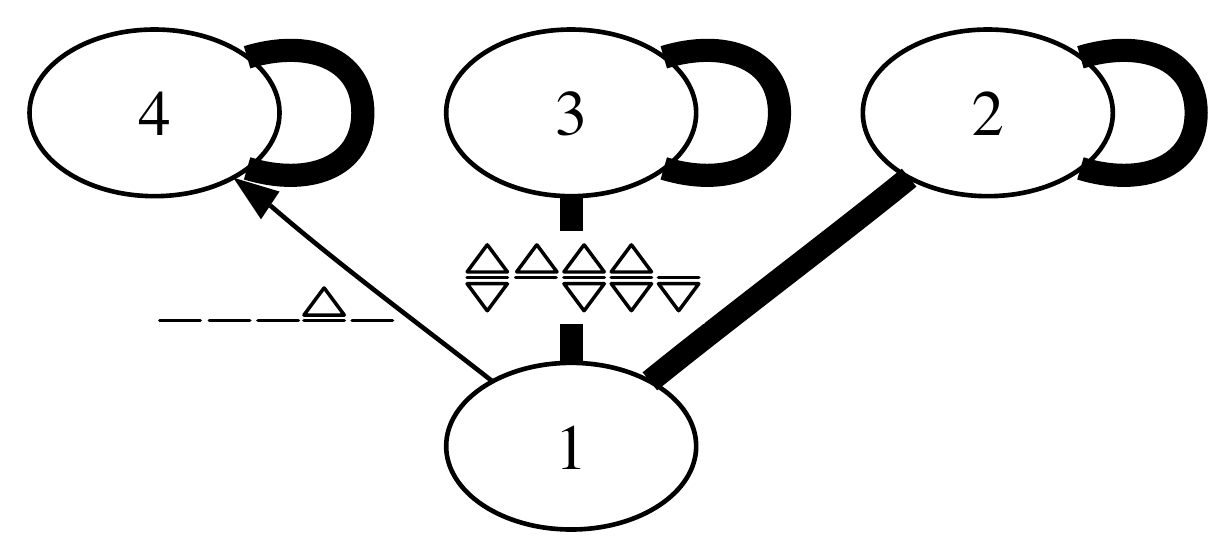}
\hspace{1cm}
\includegraphics[height=1.2cm]{Diamonds}
}\\
\includegraphics[height=3.25cm]{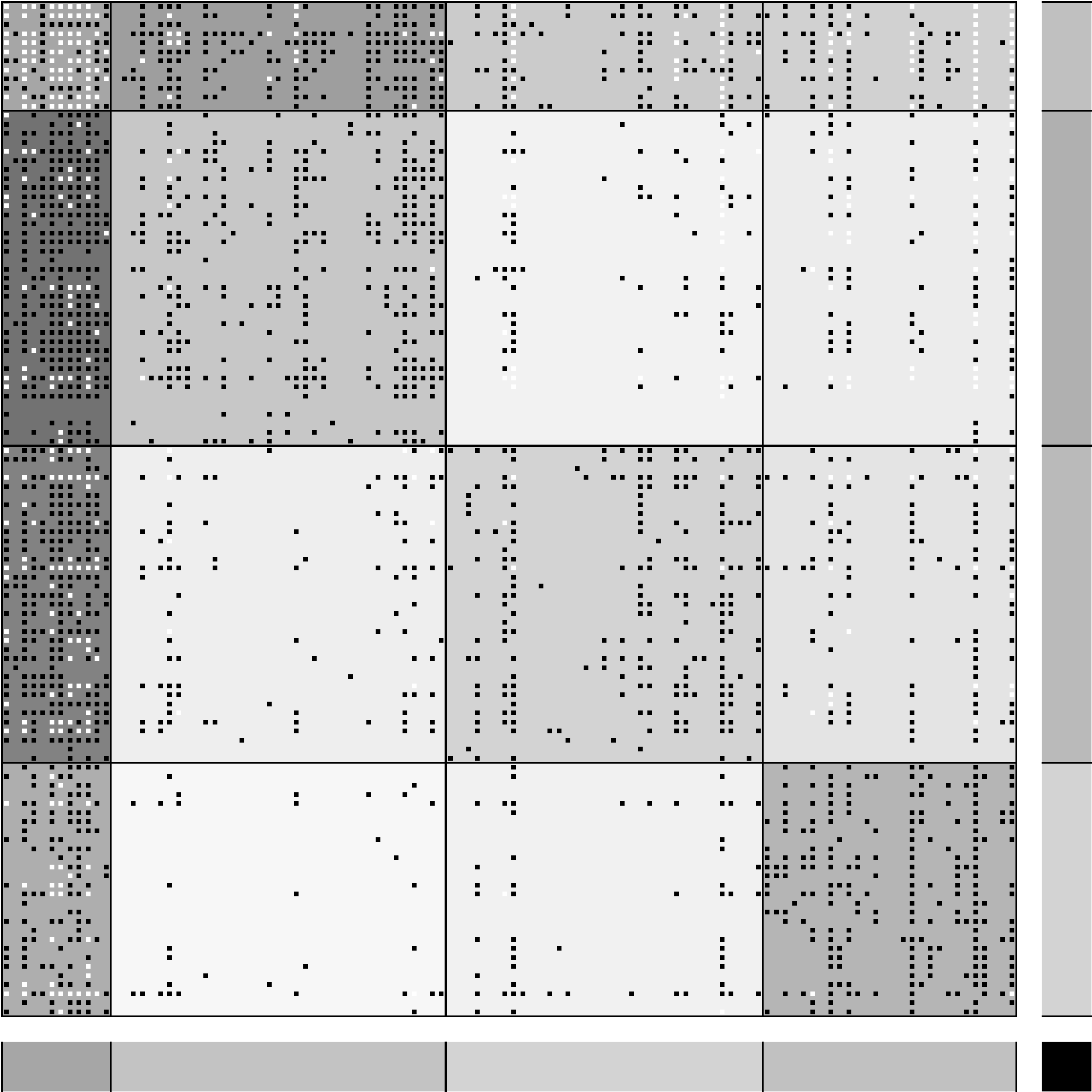} &
\includegraphics[height=3.25cm]{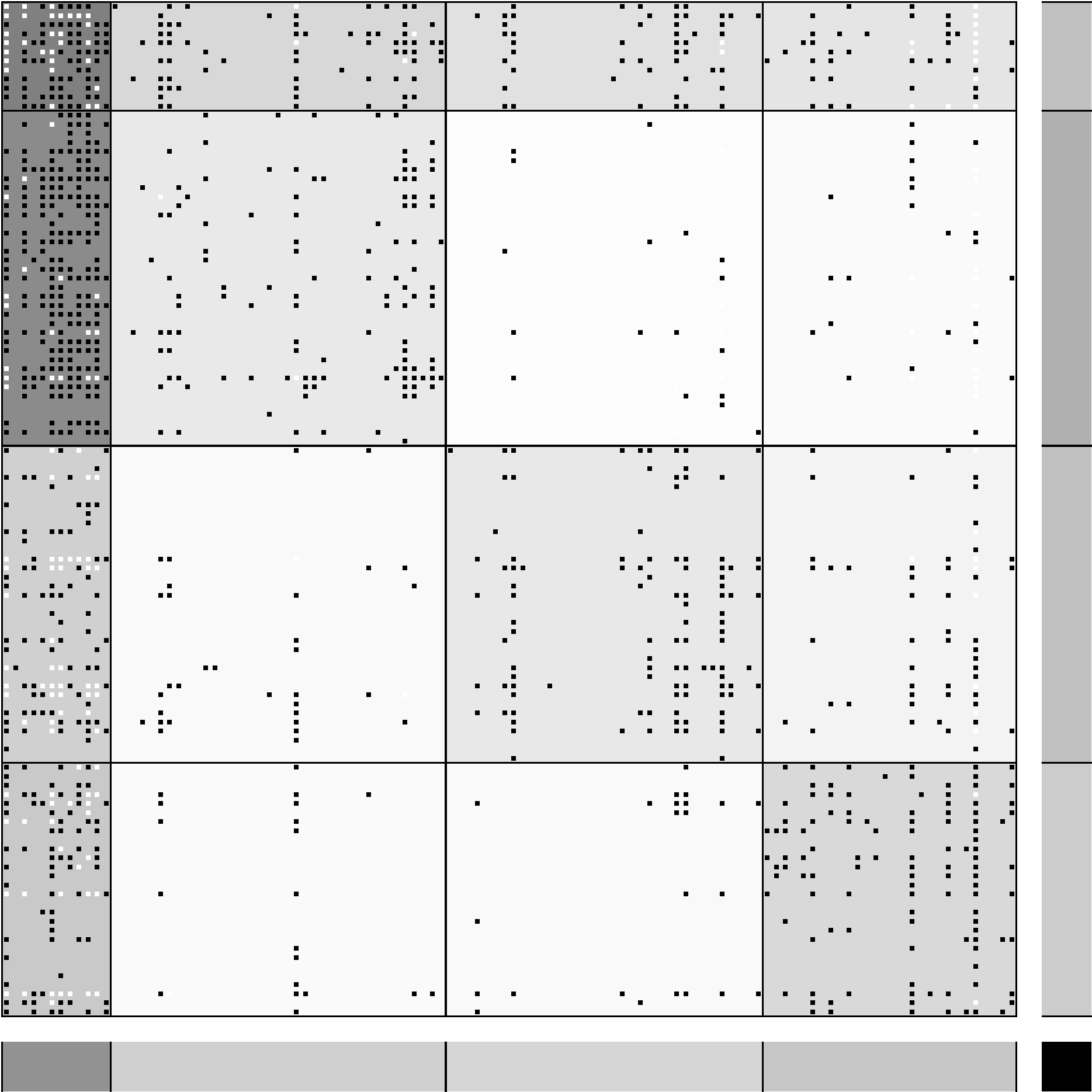} &
\includegraphics[height=3.25cm]{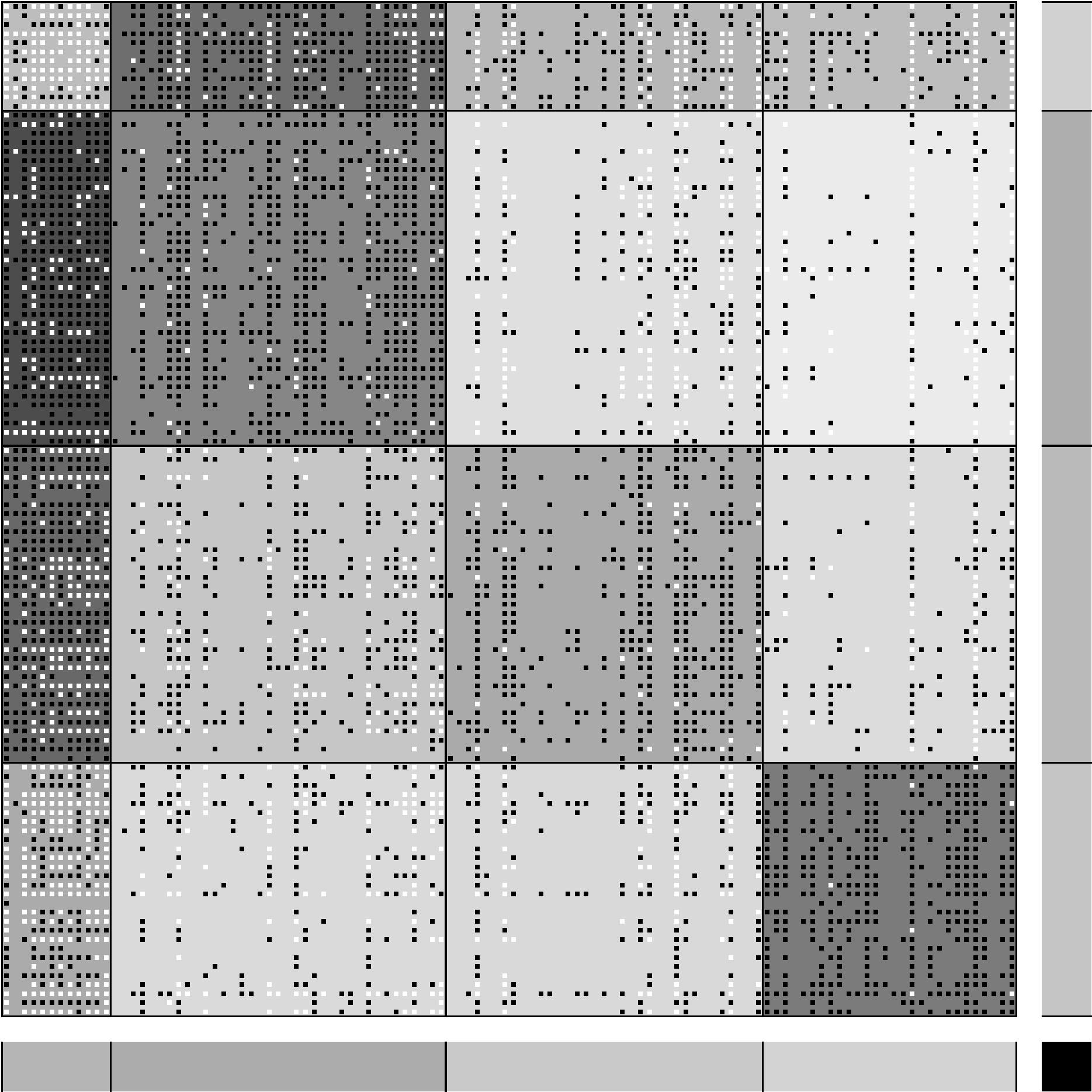} &
\includegraphics[height=3.25cm]{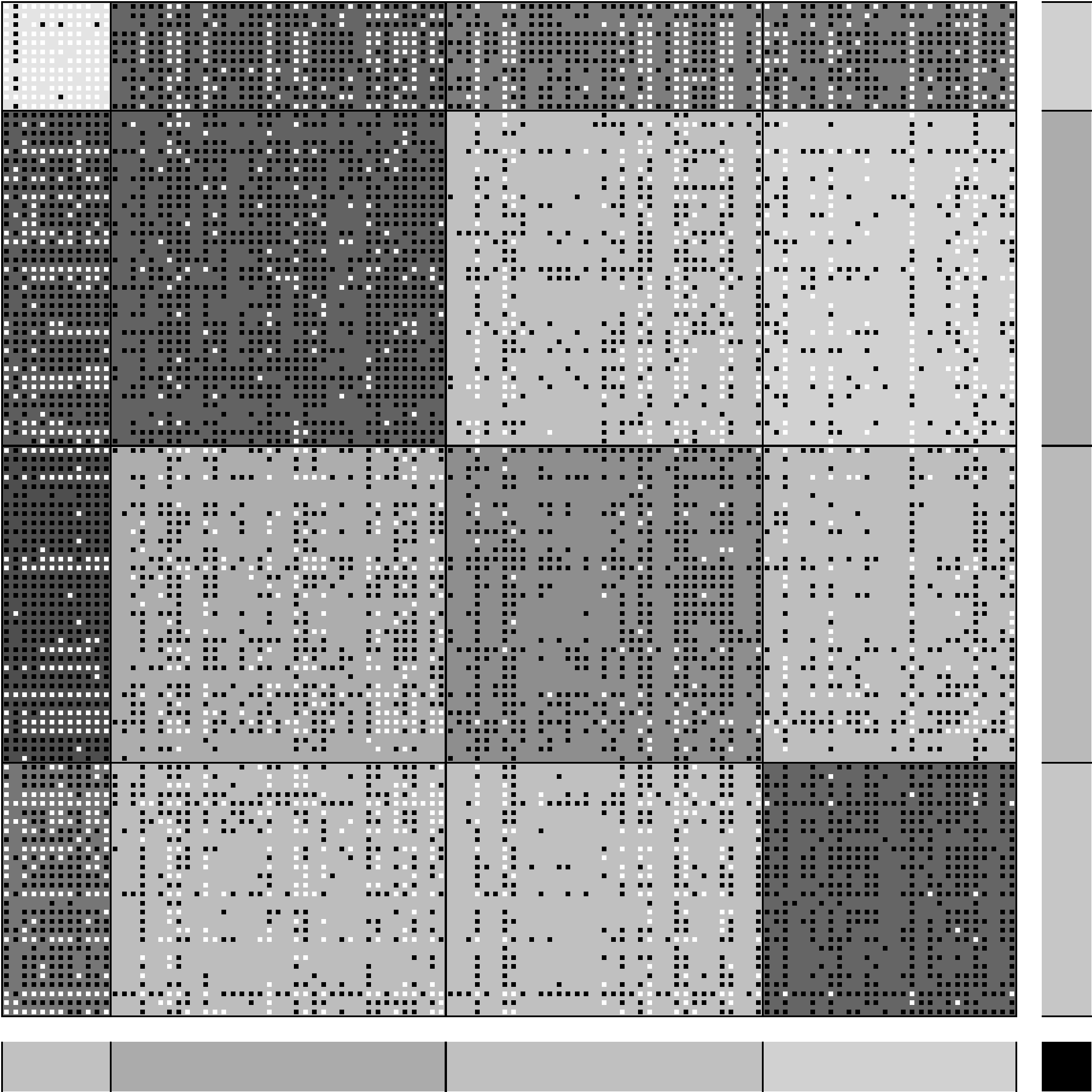} &
\includegraphics[height=3.25cm]{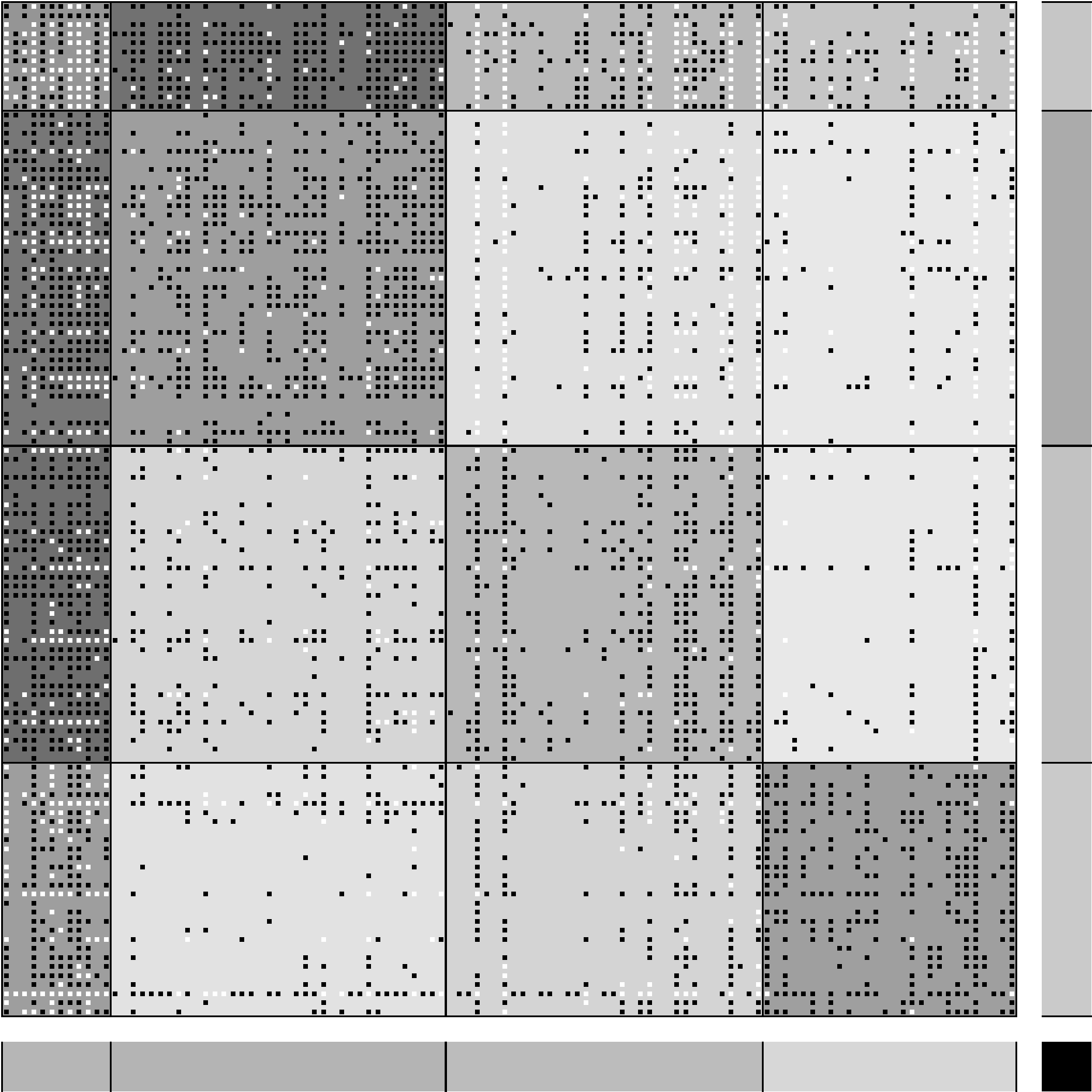} \\
meat \& meat prepartions & animal oil \& fats & paper \& paperboard & machinery & footwear\\
\end{tabular}

\caption{Consensus image graphs and matrix plots for the 5 commodities studied at $q=2$ to $q=4$ roles. Triangle labels indicate commodity and direction of the flow of goods. Labels and matrix plots from left to right: meat and meat preparations, animal oil and fat, paper and paper board, machinery, footwear. Each of these is representative of one of the five factor clusters of
55 matrices for all major international trade commodities \cite{SmithNemeth,SmithWhite}. Unlabeled links carry all five commodities in both directions. Blocks indexed from left to right and top to bottom. Side and bottom bars encode the marginal fraction of import and export of the total traded volume for each block in gray scale, respectively. Black dots indicate trade greater than expected from the marginals for pairs of countries, white dots smaller than expected. Background shading of blocks corresponds to density of black dots in block.  See Table 1 
for individual countries grouped in blocks.}
\label{SuppRBM1}
\end{figure*}
\end{samepage}

\clearpage

\begin{figure*}[t]

\begin{tabular}{ccccc}
\multicolumn{5}{r}{
\includegraphics[height=3cm]{q5_BW_Diamonds}
\hspace{1cm}
\includegraphics[height=1.2cm]{Diamonds}
}\\
\includegraphics[height=3.35cm]{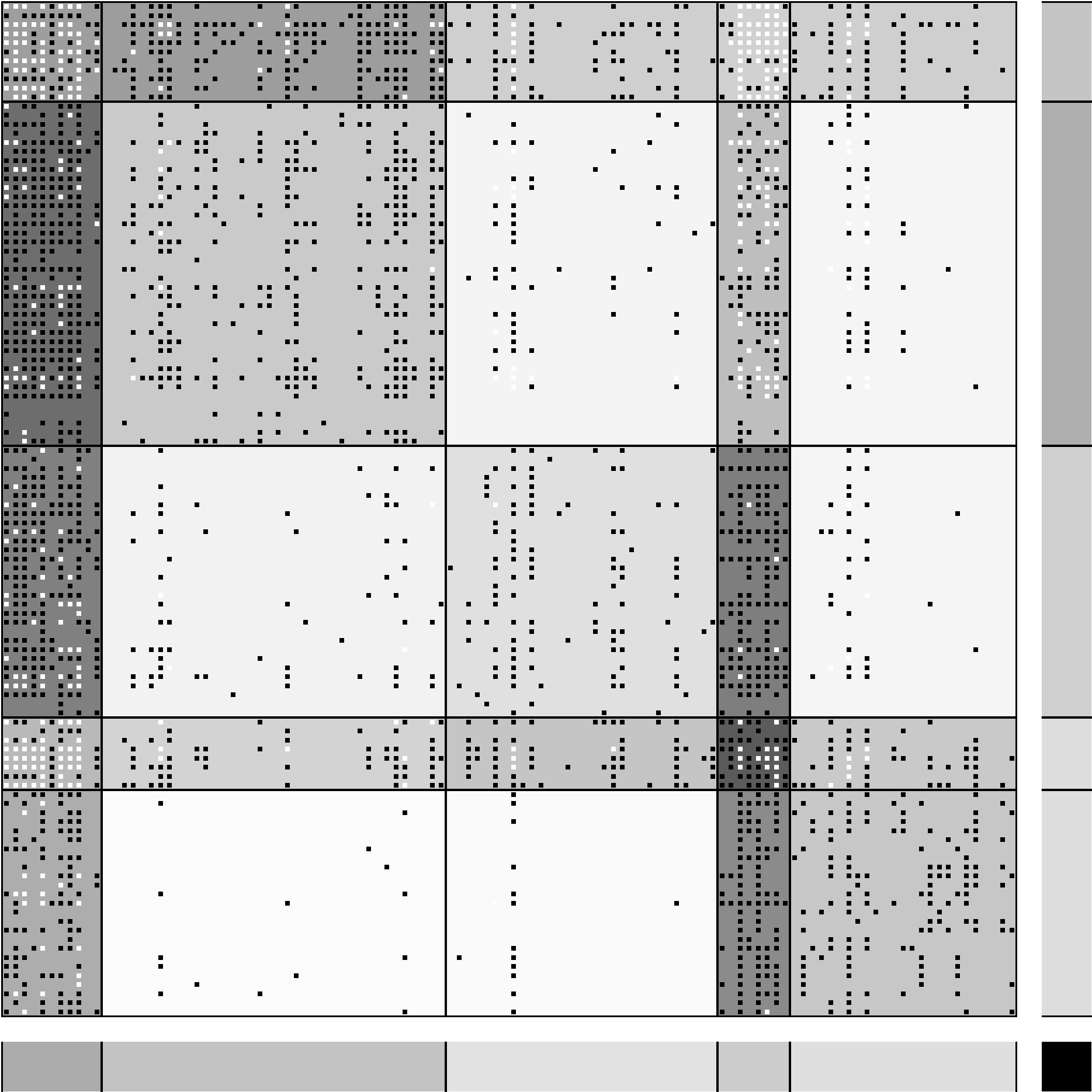} &
\includegraphics[height=3.25cm]{c41_q5} &
\includegraphics[height=3.25cm]{c64_q5} &
\includegraphics[height=3.25cm]{c71_q5} &
\includegraphics[height=3.25cm]{c85_q5} \\
meat \& meat prepartions & animal oil \& fats & paper \& paperboard & machinery & footwear\vspace{0.25cm}\\
\multicolumn{5}{r}{
\includegraphics[height=2.5cm]{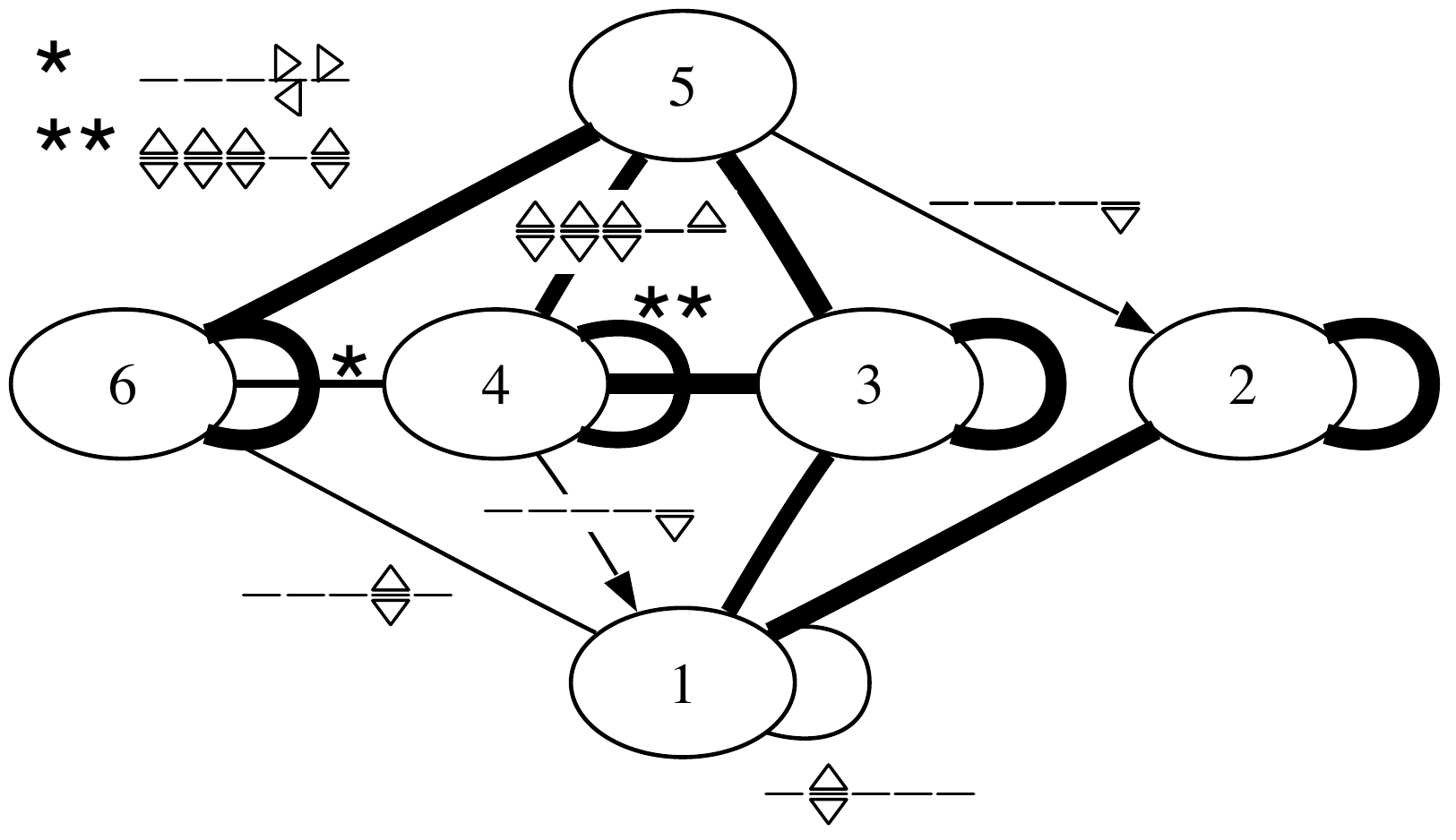}
\hspace{0.5cm}
\includegraphics[height=1.2cm]{Diamonds}
}\\

\includegraphics[height=3.25cm]{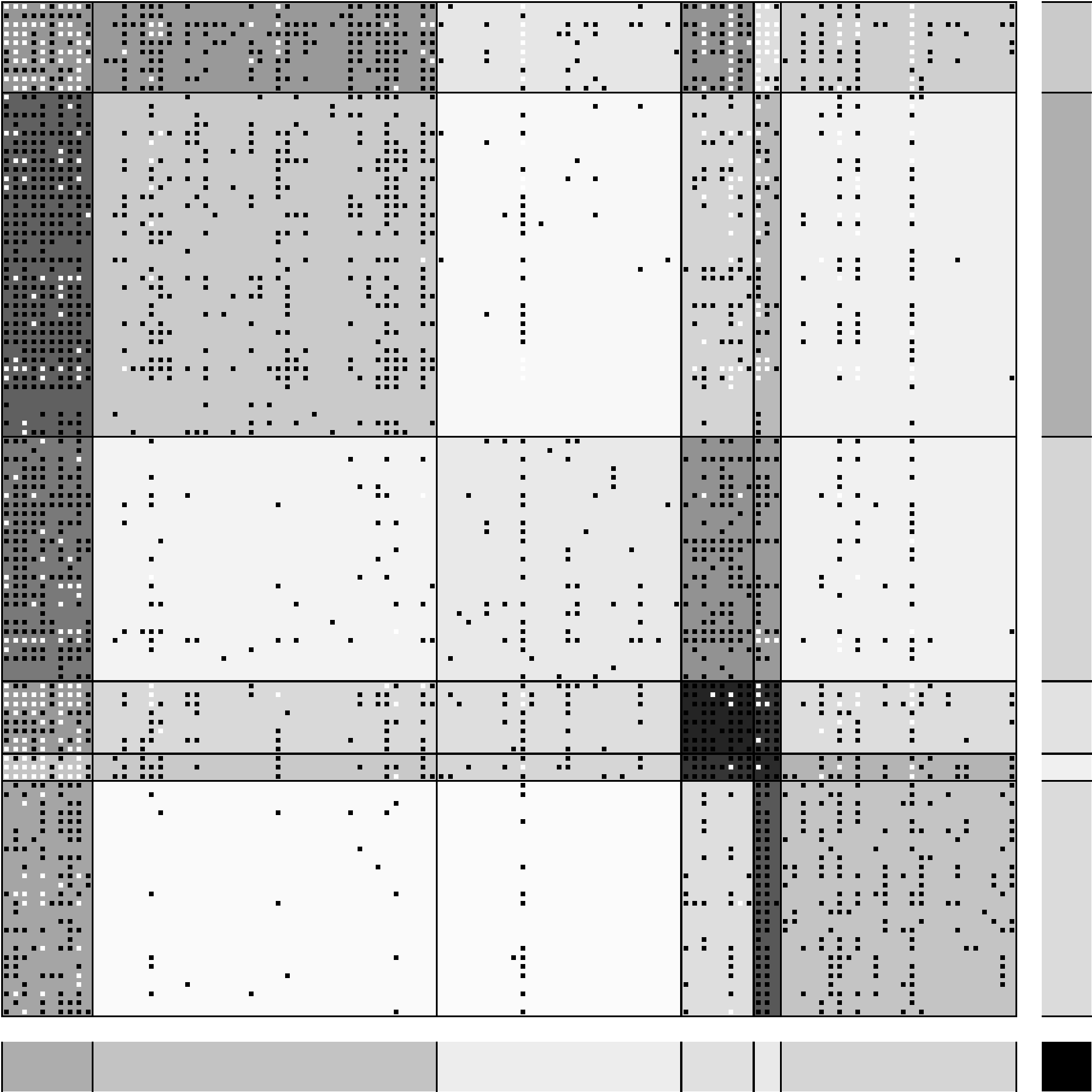} &
\includegraphics[height=3.25cm]{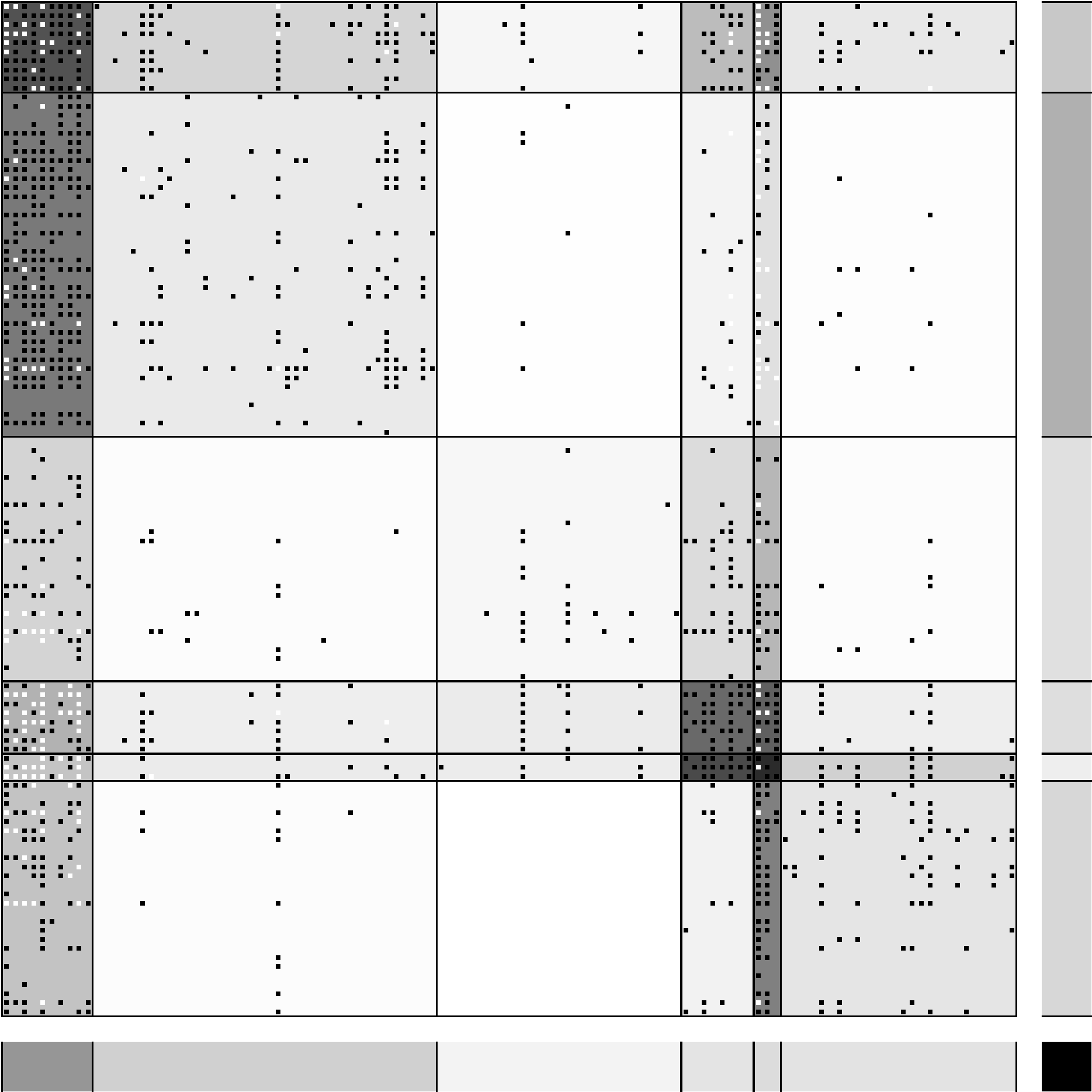} &
\includegraphics[height=3.25cm]{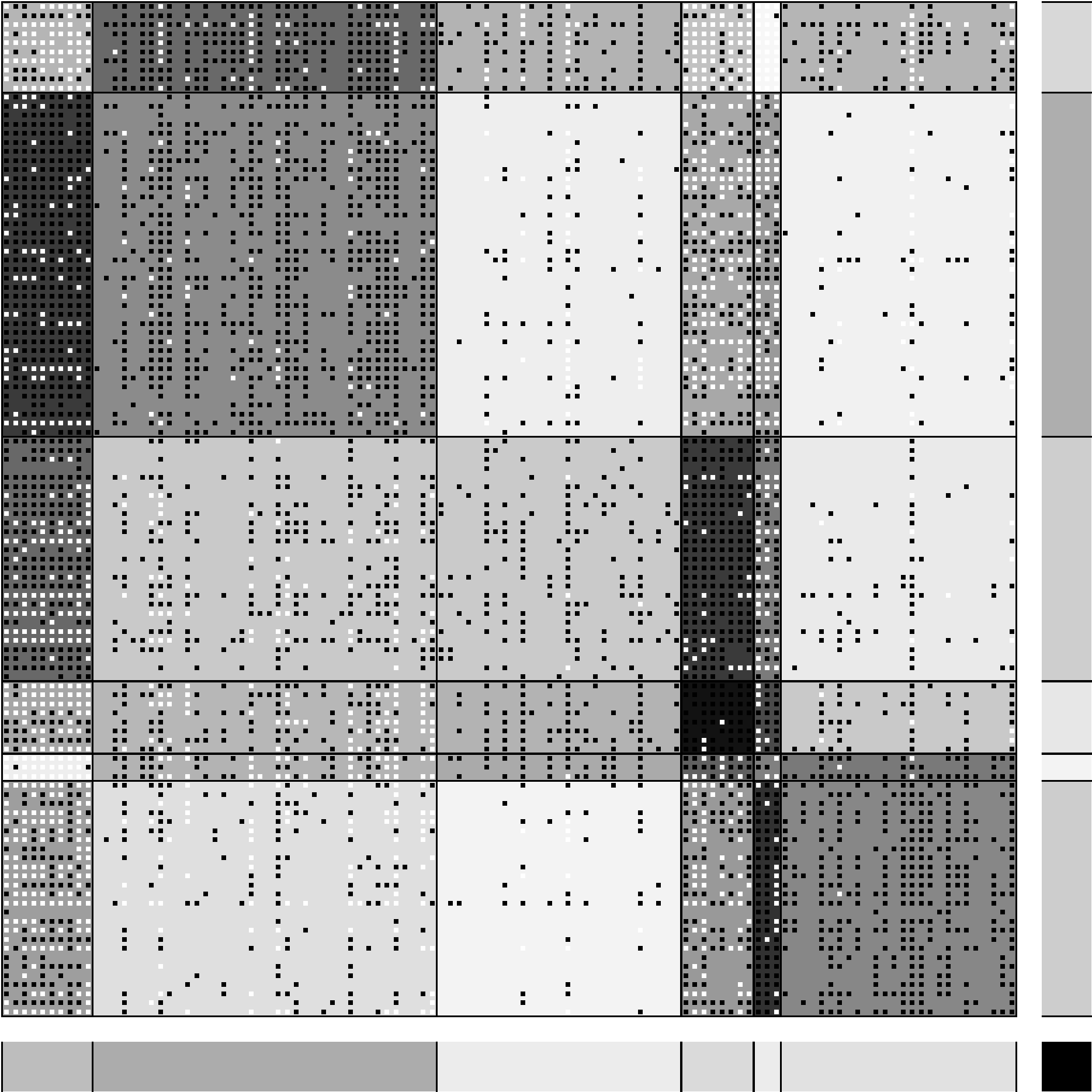} &
\includegraphics[height=3.25cm]{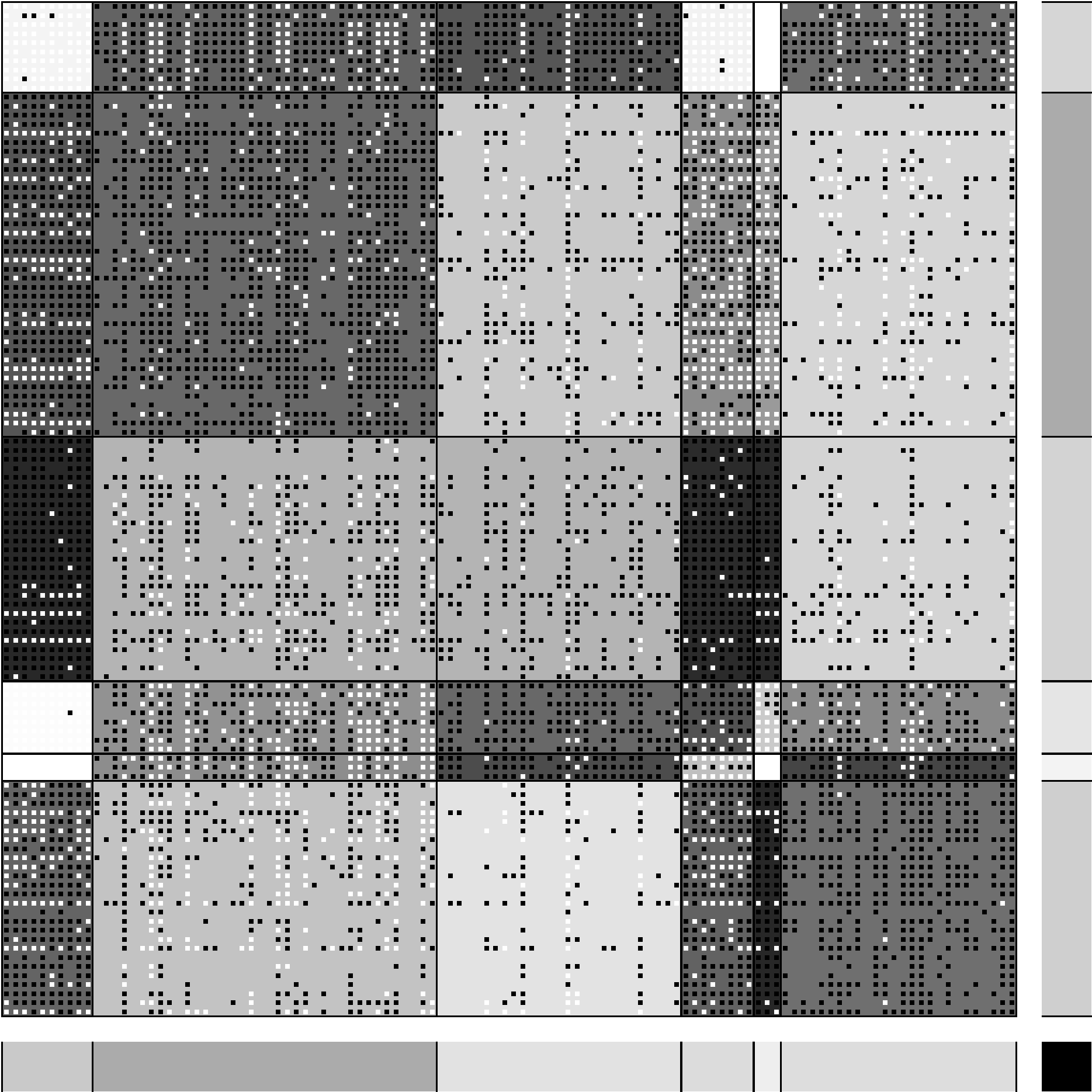} &
\includegraphics[height=3.25cm]{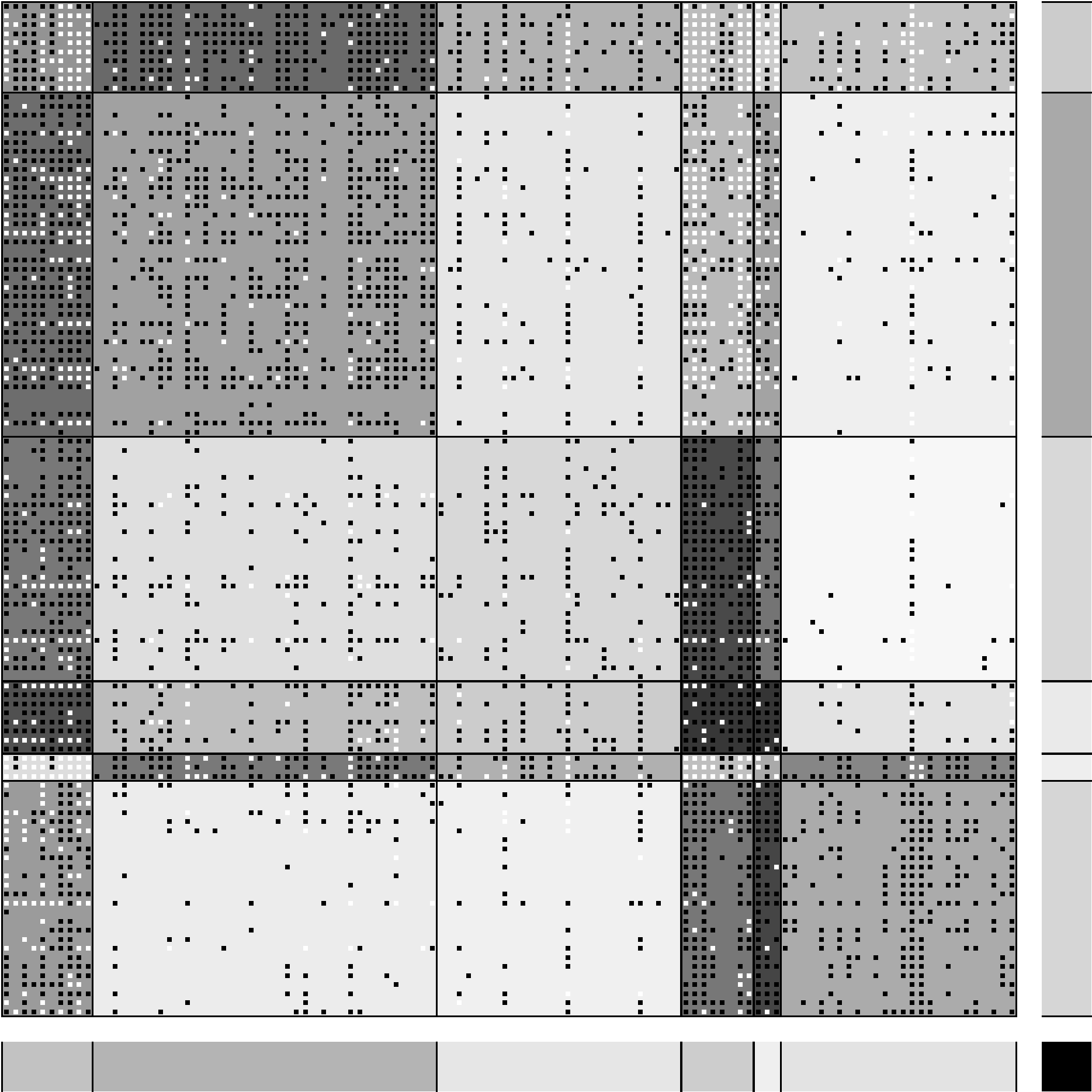} \\
meat \& meat prepartions & animal oil \& fats & paper \& paperboard & machinery & footwear\vspace{0.25cm}\\
\multicolumn{5}{r}{
\includegraphics[height=4.5cm]{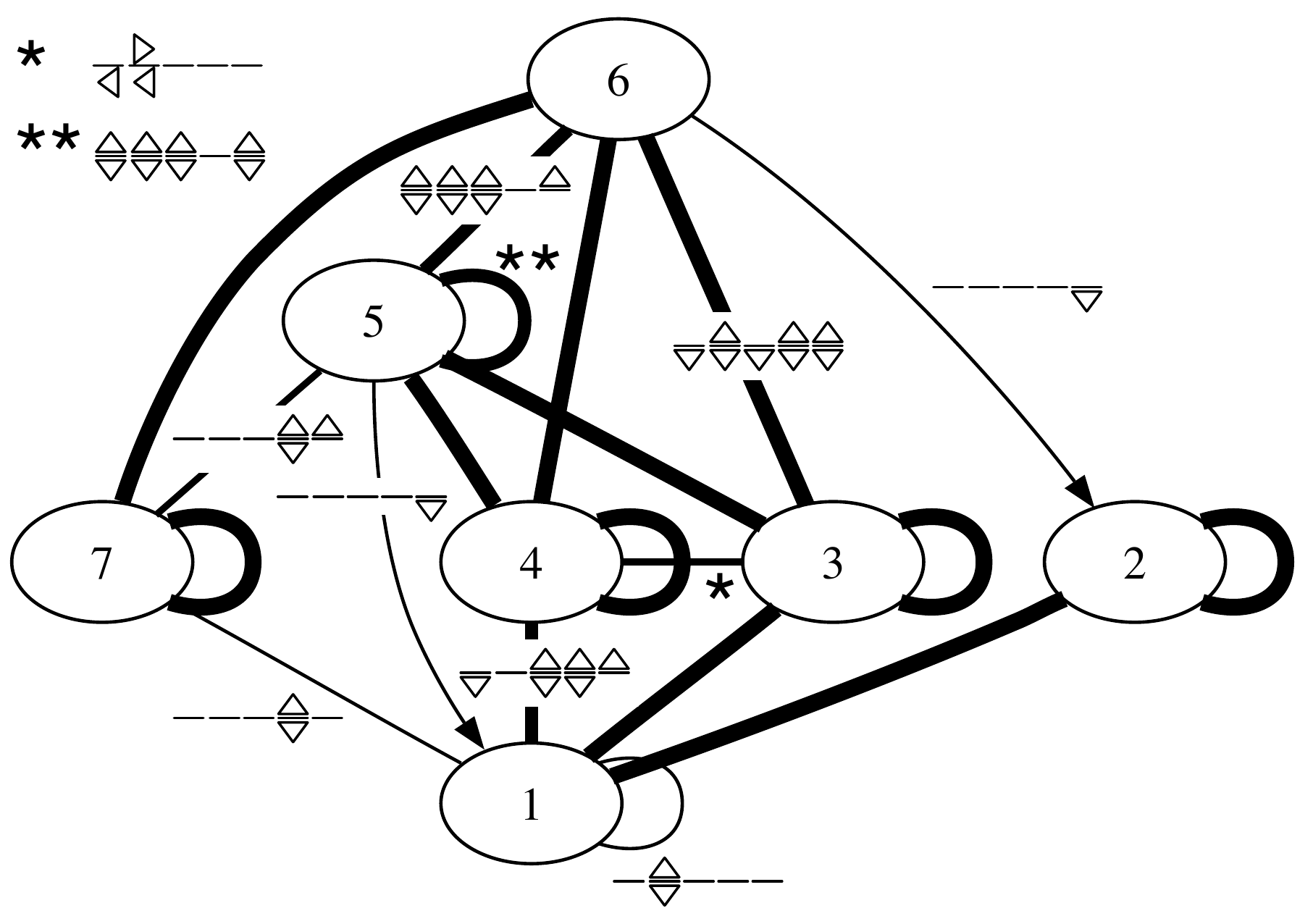}
\includegraphics[height=1.2cm]{Diamonds}
}\\

\includegraphics[height=3.25cm]{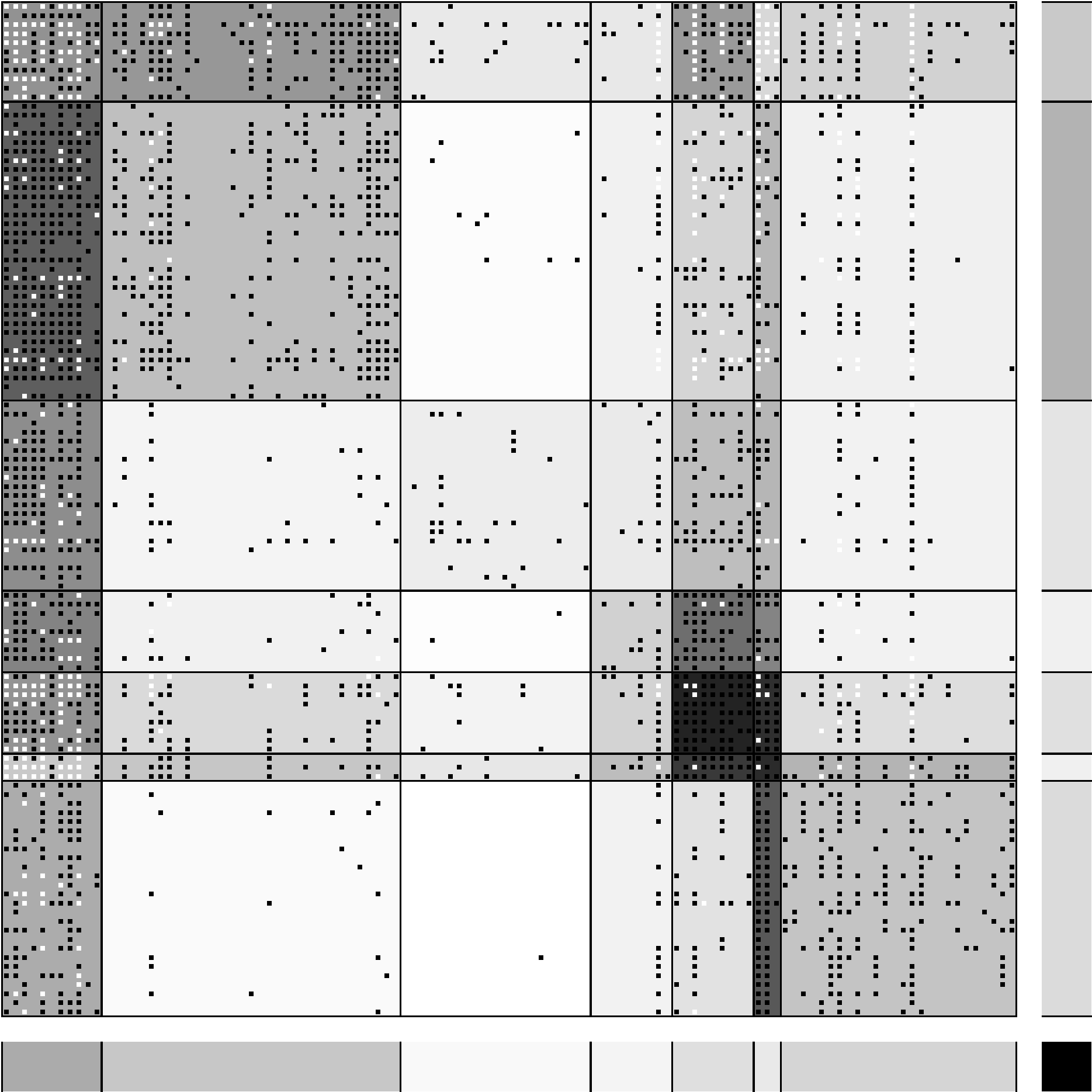} &
\includegraphics[height=3.25cm]{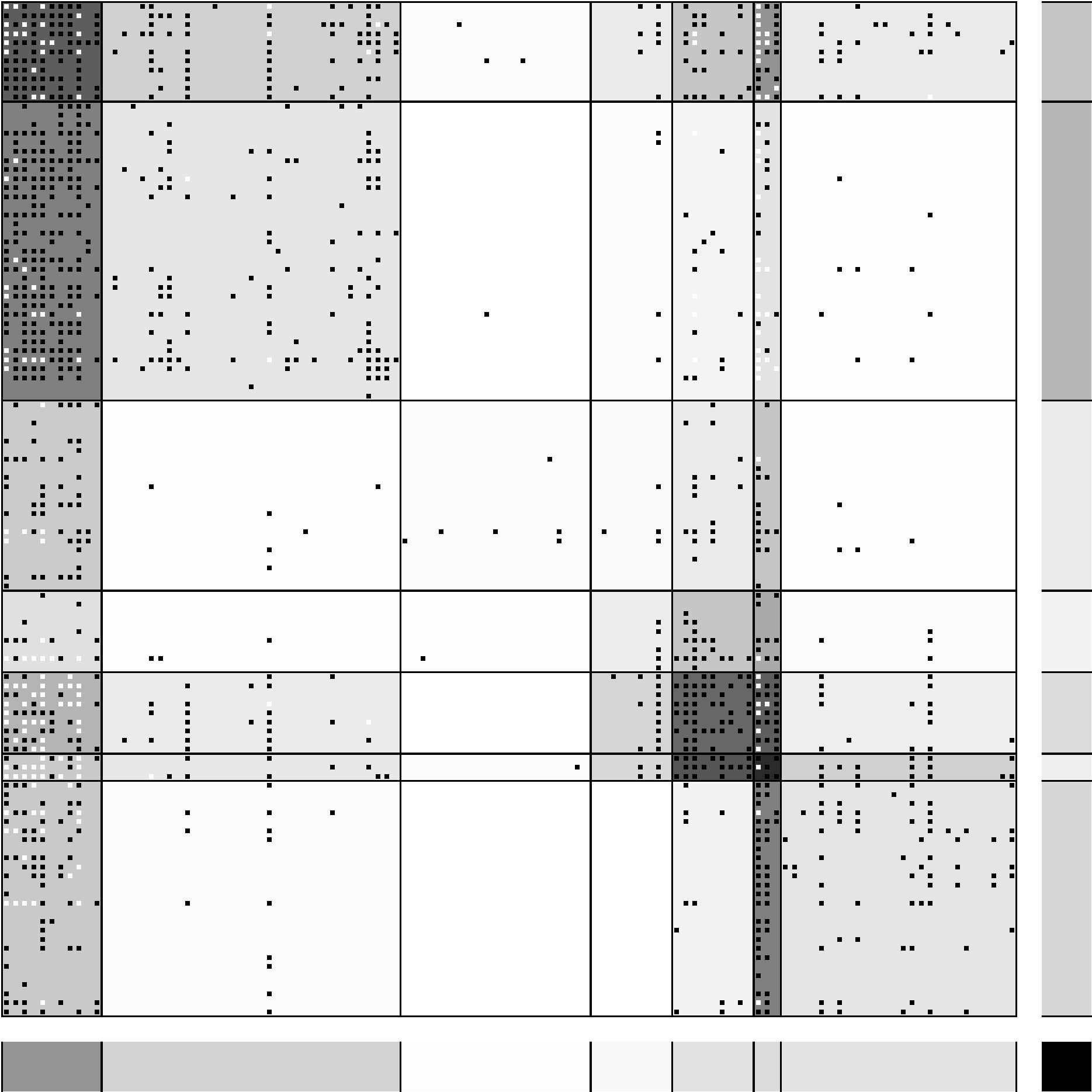} &
\includegraphics[height=3.25cm]{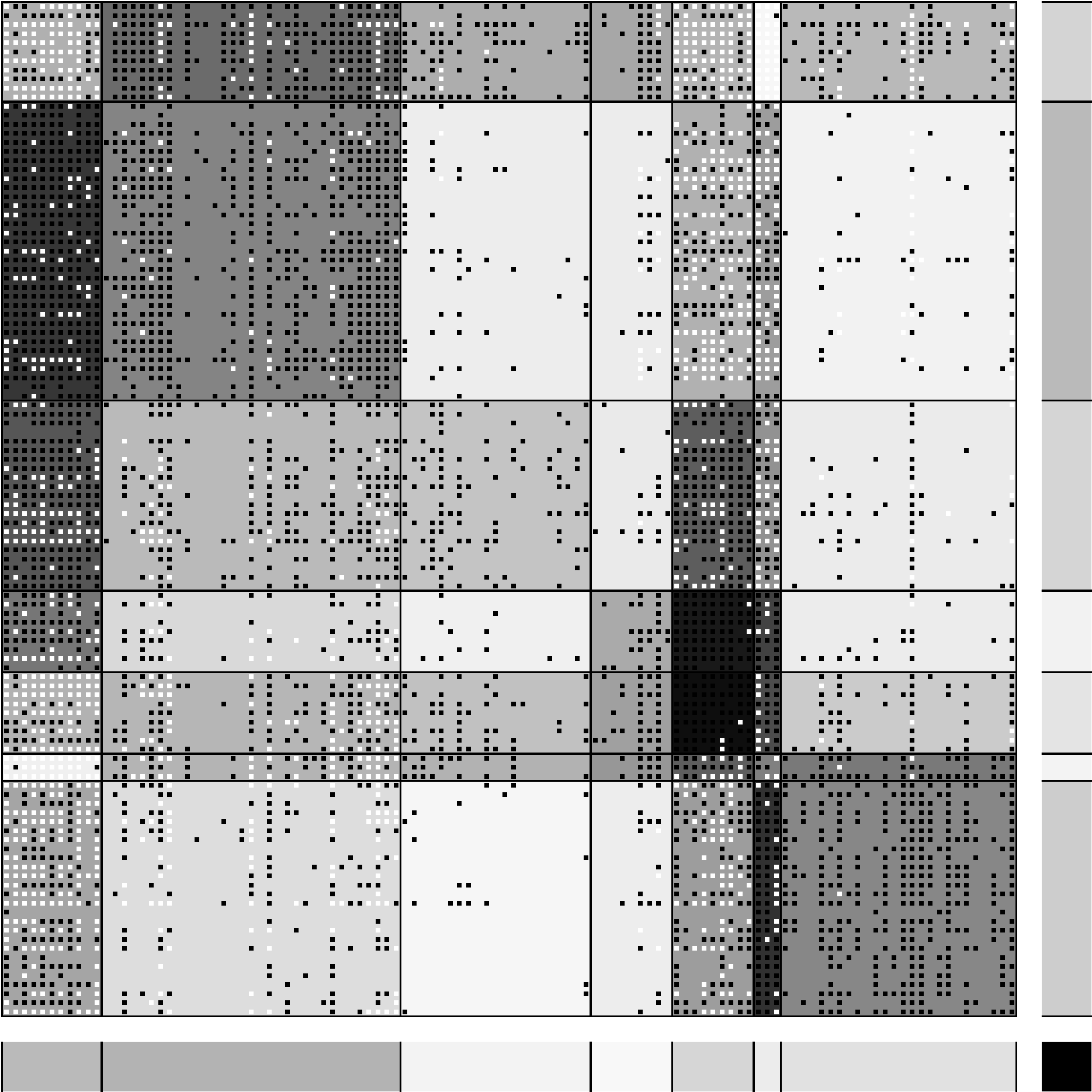} &
\includegraphics[height=3.25cm]{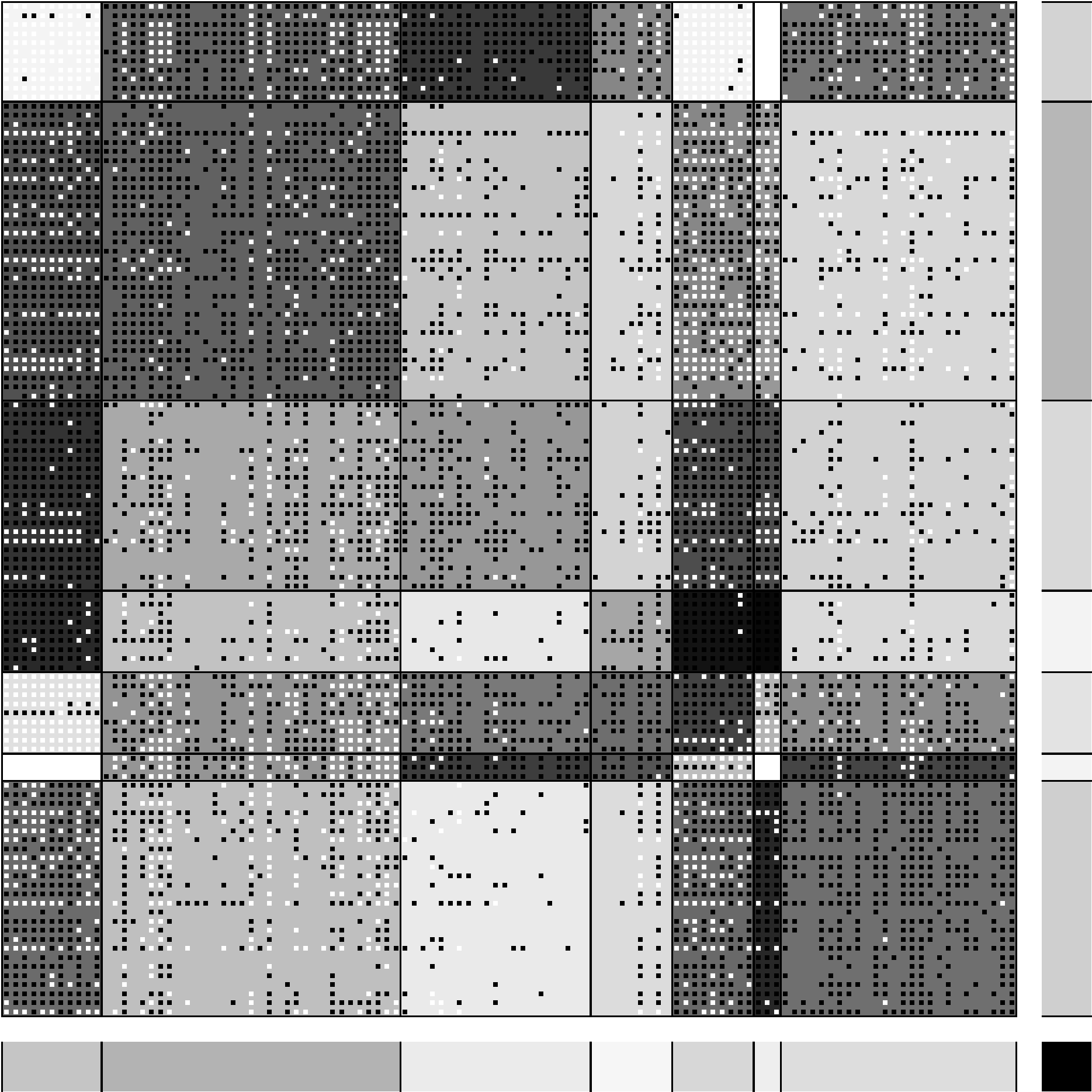} &
\includegraphics[height=3.25cm]{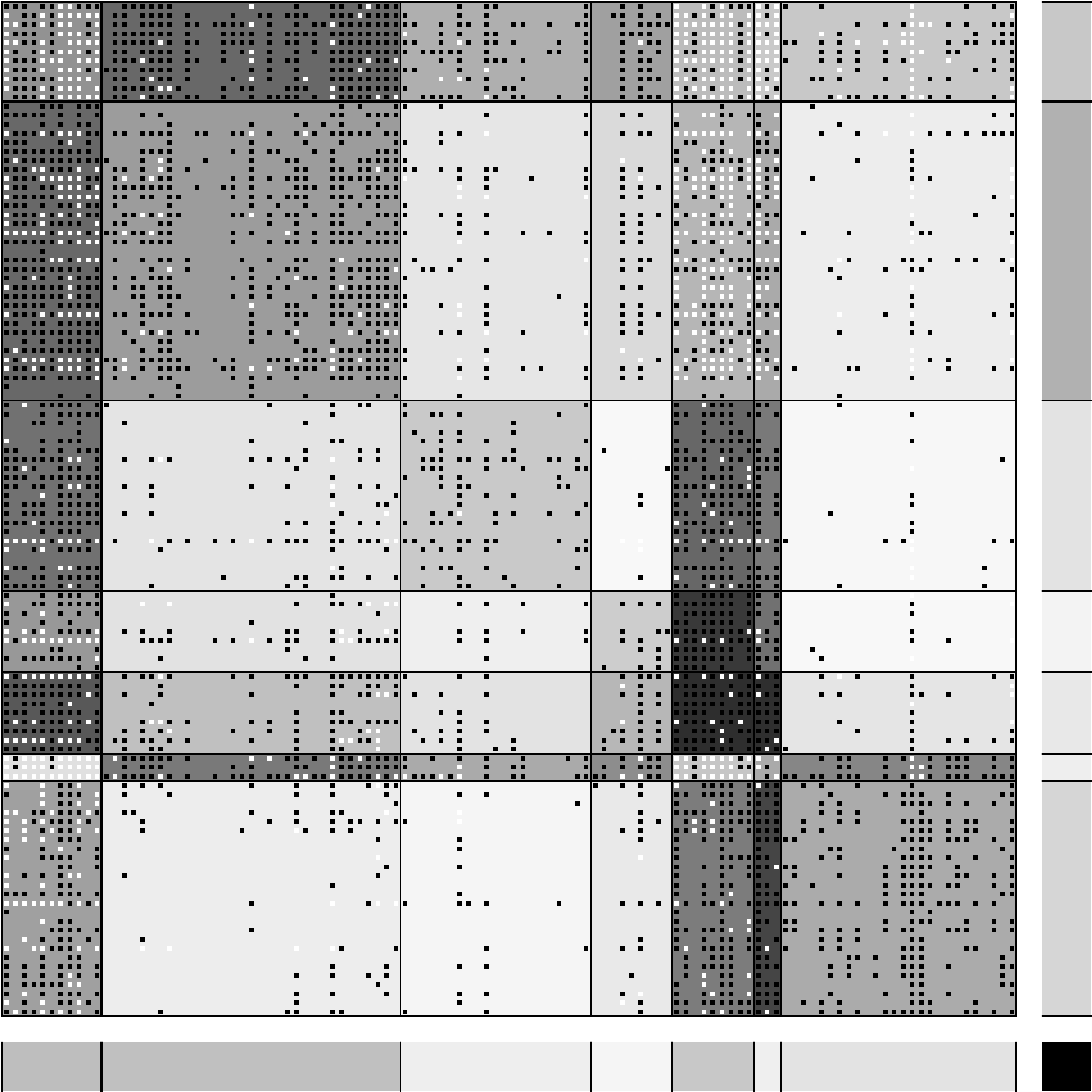} \\
meat \& meat prepartions & animal oil \& fats & paper \& paperboard & machinery & footwear\\
\end{tabular}

\caption{Consensus image graphs and matrix plots for the 5 commodities studied at $q=5$ to $q=7$ roles. Triangle labels indicate commodity and direction of the flow of goods. Labels and matrix plots from left to right: meat and meat preparations, animal oil and fat, paper and paper board, machinery, footwear. Each of these is representative of one of the five factor clusters of
55 matrices for all major international trade commodities \cite{SmithNemeth,SmithWhite}. Unlabeled links carry all five commodities in both directions. Blocks indexed from left to right and top to bottom. Side and bottom bars encode the marginal fraction of import and export of the total traded volume for each block in gray scale, respectively. Black dots indicate trade greater than expected from the marginals for pairs of countries, white dots smaller than expected. Background shading of blocks corresponds to density of black dots in block. See Table 1 
for individual countries grouped in blocks.}
\label{SuppRBM2}
\end{figure*}

\clearpage

\begin{figure*}[t]

\begin{tabular}{ccccc}
\multicolumn{5}{r}{
\includegraphics[height=6cm]{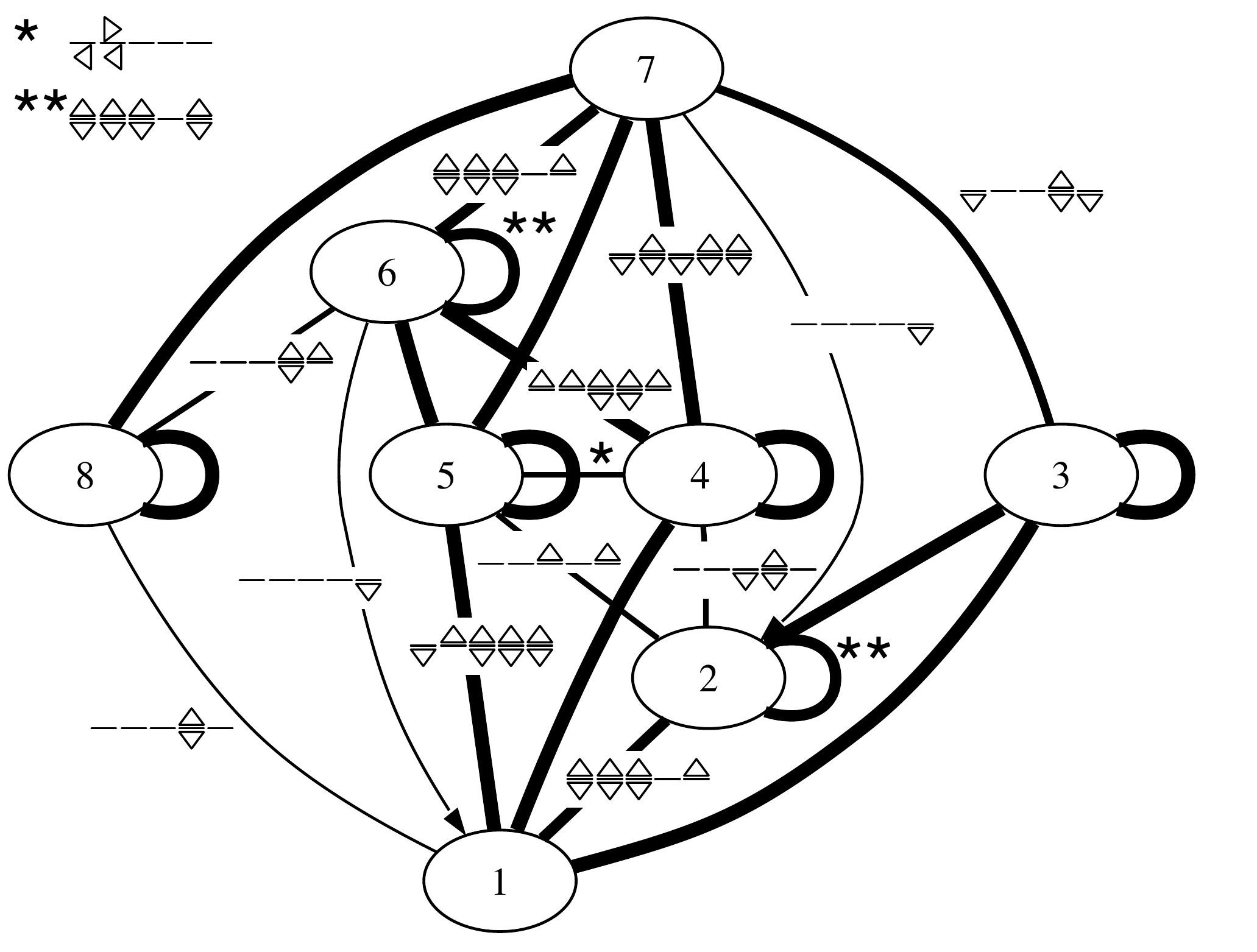}
\includegraphics[height=1.2cm]{Diamonds}
}\\
\includegraphics[height=3.25cm]{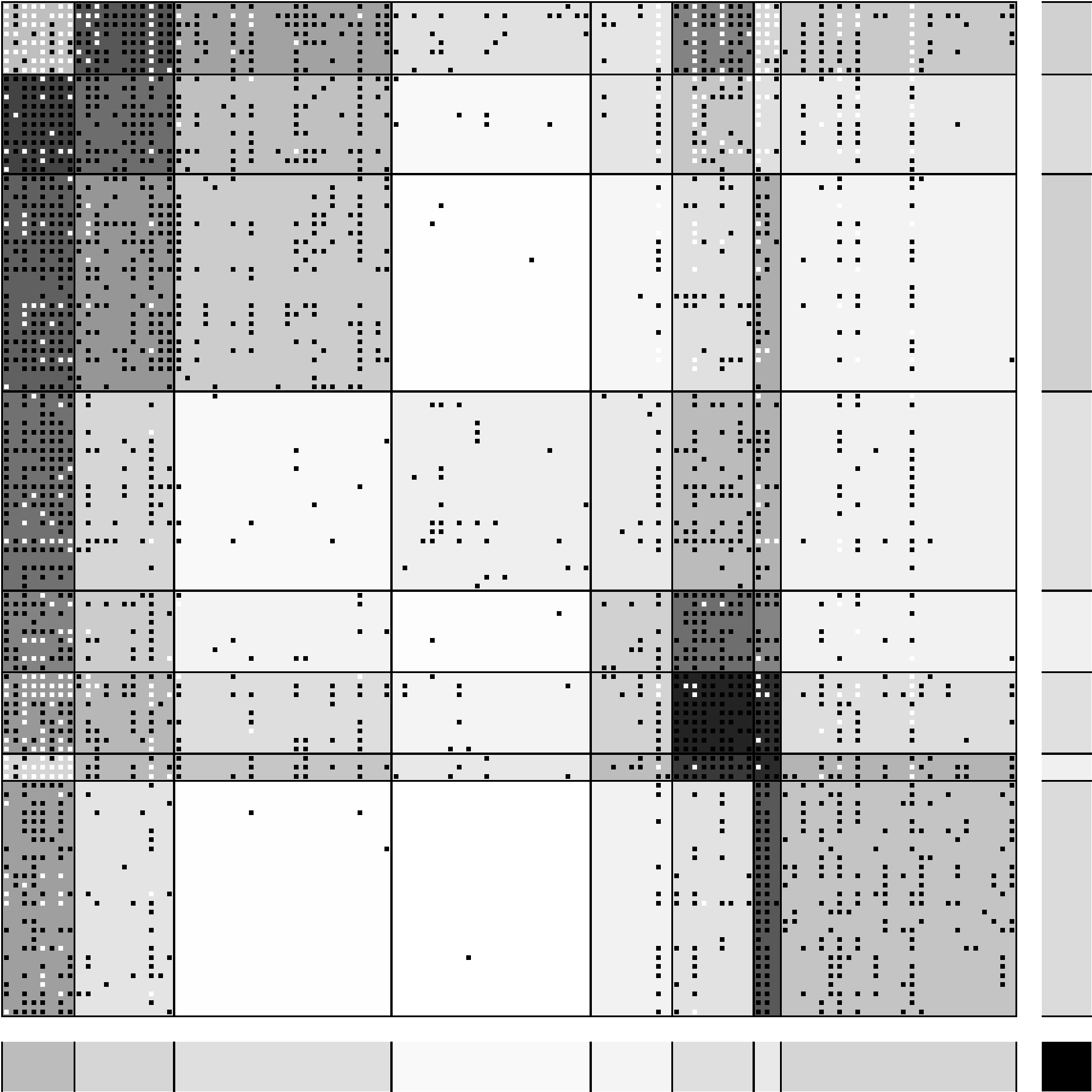} &
\includegraphics[height=3.25cm]{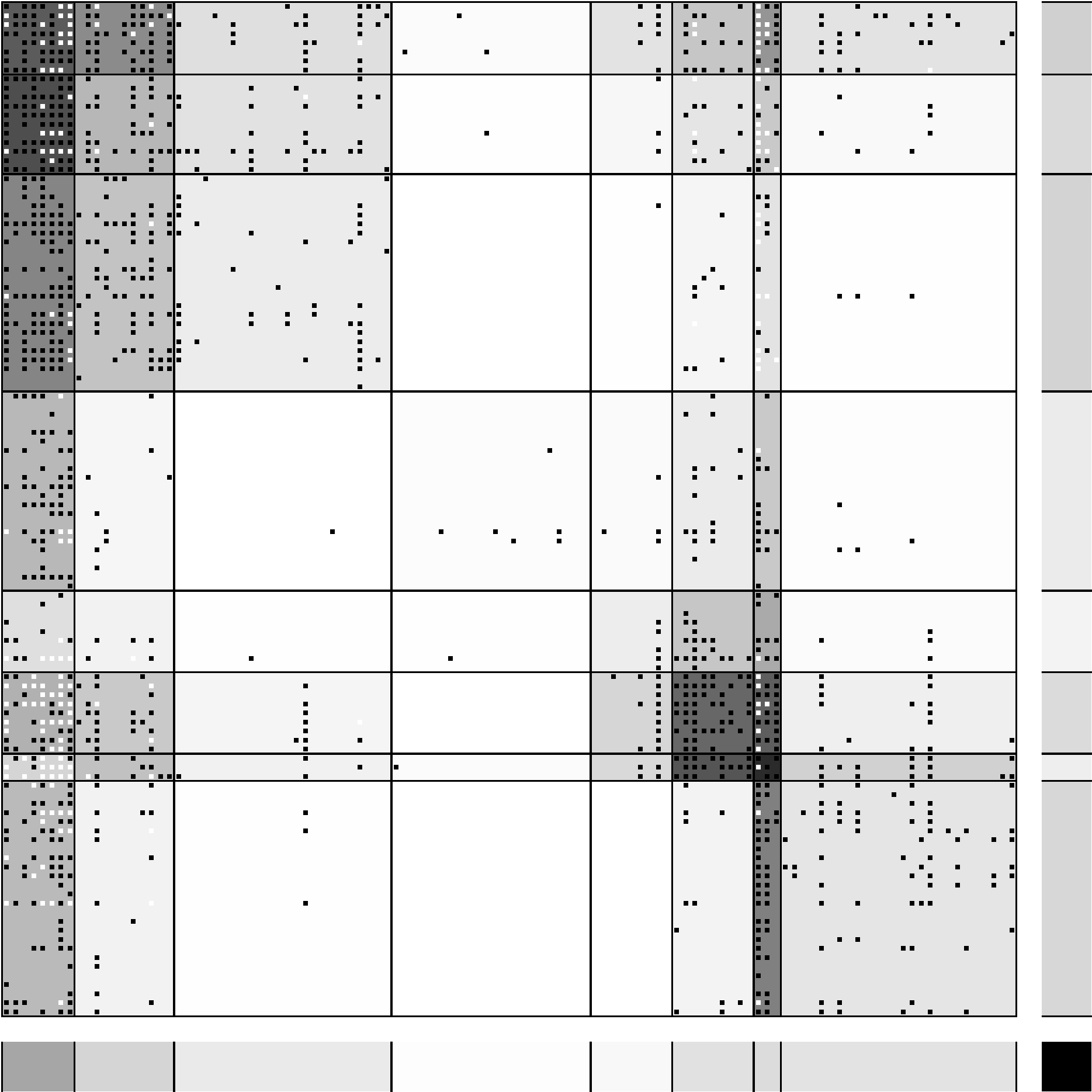} &
\includegraphics[height=3.25cm]{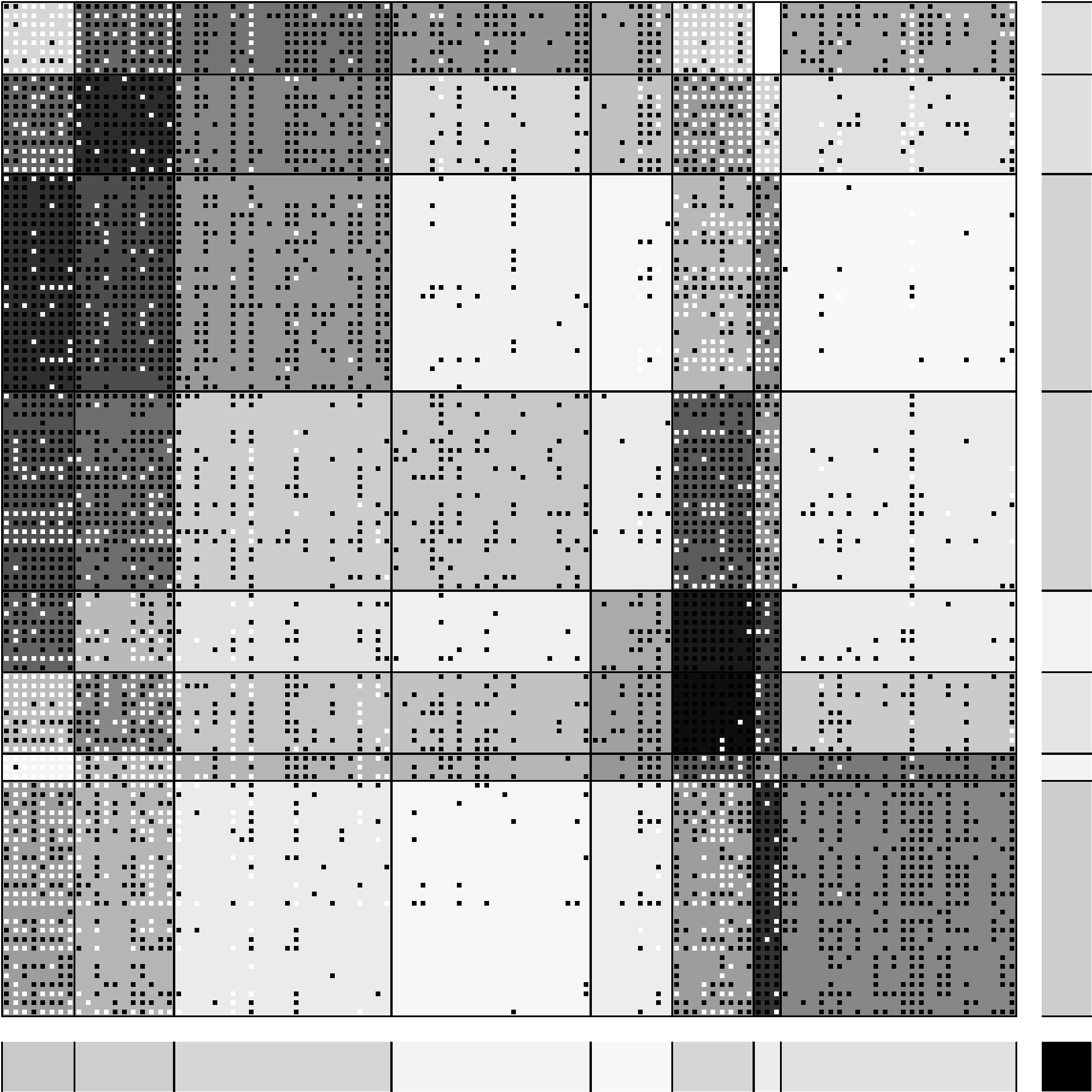} &
\includegraphics[height=3.25cm]{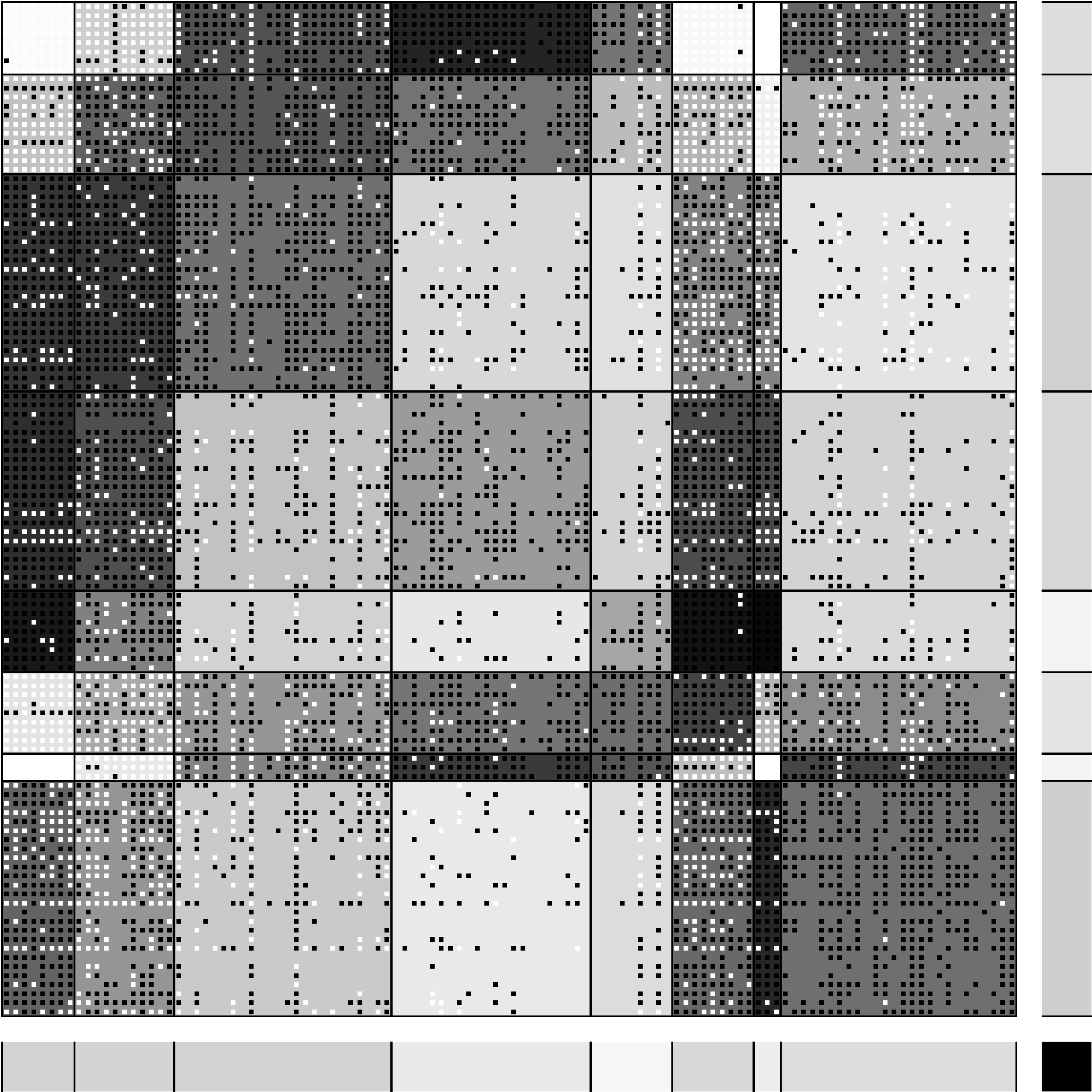} &
\includegraphics[height=3.25cm]{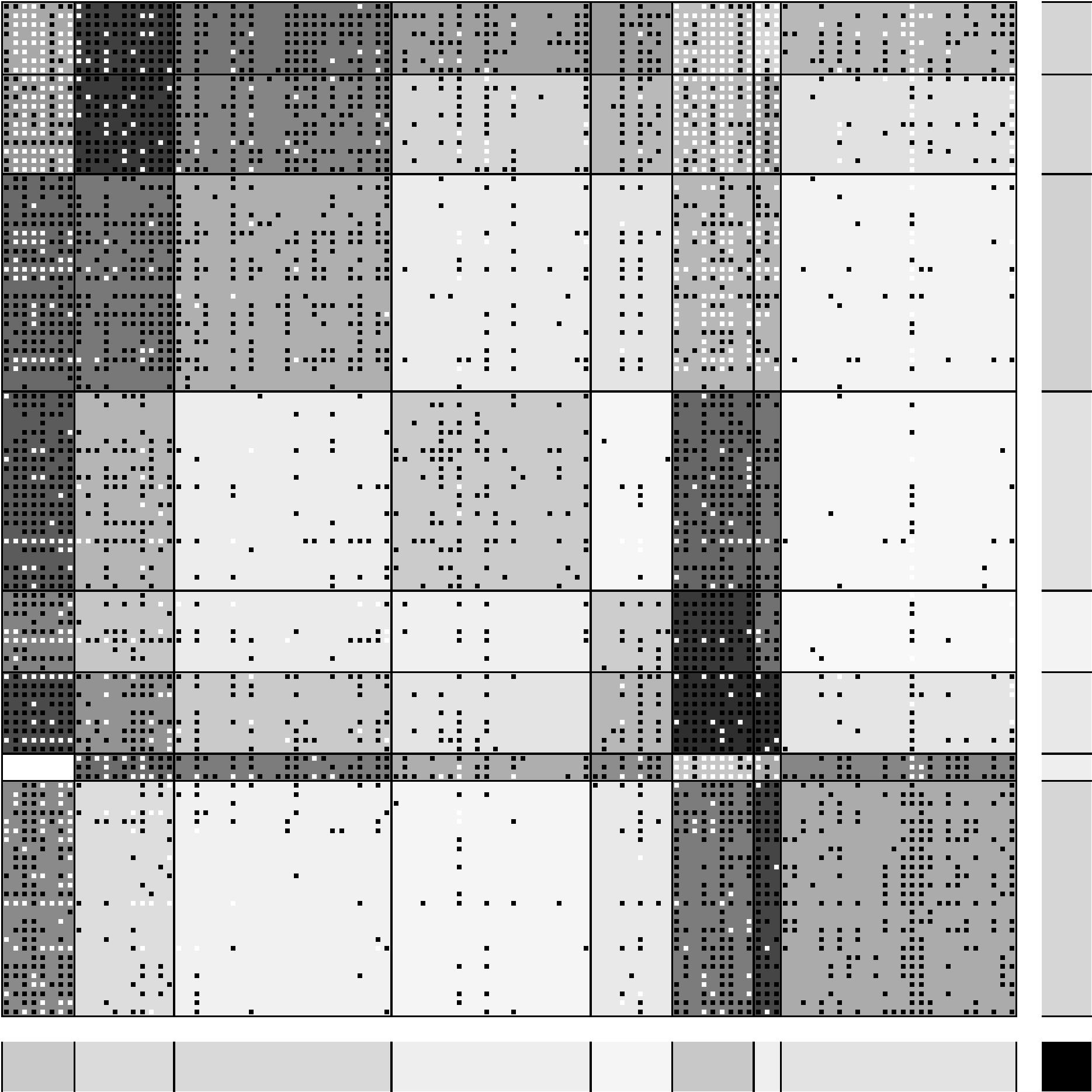} \\
meat \& meat prepartions & animal oil \& fats & paper \& paperboard & machinery & footwear\vspace{0.5cm}\\

\multicolumn{5}{r}{
\includegraphics[height=6cm]{q9_BW_Diamonds}
\includegraphics[height=1.2cm]{Diamonds}
}\\
\includegraphics[height=3.25cm]{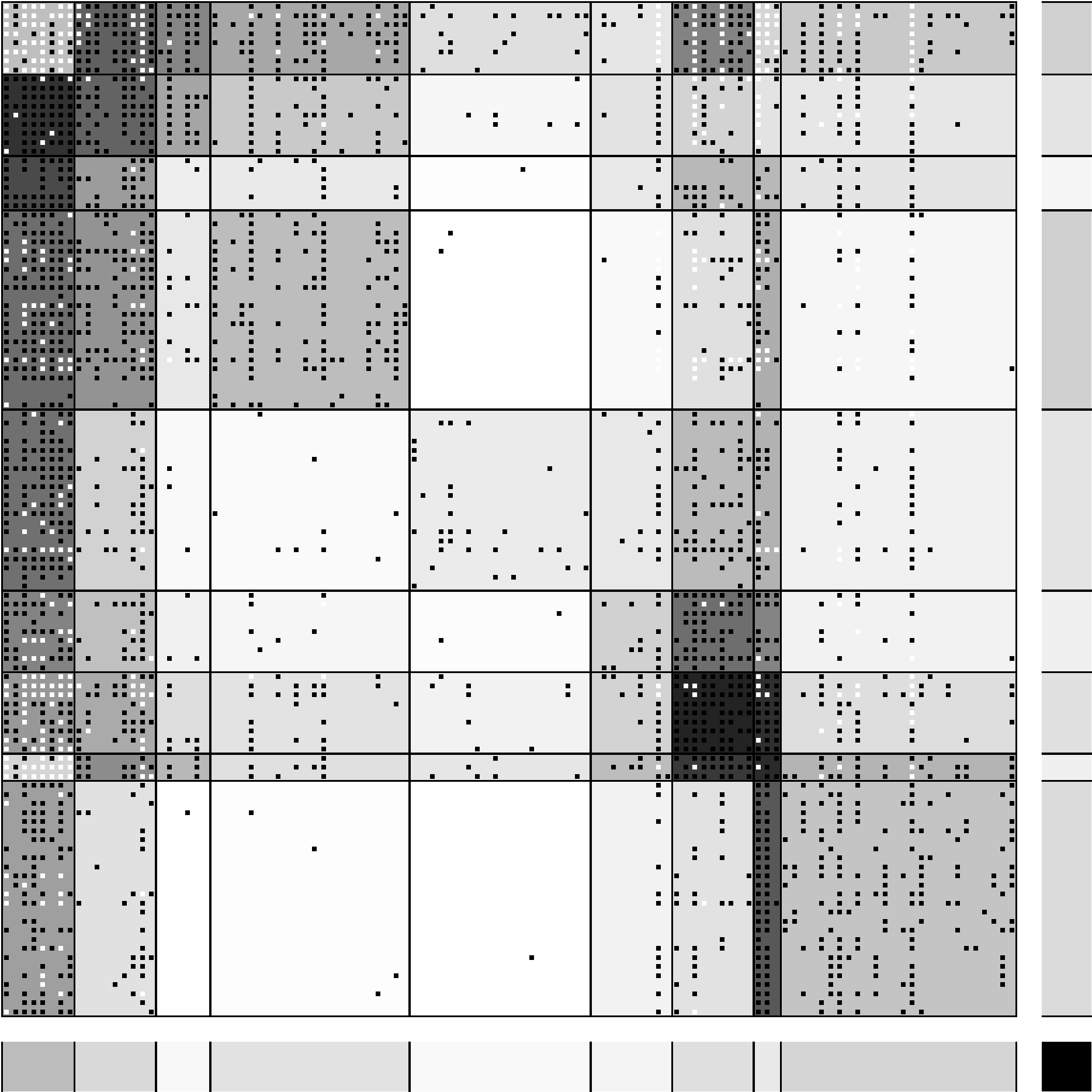} &
\includegraphics[height=3.25cm]{c41_q9} &
\includegraphics[height=3.25cm]{c64_q9} &
\includegraphics[height=3.25cm]{c71_q9} &
\includegraphics[height=3.25cm]{c85_q9} \\

\end{tabular}

\caption{Consensus image graphs and matrix plots for the 5 commodities studied at $q=8$ to $q=9$ roles. Triangle labels indicate commodity and direction of the flow of goods. Labels and matrix plots from left to right: meat and meat preparations, animal oil and fat, paper and paper board, machinery, footwear. Each of these is representative of one of the five factor clusters of
55 matrices for all major international trade commodities \cite{SmithNemeth,SmithWhite}. Unlabeled links carry all five commodities in both directions. Blocks indexed from left to right and top to bottom. Side and bottom bars encode the marginal fraction of import and export of the total traded volume for each block in gray scale, respectively. Black dots indicate trade greater than expected from the marginals for pairs of countries, white dots smaller than expected. Background shading of blocks corresponds to density of black dots in block. See Table 1 
 for individual countries grouped in blocks.}
 \label{SuppRBM3}
\end{figure*}

\clearpage

\noindent An advantage of marginal density blockmodeling developed here is that as the number of roles increases, their memberships may merge as well as split. Successive partitions are not always subdivisions forming hierarchical clusters, although there is a strong tendency for that to occur. The five rectangles enclosing pairs of roles in Figure 8 show where subdivisions tend to be hierarchical. In each case, however, some other countries also join the new sub-roles, as, for example, when the less developed periphery of the two-role model splits into two sub-roles that are also joined by some countries from the core.
\begin{figure*}[h]
\begin{center}
\includegraphics[height=15cm]{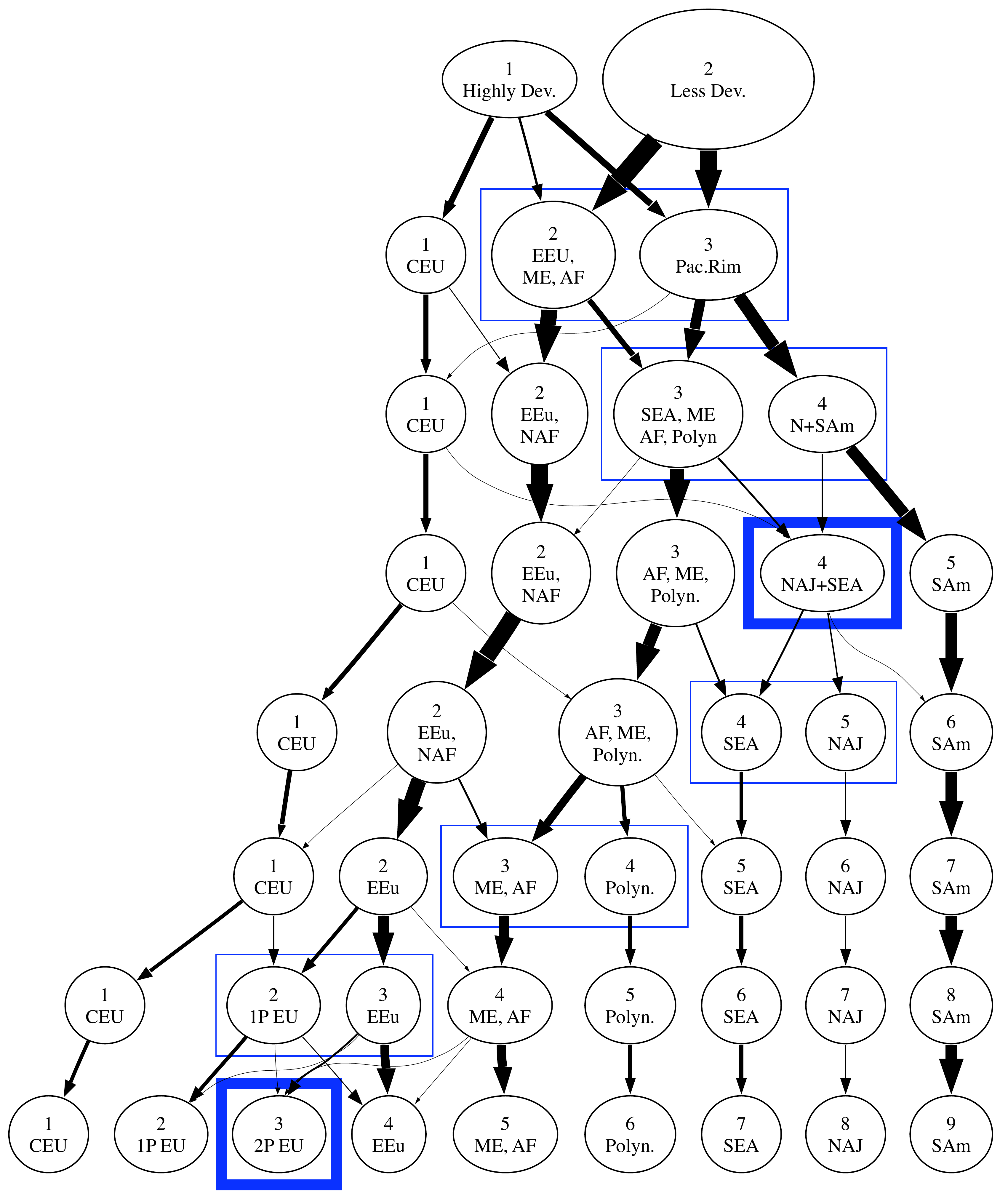}
\end{center}
\caption{ Splitting and merging diagram of the assortment of countries into roles as the number of roles increases. The width of the arrows is proportional to the number of countries that pass from role to role as the number of classes in increased by one. Rectangles indicate the major split  on each level and squares show new roles formed from overlap or merging. See Table 1 
 for the individual countries in each role at each level. The compact layout of this splitting/merging diagram show how splits tend to distribute countries to smaller blocks that are adjacent in the partition order. This suggested the compact order of the blocks in Table 1 
and Figures \ref{NineRoleBMs}, \ref{SuppRBM1},  \ref{SuppRBM2} and \ref{SuppRBM3}. The only three exceptions to compactness are China's realignment to block 1 at level 3 and back to block 4 at level 4, and Saudi Arabia's realignment to block 3 at level 5. As already noted in the matrix plots and image graphs, differentiation first happens around the Pacific and then in Europe, Africa and the Middle East. Labels are CEU: Central Europe, EEU: Eastern Europe, ME: Middle East, AF: Africa, NAF: Northern Africa, SEA: South East Asia, Polyn: Polynesia, N+SAm: North and South and Middle America, SAm: South and Middle America, NAJ: North America and Japan, 1P EU: 1st periphery EU, 2nd Periphery EU.}
\label{SupportSplitting}
\end{figure*}

\end{document}